\documentclass{aa}

\pdfoutput=1

\usepackage{graphicx}
\usepackage{subfigure}
\usepackage{txfonts}

\usepackage{natbib}

\usepackage{amsmath}
\usepackage{amsbsy}

\begin{document}

\title{Simulations of ram-pressure stripping in galaxy--cluster
  interactions}

\author{Dominik Steinhauser\inst{1}, Sabine Schindler\inst{1}, and
  Volker Springel\inst{2,3}}

\institute{Institut f\"{u}r Astro- und Teilchenphysik Universit\"{a}t
  Innsbruck, Technikerstrasse 25/8, 6020 Innsbruck, Austria \and
  Heidelberger Institut f\"{u}r Theoretische Studien,
  Schloss-Wolfsbrunnenweg 35, 69118 Heidelberg, Germany \and Zentrum
  f\"ur Astronomie der Universit\"at Heidelberg, Astronomisches
  Recheninstitut, M\"{o}nchhofstr.~12-14, 69120 Heidelberg}
 

 \titlerunning{Ram-pressure stripping in galaxy--cluster interactions}

\authorrunning{D.~Steinhauser, S.~Schindler, V.~Springel}


  \abstract
  { Observationally, the quenching of star-forming galaxies appears to
    depend both on their mass and environment. The exact cause of the
    environmental dependence is still poorly understood, yet
    semi-analytic models (SAMs) of galaxy formation need to
    parameterise it to reproduce observations of galaxy properties.  }
  {In this work, we use hydrodynamical simulations to investigate the
    quenching of disk galaxies through ram-pressure stripping (RPS) as
    they fall into galaxy clusters with the goal of characterising the
    importance of this effect for the reddening of disk galaxies. In
    particular, we compare our findings for the mass loss and evolution of
    the star formation rate in our simulations with
    prescriptions commonly employed in SAMs. We also analyse 
    the gaseous wake of the galaxy, focusing on gas mixing and  metal enrichment
    of the intracluster medium (ICM).}
  {Our set-up employs a live model of a galaxy cluster that interacts
    with infalling disk galaxies on different orbits. We use the
    moving-mesh code AREPO, augmented with a special refinement
    strategy to yield high resolution around the galaxy on
    its way through the cluster in a computationally efficient
    way. Cooling, star formation, and stellar feedback are included
    according to a simple sub-resolution model. Stellar light maps and
    the evolution of galaxy colours are computed with the stellar
    synthesis code FSPS  to draw conclusions about quenching
    timescales of our model galaxies.
    }
  {We find that the stripping models employed in current SAMs often
    differ substantially from our direct simulations. In most cases,
    the actual stripping radius of the simulated disk galaxies is
    larger than assumed in the SAMs, corresponding to an
    over prediction of the mass loss in SAMs. As long as the disk
    is not completely stripped in peaks of RPS during pericentre
    passage, some gas that remains bound to the galaxies is
    redistributed to the outer parts of disks as soon as the ram
    pressure becomes weaker again, an effect that is not 
    captured in simplified treatements of RPS.
    Star formation in our
    model galaxies is quenched mainly because the hot gas halo is
    stripped, depriving the galaxy of its gas supply. 
    The cold gas disk is only stripped completely in extreme cases, leading to full
    quenching and significant reddening on a very short timescale.  
    Depending on the inclination angle, this can light up a
    galaxy for a few hundred Myrs until all of the gas is stripped or
    consumed and star formation drops to almost zero, suggesting a
    typical quenching timescale of $\sim$200 Myr. 
    On the other hand, galaxies experiencing only mild ram pressure
    actually show an enhanced star formation rate that is consistent with 
    observations.
    Stripped gas in the
    wake is mixed efficiently with intracluster gas already a few
    tens of kpc behind the disk, and this gas is free of residual star
    formation.}
   {}
  
   \keywords{galaxies:clusters:general - galaxies:evolution -
     galaxies:interactions - galaxies:star formation -
     methods:numerical}

   \maketitle


\section{Introduction}

Galaxy surveys have shown that the quenching of disk galaxies mainly
depends on their mass and environment \citep{Peng2010,
  Schawinski2014}. However, the question of the physical mechanisms
behind quenching still persists, and especially the environmental
dependence, which also gives rise to phenomena such as `galaxy
conformity' \citep{Weinmann2006, Kauffmann2013}, is not well
understood. Since the density of the environment seems to play an
important role, and the merger cross section is small in galaxy
clusters owing to the high relative velocity dispersion, classic
ram-pressure stripping \citep[RPS, e.g.][]{Gunn1972}, the removal and possible compression
of parts of the gas disk due to interaction with the intracluster
medium (ICM), remains one of the main candidates responsible 
for environmental quenching of disk
galaxies. Indirect evidence for RPS in action are asymmetries and
truncated radial density profiles \citep[e.g.][]{Merluzzi2013,
    Chung2009} observed in many systems, which cannot be easily
reproduced by pure `starvation' scenarios (the removal of the
  extended gas reservoir that refuels the disk with gas available for
  star formation (SF); e.g. \citealt{Larson1980}).
  Furthermore,
starvation cannot modify the colours of galaxies on sufficiently short
timescales, unlike RPS \citep{Boselli2009}.

It hence is plausible that the evolution of a galaxy strongly depends on
the strength of ram pressure (RP) it experiences.  Observations
\citep[e.g.][]{Boselli1997, Vollmer2001, Scott2010} often suggest only
rather mild RPS acting on galaxies with atomic
hydrogen only partly removed and slightly displaced from the stellar
disk. Correspondingly, SF is slowly quenched and only
slightly enhanced at the compressed interface between the intracluster
and intrastellar media.  High resolution observations
\citep[e.g.][]{Abramson2014} have shown that dense gas clouds can stay
in place even beyond the stripping radius, whereas the surrounding,
diffuse and less dense gas is stripped.  However, those clouds
continue to form stars at a very low rate, contributing at most a few
percent of the pre-stripping SFR, meaning that SF
effectively ceases as soon as the gas with low density
is stripped.

On the other hand, in very massive clusters (with
$\rho\sim10^{-24}\;\mathrm{g}\,\mathrm{cm}^{-3}$ and galaxy velocities
$v>1000\;\mathrm{km}\,\mathrm{s}^{-1}$), extreme RP values
can occur. In such cases, galaxies are significantly deficient
in HI and have asymmetric morphologies
  \citep[e.g.][]{Boselli2006, Fumagalli2014, McPartland2016}. Those
  galaxies are sometimes increasing their SF, becoming
  temporarily brighter than even the BCG of the cluster
  \citep[e.g.][]{Ebeling2014}. Although this is only expected to occur
rarely and only in very massive clusters, several such cases have
been discovered \citep[e.g.][]{Owen2006, Cortese2007, Owers2012,
Ebeling2014}. This indicates that shock compression of the
interstellar medium (ISM) may induce starbursts.  Also, the
  stripped gas can form stars in a knot-like structure in the wake of
  the galaxy, hence a so-called ``fireball'' galaxy is formed. A few
  prominent examples of such fireball galaxies have been observed
  \citep[e.g.][]{Kenney2014, Yoshida2008, Hester2010, Yagi2010}. They
all show bright star-forming knots in the stripped wake of the galaxy.
However, detailed observations of the gaseous tails
\citep{Boissier2012} show that the SF efficiency in the
stripped tail is ten times lower than in the disk itself and, in most
cases, close to the detection limit. This is broadly consistent with
the expectation that stripped gas should not become dense enough to
form stars, but it is not clear whether the higher density and
pressure encountered in more massive clusters could change this and
provide enough compression to stimulate SF in the wake
\citep[e.g.][]{Kapferer2009, Tonnesen2012}.

Apart from being interesting in its own right, RPS
is likely to be an environmental effect that is very important for galaxy
formation. Semi-analytical models (SAM) of galaxy formation
\citep[e.g.][]{DeLucia2004, Guo2011, Benson2012} need to parameterise
many physical processes to correctly reproduce basic galaxy properties
such as the observed stellar mass function. One of these processes is
RPS, otherwise the colour distribution of galaxies
in clusters or the satellite abundance in Milky Way-sized galaxies
cannot be explained. Star formation in SAMs is usually implemented by
transforming the cold gas on a characteristic timescale into
stars. The timescale is effectively determined by translating a gas
surface density into a SF density, guided by the
phenomenological relation of \citet{Kennicutt1998}. Parameterising RPS 
is however much more uncertain. Often, only the
stripping of the hot halo gas is taken into account
\citep[e.g.][]{Guo2011}, but implementations that consider the
stripping of gas disks also exist \citep[e.g.][]{Tecce2010}.  In most
cases, these prescriptions, therefore, rely on the standard \citet{Gunn1972}
criterion or slight variations for calculating the
stripping radius and corresponding removal of gas from the disk.

In this study, we carry out hydrodynamical simulations of galaxies
undergoing RPS in a realistic cluster environment,
as well as control simulations of the same galaxies in isolation.
We do this, in particular, to compare the stripping radius
  obtained from our simulations with the theoretical prescriptions
  used in SAMs.  These models are commonly producing a fraction of 
  red satellite galaxies that is too high, possibly related to an
  overestimation of environmental effects and, hence, too rapid
  quenching \citep[e.g.][]{Kimm2009, Guo2013, Wang2014}.  Here, we
  provide an important test and possible improvements for gas removal
  and quenching models. In this context, we investigate how fast
galaxies, depending on their mass as well, are quenched purely by
RP.  Using the colour evolution of our model galaxies, we
determine quenching timescales that can be compared to observational
constraints.  We are also interested in the question of whether the
observed cases of ``fireball'' galaxies undergoing extreme RPS can be
reproduced. Finally, we analyse the stripped, turbulent gaseous wake
of the galaxies. In principle, the distribution of metals in the
stripped wake should provide clues about the metal enrichment of the
ICM and, in particular, how fast gas is mixed
with the hot and thin ICM and hence if it is still possible to form
stars there.

Previous work along this line has been largely carried out with
wind tunnel-like set-ups \citep[e.g.][]{Roediger2006, Roediger2006a,
  Kronberger2008, Bekki2009, Tonnesen2010,
  Tonnesen2012, Ruszkowski2014}, but a number of works also addressed
the more realistic problem of following a galaxy on an orbit through a
galaxy cluster \citep[e.g.][]{Vollmer2001a, Roediger2007, Jachym2007, 
  Roediger2008, Tonnesen2008, Hess2012}. Compared to these earlier works, our
simulations improve the self-consistency of the simulations (for
example by using a live cluster model instead of an analytic
potential), the hydrodynamical technique (with a moving mesh and a
refinement criterion tailored to the problem at hand), and/or the
numerical resolution.

This paper is structured as follows. In Sect.~\ref{secintro}, we
introduce our numerical methodology and describe our different
numerical simulations and their parameters. In
  Sect.~\ref{sec:results}, we present and interpret our results,
  particularly the evolution of the model galaxies in isolation
  (Sect.~\ref{sec:results_isolation}), stripping of gas and metal
  enrichment of the ICM (Sect.~\ref{sec:rps}), SFR and colour
  evolution in different environments (Sects.~\ref{sec:sfr} and
  \ref{sec:colour_evolution}) and a comparison with models used in
  SAMs (Sect.~\ref{sec:samcomparison}).
In Sect.~\ref{sec:app_res_study}, we provide tests of numerical
convergence, and in Sect.~\ref{secconclusions} we give a discussion
and summary of our conclusions.

\section{Numerical methodology} \label{secintro}

In the following, we present our numerical set-up used in this study.
First, we introduce the model for isolated galaxies and we detail how
the background galaxy cluster and its intracluster medium is represented.
Furthermore, we choose merging orbits for the galaxies in the cluster
and finally discuss the numerical code and simulation set-up used
for this study.

\subsection{Isolated galaxy and cluster models}
\label{sec:galaxy_cluster_models}

For our RPS simulations, we construct composite
models of disk galaxies that are approximately in equilibrium.  These
model galaxies are then either evolved in isolation or dropped into
our galaxy cluster model. The two types of simulations allow us to compare
cluster and field galaxies, i.e.~galaxies without any interaction evolving 
in isolation and galaxies that are influenced by the potential of the cluster 
and undergoing RPS.

The model galaxies are constructed following the approach described in
\citet{Springel1999} and \citet{Springel2005a} with structural properties
derived from the theoretical work of \citet{Mo1998}.  The dark-matter (DM)
mass distribution is modelled with a \citet{Hernquist1990} profile,
\begin{equation}
  \rho_\mathrm{DM}\left( r \right) =  \frac{M_\mathrm{DM}}{2\pi} \frac{a}{r \left( r+a \right)^3},
  \label{eq:hernquist}
\end{equation}
where the total mass $M_\mathrm{DM}$ is related to the virial velocity
$v_{200}$ by
\begin{equation}
  M_\mathrm{DM} = \frac{v_{200}^3}{10GH_0}.
  \label{eq:mass_v200}
\end{equation}
The chosen scale length $a$ of the Hernquist profile can be set by
demanding that the corresponding NFW profile \citep{Navarro1996} of a
halo with scale radius $r_s$ and the given circular velocity has the
same inner density profile. For the scale radius, we take a value
derived from the expected concentration index $c = r_{200}/r_s$ for
halos of this mass based on cosmological N-body simulations. The
advantage of using a Hernquist instead of a NFW profile is that no
sharp edge for the halo needs to be assumed and that the total halo mass
converges.

The baryonic matter of the model galaxy is represented by a stellar
bulge, an exponential stellar disk, and an exponential gas disk as
described in \citet{Springel2005a}. The surface density profile of the
disk is
\begin{equation}
  \Sigma_{\mathrm{gas}/\star} = \frac{M_{\mathrm{gas}/\star}}{2\pi r_0^2}\exp \left(-r/r_0 \right),
  \label{eq:surf_dens_disk}
\end{equation}
\noindent 
where $r_0$ is the common scale length of both stellar and gaseous
components. The bulge is modelled similar to the DM halo with a
Hernquist profile.

The scale length of the disk is related to its angular momentum,
whereas the bulge scale length is set to a fraction of the disk scale
length. The vertical structure of the stellar disk is modelled as an
isothermal sheet with radially constant scale length. For the gas
disk, however, the vertical structure is determined by self-gravity
and ISM pressure, which is modelled with a sub-resolution
effective equation of state. Finally, the velocity structures of the
DM halo and bulge are approximated by locally triaxial Gaussians
with dispersions computed with the Jeans equations. More sophisticated
methods to compute the velocity distribution function
\citep[e.g.][]{Yurin2014} could also be used, but the quality of the
initial equilibrium reached here with the Jeans' approximation is good
enough for the purposes of this study.

Additionally, we include a hot gas halo in the galaxy model similar to
\citet{Moster2011}. Motivated by observations of hot gas in clusters, we
adopt a $\beta$ profile \citep[e.g.][]{Eke1998} for the hot gas halo
of the galaxy with the radial density profile
\begin{equation}
 \rho\left( r \right) = \rho_0 \left[ 1 + \left( \frac{r}{r_c} \right)^2 \right]^{-\frac{3}{2}\beta}.
\end{equation}
\noindent
As in \citet{Moster2011}, we use $\beta={2}/{3}$ and $r_c = 0.22\, a$.
The central density is then defined by the adopted total mass in the
hot gas halo.  Assuming hydrostatic equilibrium, the radial
temperature profile is given by
\begin{equation}
 T\left( r \right) = \frac{\mu
   m_p}{k_\mathrm{B}\,\rho\left(r\right)}\int_r^\infty
 \rho\left(r'\right)\frac{GM\left(r'\right)}{r'^2}{\rm d}r',
\end{equation}
\noindent
where $\rho\left(r\right)$ is the density of the hot gas halo given
above, $M\left(r\right)$ the total cumulative mass within a radius
$r$, $\mu$ the mean molecular weight, $m_p$ the proton mass, and $G$ and
$k_\mathrm{B}$ denoting the gravitational and Boltzmann's constant,
respectively. Furthermore, the hot gas halo has a rotational velocity
around the spin axis of the disk. Assuming that the hot gas halo has
the same specific angular momentum $j={J}/{M}$ as the DM
halo, the total angular momentum of the hot gas halo is given by
$J_\mathrm{gas}=({J_\mathrm{DM}} / {M_\mathrm{DM}}) M_\mathrm{gas}$,
which is distributed into shells of the hot gas halo. Assuming these
shells are in solid body rotation, a rotational velocity can be
assigned.

In order to investigate the effects of RPS and
quenching on galaxies of different mass, we are using two galaxies that are different in size
with properties given in Table~\ref{tab:galaxy}.  The
first model galaxy, denoted G1, has a total mass of
$10^{12}h^{-1}\mathrm{M}_\odot$ and contains a bulge. Moreover, we use
three different hot gas halos with this model galaxy, using the same
(G1a), half (G1b) or one percent (G1c) of the disk's baryonic mass for
the hot gas halo.  The latter configuration with very little hot gas is
meant to simulate a case with effectively no gaseous halo while
concurrently avoiding technical problems that would occur for strictly
empty space in the moving-mesh code, which always needs to tessellate
the whole simulation domain.

The second model galaxy, G2, has a third
($3\times10^{11}h^{-1}\mathrm{M}_\odot$) of the total mass of G1, 
a hot gas halo that contains 10\% of the baryonic mass of the disk 
and does not have a bulge. In Table~\ref{tab:galaxy},
we list the chosen free parameters of this galaxy model and its corresponding
properties. We note that in our cluster simulations
the hot halo of the galaxies is truncated at a radius where its density
falls below the background density of the ICM, reducing the mass in
this component slightly.

\begin{table}[tb]
  \centering
  \caption{Initial properties of the two primary galaxy models G1 and G2 used in this study.}
  \begin{tabular}{ll|cc}
    \hline \hline
     & & G1a (G1b, G1c) & G2 \\
    \hline \hline
    $c$ & & 9 & 10\\
    $v_{200}$ & $\Big[\mathrm{km}\,\mathrm{s}^{-1}\Big]$ & 170 & 110 \\
    $\lambda$ & & 0.033 & 0.04\\
    $m_d$ & & 0.041 & 0.041 \\
    $m_b$ & & 0.01367 & 0 \\
    $f_{\mathrm{gas}}$ & & 0.35 & 0.35 \\
    $z_0$ & & 0.2 & 0.2 \\
    $M_\mathrm{DM}$ & $\Big[ h^{-1}\,\mathrm{M}_\odot \Big]$& $1.02\times10^{12}$ & $2.95\times10^{11}$ \\
    $M_\mathrm{gas, d}$ & $\Big[ h^{-1}\,\mathrm{M}_\odot \Big]$ & $1.64\times10^{10}$ & $4.44\times10^{9}$ \\
    $M_\mathrm{\star, d}$ & $\Big[ h^{-1}\,\mathrm{M}_\odot \Big]$ & $3.04\times10^{10}$ & $8.25\times10^{9}$ \\ 
    $M_\mathrm{\star, b}$ & $\Big[ h^{-1}\,\mathrm{M}_\odot \Big]$ & $1.56\times10^{10}$ & $0$ \\ 
    $M_\mathrm{gas, h}$ & $\Big[ h^{-1}\,\mathrm{M}_\odot \Big]$ & $6.24(3.12, 0.0625)\times10^{10}$ & $1.26\times10^{9}$ \\ 
    $r_0$ & $\Big[ h^{-1}\,\mathrm{kpc} \Big]$ & 2.60943 & 2.2063 \\
    $a_b$ & ${\Big[ h^{-1}\,\mathrm{kpc} \Big]}$ & 0.5219 & - \\
    \hline \hline
  \end{tabular}
  \vspace*{0.2cm}
  \tablefoot{We specify the principal structure of these models (in particular total mass and disk)  with choices for the concentration parameter
    $c$,  rotation velocity $v_{200}$  at the virial radius $r_{200}$, 
    and  spin parameter $\lambda$. The values $m_d$ and $m_b$ are the disk and bulge mass fractions, respectively, with $f_\mathrm{gas}$ specifying 
    the initial amount of gas in the disk and $z_0$ giving the disk height as a fraction of the disk scale length $r_0$
    (equal for both the stellar and gaseous disk). 
    The latter is determined by the spin parameter. The value ${a_b}$ is the scale factor of the Hernquist profile describing the bulge. 
    These parameters also determine other resulting properties of the corresponding model galaxy,
    in particular the masses of the different components, as listed in the table. 
    The amount $M_{\rm gas,h}$ of gas in the hot corona around the galaxies is chosen freely. Here we investigate three different choices (G1a, G1b, G1c) for galaxy model G1, and only one choice for G2.}
  \label{tab:galaxy}
\end{table}

\begin{table}[tb]
  \centering
  \caption{Basic properties of the three galaxy cluster models studied here.}
  \begin{tabular}{ll|c|c|c}
    \hline \hline
    model & & A & B & C \\ 
    \hline \hline
    $c$ & & 4.28 & 4.57 & 5.63 \\
    $v_{200}$ & $\Big[\mathrm{km}\,\mathrm{s}^{-1} \Big]$ & 1500 & 1200 & 600 \\
    $f_\mathrm{gas}$ &  & 0.14 & 0.10 & 0.07 \\
    $M_{\mathrm{tot}}$ & $\Big[ h^{-1}\,\mathrm{M}_\odot\Big]$ & $7.85\times10^{14}$ & $4\times10^{14}$ & $5\times10^{13}$ \\
    $M_\mathrm{ICM}$ & $\Big[ h^{-1}\,\mathrm{M}_\odot\Big]$ & $1.10\times10^{14}$ & $4\times10^{13}$ & $3.5\times10^{12}$ \\
    \hline \hline
  \end{tabular}
  \vspace*{0.2cm}
  \tablefoot{The concentration parameter $c$ and the circular velocity $v_{200}$ at the
    virial radius define the density structure with the gas 
    fraction $f_\mathrm{gas}$ set to representative observational values. The values $M_{\mathrm{tot}}$ and $M_\mathrm{ICM}$ give the resulting total and ICM mass, respectively.}
  \label{tab:cluster}
\end{table}

To simulate the behaviour of the galaxies while falling into a galaxy
cluster, we generate idealised galaxy cluster models in the following
way. We use a spherically symmetric DM and gas halo,
which both follow a Hernquist profile. As with the model galaxies, some
fraction of the mass in the halo is assigned to the gas that
represents the ICM.  We use three different
model galaxy clusters, as shown in Table~\ref{tab:cluster}.  The
selected cluster masses are somewhat arbitrary, but for the sake of
definiteness we refer to models A and B as high- and low-mass clusters
(comparable to Virgo and Coma cluster), respectively, and to model C
as a galaxy group.  

For all models, we select further properties according to
observational constraints and cosmological simulations.  We calculate
the concentration parameter from the concentration-mass relation as
measured in \citet{Neto2007} based on the Millennium simulation. The
virial velocity $v_{200}$ can be calculated using
Eq.~\ref{eq:mass_v200}. Finally, we adopt the baryon fraction
in the models as suggested by observations. Here we adopt the best fit
to different X-ray observations of galaxy groups and clusters from
\citet{Sun2012}.  The resulting parameters and properties for our
model clusters are given in Table~\ref{tab:cluster}.

\begin{figure}[tbp]
  \centering
  \subfigure[]{\includegraphics[width=\linewidth]{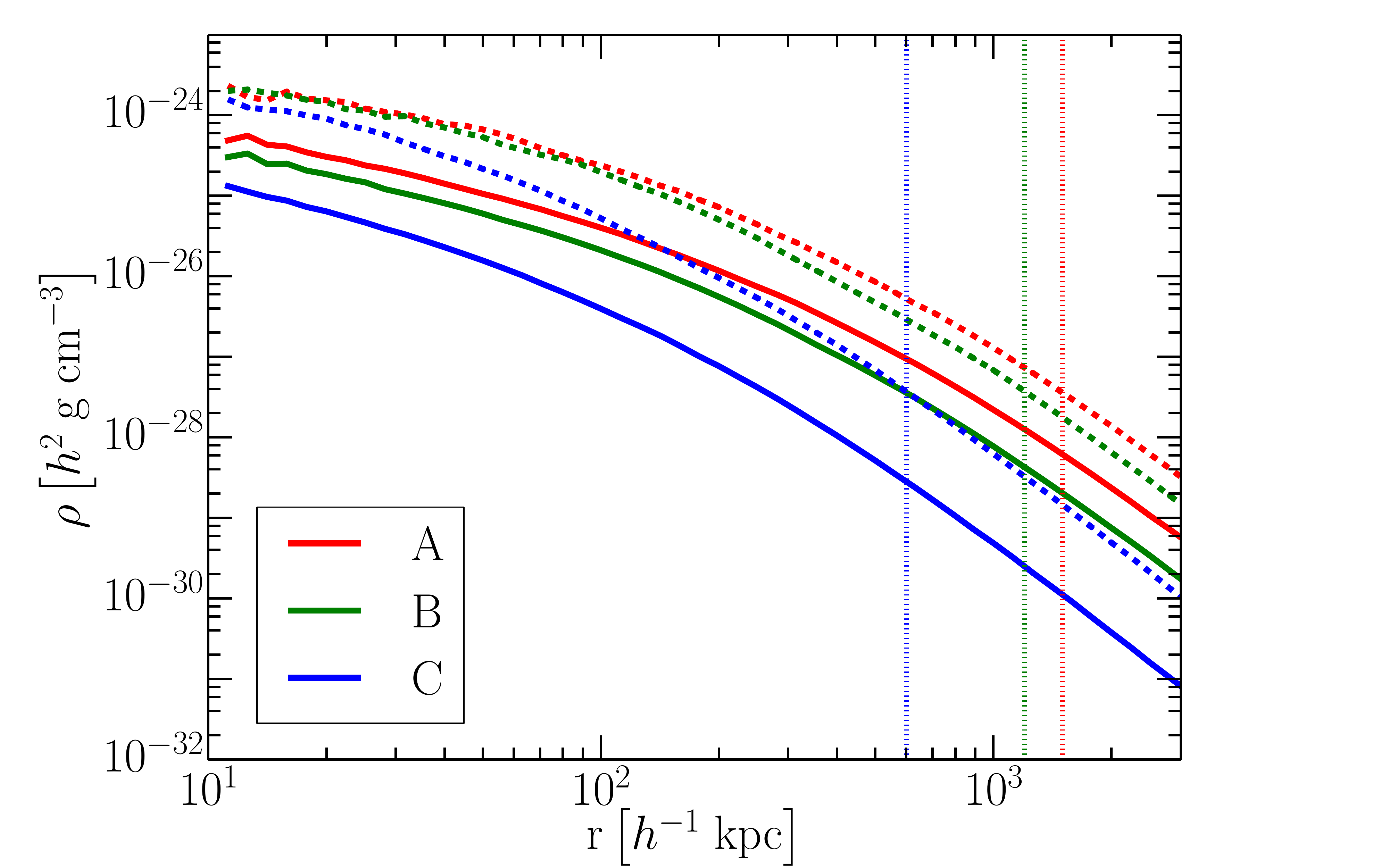}\label{fig:cluster_rad_density_a}}
  \subfigure[]{\includegraphics[width=\linewidth]{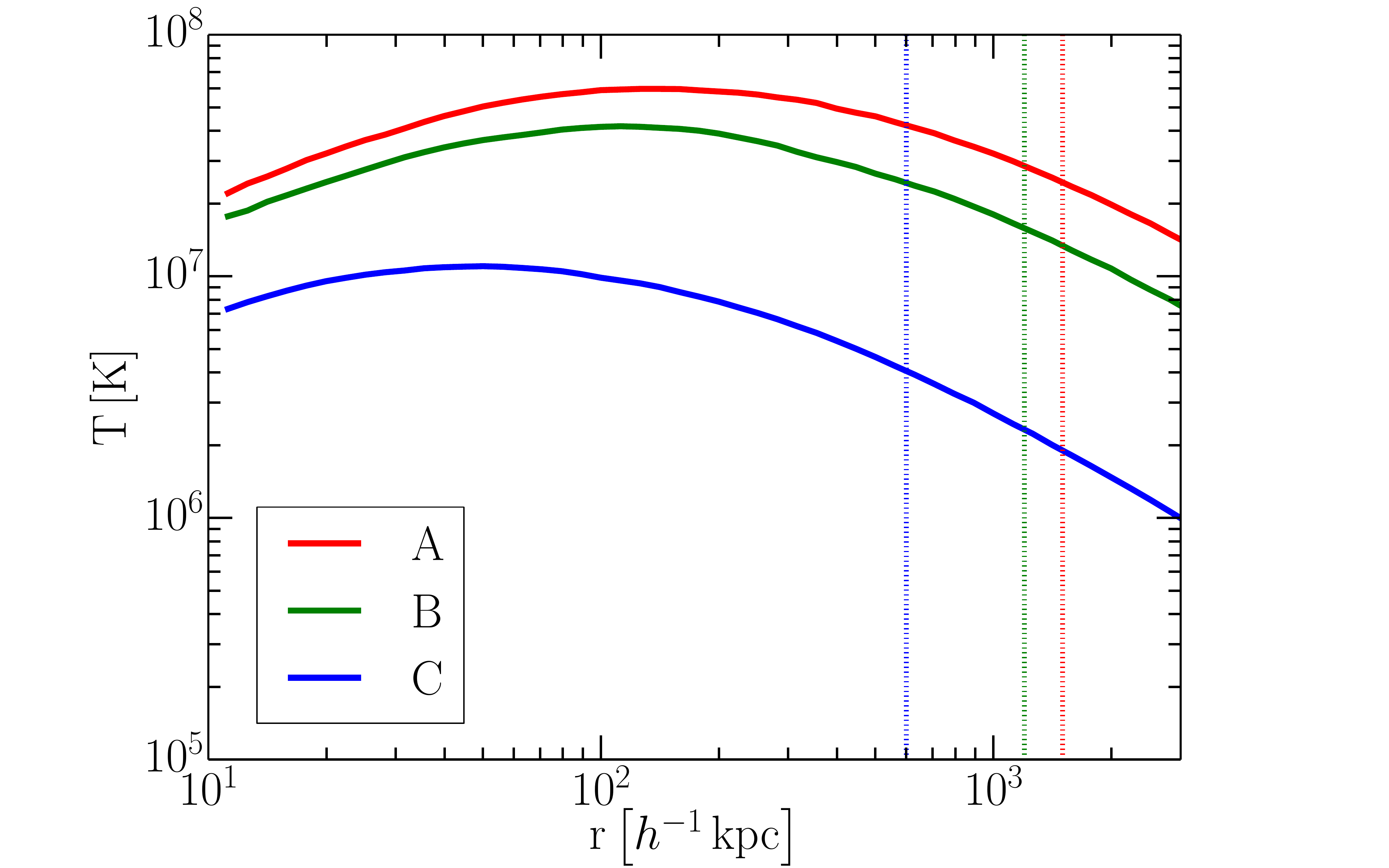}\label{fig:cluster_rad_density_b}}
  \caption{(a) Radial density profiles of the gas (solid lines) and
    DM (dashed lines) of the galaxy cluster models used in
    this study. The profiles are shown at the beginning of the galaxy
    interaction, after the cluster models were allowed to relax for
    $\sim 1\,{\rm Gyr}$.  Vertical dotted lines indicate the virial radius
    $r_{200}$ for clusters A, B, and C, from right to left. (b)
    Corresponding temperature profiles of the ICM.}
  \label{fig:cluster_rad_density}
\end{figure}

Because the initial approximation of the velocity distribution
function is not exact (especially in the centre), the constructed
cluster models are not in perfect equilibrium at the beginning. 
Already after a few Myrs of dynamical evolution, however, clusters relax to
a slightly softer inner density profile with a flatter gas distribution
in the centre and show no significant secular evolution thereafter.
The radial density profiles of DM (dashed line) and ICM gas
(solid line), and ICM temperature are shown in
Fig.~\ref{fig:cluster_rad_density} for our three different
clusters. We note that the flat entropy cores could be related to
numerical effects, such as artificial heating caused by softening
effects or N-body noise, as investigated by
\citet{Vazza2011}. However, the effect is limited to the very centre
of the clusters and does not affect our study of RPS,
as we have no trajectories of galaxies passing through the
innermost region.

\subsection{Merger orbits}
\label{sec:merger_orbits}

One of the model galaxies and one of the live galaxy clusters
described in the previous section are put on a parabolic merging
orbit.  The initial conditions for the orbits are set up the same way
as in \citet[see Fig.~8 therein for a sketch of the
configuration]{Duc2000}, and, in all of the simulations in this work, the orbital
angular momentum is oriented parallel to the $z$-axis. 
Instead of
using two galaxies, galaxy one in the binary collision set-up of
\citet{Duc2000} is replaced with the cluster model, without setting
any inclination, hence
$\left( \Phi_1, \phi_1\right) = \left(0,0\right)$.  The second galaxy
is the model galaxy orbiting in the cluster.  The different parameters
used for the distinct simulations are listed in Table~\ref{tab:orbits}.
The impact parameter $b$ corresponds to the minimum separation of the
cluster and galaxy if they were point masses and moved on
Keplerian orbits.  Different orbits of the galaxy through the cluster
and hence different RP scenarios can be obtained by altering
the impact parameter and the initial separation of galaxy and cluster
centre.

\begin{table}[tb]
  \centering
  \caption{Primary galaxy--cluster interaction simulations 
    performed in this work.}
  \begin{tabular}{c|ccccc}
    \hline \hline
    run & cluster & galaxy & $b$  & max v  & max $\rho$ ICM  \\
    label & & & $\left[ h^{-1}\,\mathrm{kpc} \right]$ & $\left[\mathrm{km}\,\mathrm{s}^{-1} \right]$ & $\left[h^2\,\mathrm{g}\,\mathrm{cm}^{-3} \right]$ \\
    \hline \hline

    S1 & A & G1a & 100 & 3151 & $6.4\times10^{-27}$ \\
    S2 & A inc & G1a & 100 & 3141 & $6.3\times10^{-27}$ \\
    S3 & B & G1a & 500 & 1719 & $1.5\times10^{-28}$ \\
    S4 & C & G1a & 100 & 1255 & $2.6\times10^{-27}$ \\
    
    S5 & A & G2 & 100 & 3173 & $6.4\times10^{-27}$ \\
    S6 & B & G2 & 500 & 1780 & $1.7\times10^{-28}$ \\
    S7 & C & G2 & 100 & 1265 & $2.7\times10^{-27}$ \\
    
    \hline \hline
  \end{tabular}
  \vspace*{0.2cm}
  \tablefoot{The initial impact parameter $b$ is chosen for trajectories 
  as if galaxy and cluster were point masses; the value actually obtained 
  in the self-consistent simulation (see Fig.~\ref{fig:orbits_a}) is different. 
    Run S2 is identical to run S1, except that the galaxy is rotated by $90^\circ$, 
    resulting in a different effective inclination throughout the orbital path 
    (see Fig. \ref{fig:orbits_c}).}
  \label{tab:orbits}
\end{table}

\begin{figure*}[tbp]
    \hspace{0cm}\subfigure[]{\includegraphics[width=0.3085\linewidth]{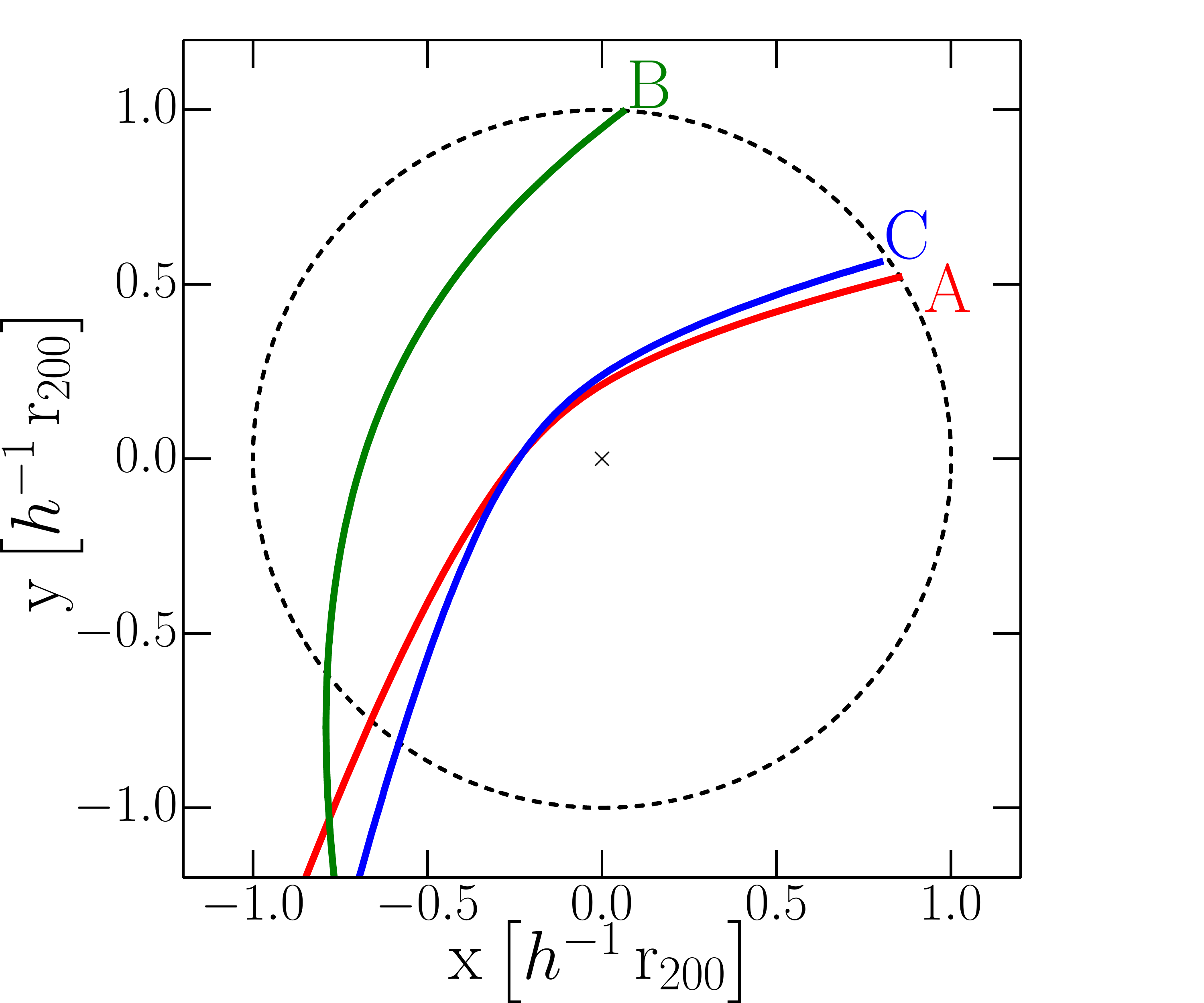} \label{fig:orbits_a}}%
    \hspace{-0cm}\subfigure[]{\includegraphics[width=0.36\linewidth]{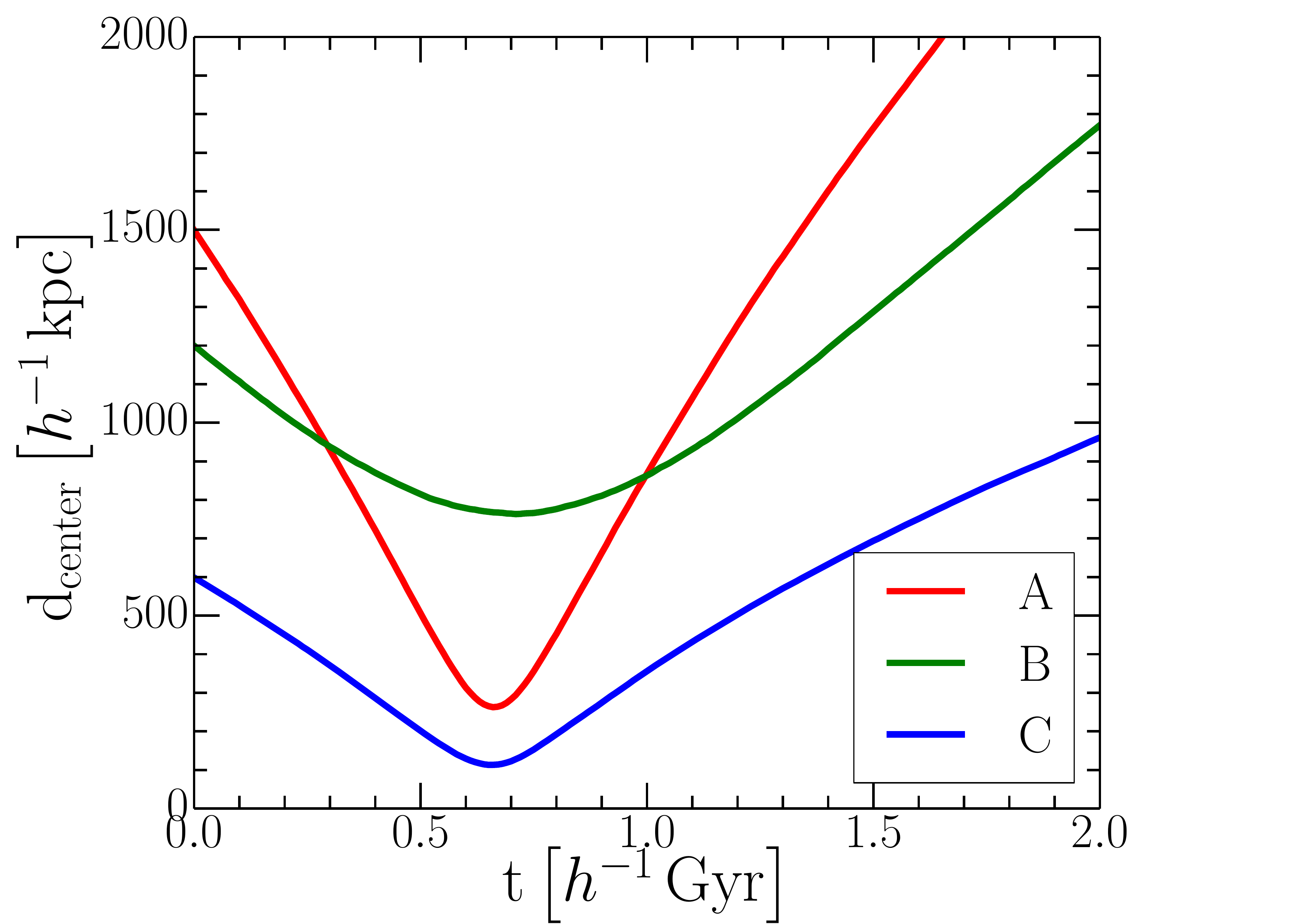} \label{fig:orbits_b}}%
    \hspace{-0.2cm}\subfigure[]{\includegraphics[width=0.36\linewidth]{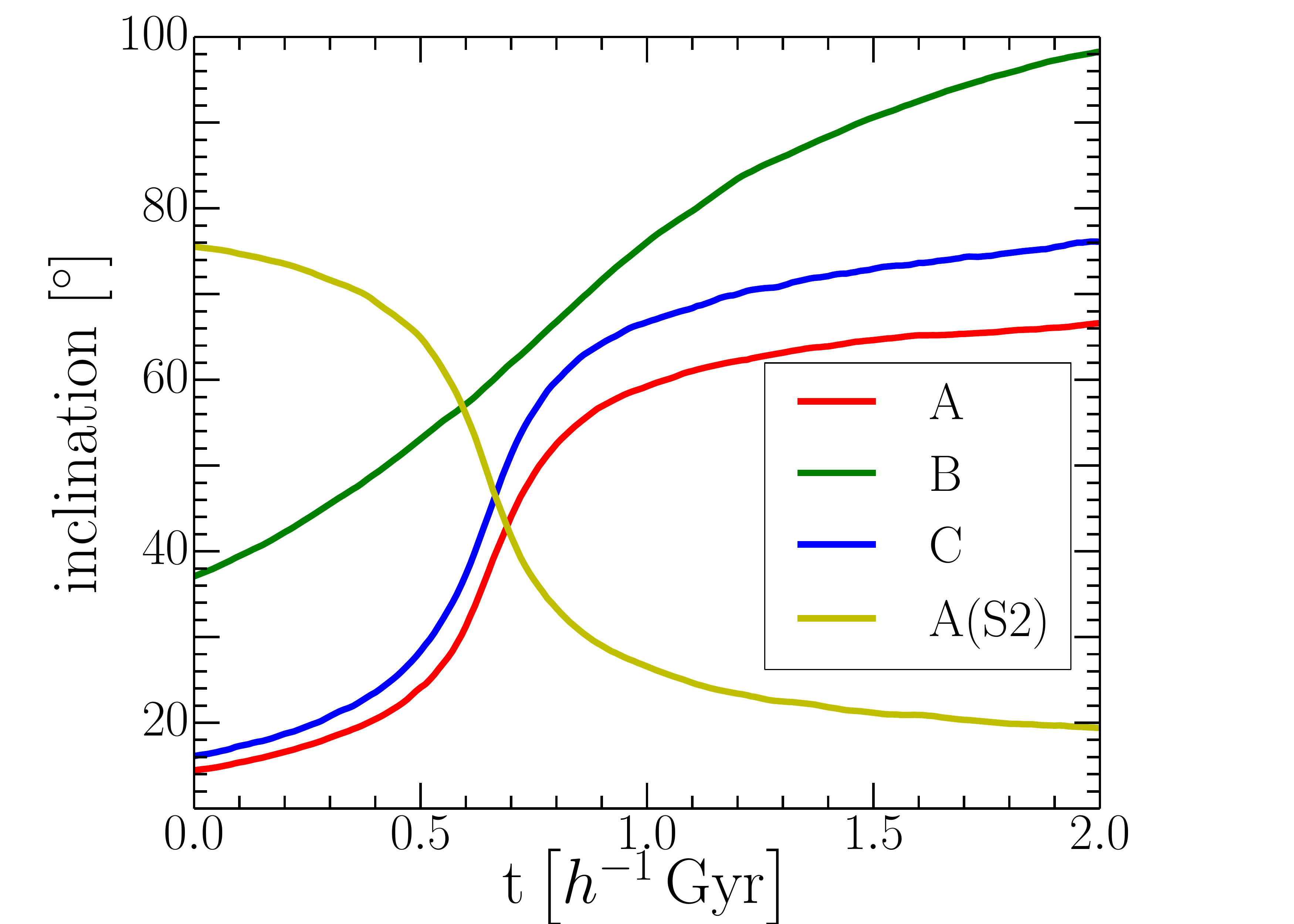} \label{fig:orbits_c}}
    \caption{Fiducial galaxy orbits adopted for studying the interaction
      of galaxy model G1a with the three different galaxy
      clusters constructed for the study. (a) Trajectories of the galaxy
      in clusters A, B, and C, scaled to the virial radius of the
      corresponding cluster. (b) Time evolution of the distance of the
      galaxy to the corresponding cluster centre. (c) Effective
      inclination angle of the galaxy (angle between the spin
      vector of the disk and its instantaneous velocity vector). 
      }
    \label{fig:orbits}
\end{figure*}

\begin{figure*}[htbp]
  \centering
  \includegraphics[width=\linewidth]{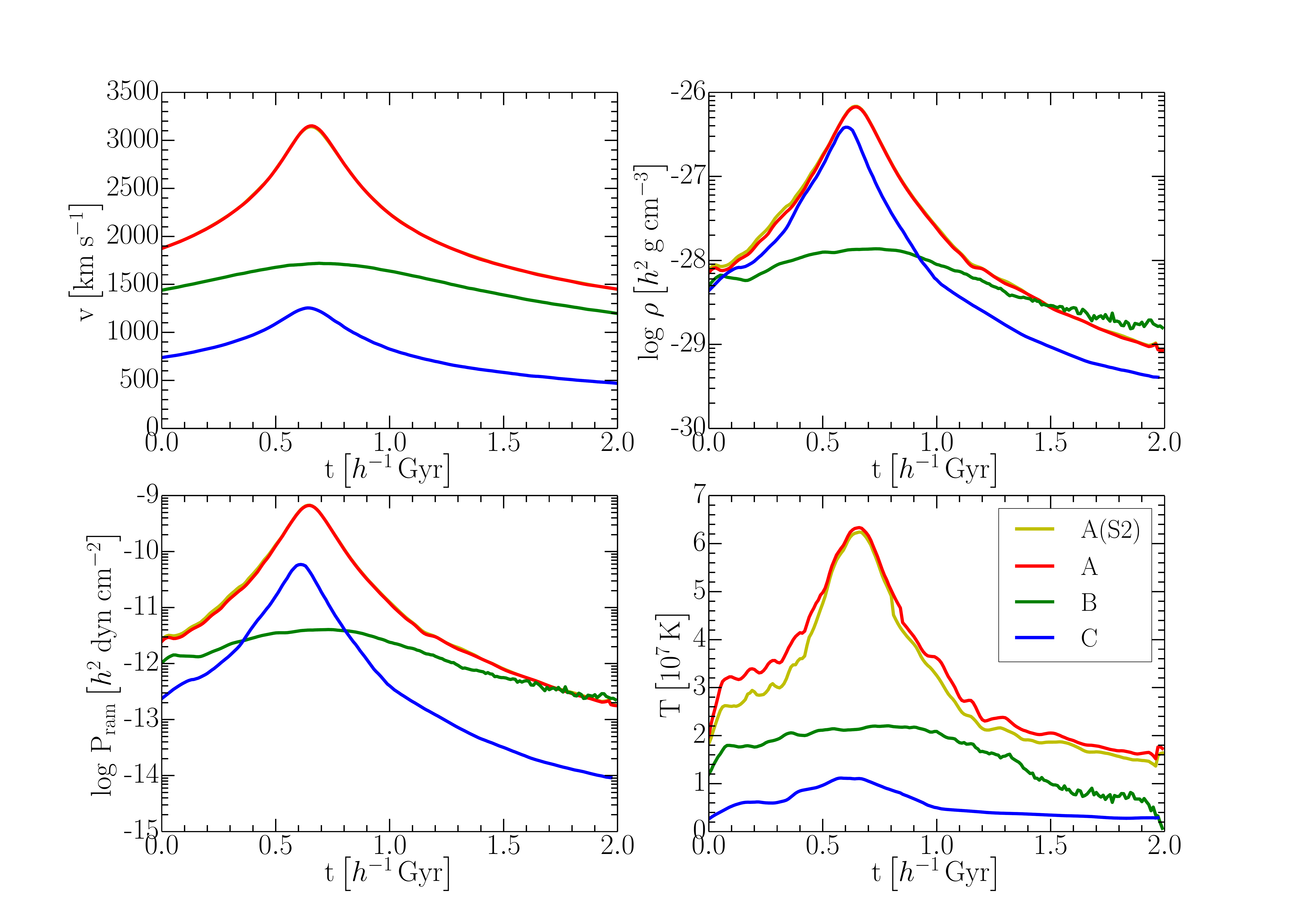}
  \caption{Physical conditions encountered by galaxy G1 as it
    falls into different galaxy clusters.  The top left panel shows
    the relative velocity of the galaxy, the top right panel shows the ICM
    density, the bottom left panel gives the resulting ram pressure
    $\mathrm{P}_\mathrm{ram}$, and the bottom right panel indicates the ICM temperature
    encountered by galaxy G1a for runs S1, S3, and S4,
    respectively. Galaxy inclination does not affect these values
    significantly, hence run S2 looks essentially the same as S1 and
    is hence not shown. Also, as in Fig.~\ref{fig:orbits}, the values
    for galaxy G2 in runs S5-S7 are very similar as the trajectories
    are nearly the same.}
  \label{fig:icm_prop}
\end{figure*}

Runs S1 to S4 were carried out with model galaxy G1 in the different clusters,
whereas galaxy G2 was used in runs S5 to S7.  The angular momentum
unit vector of the galactic disk is set to $\mathbf{\hat L}=(1,0,0)$ for runs S1, and
S3 - S7, and to $\mathbf{\hat L}=(0,1,0)$ for run S2.  The actual orbits
of the different runs are shown in Fig.~\ref{fig:orbits_a} in units of
$r_{200}$ of the corresponding cluster.  Furthermore, the distance of
the galaxy to the centre of the cluster and the effective inclination
(angle between angular momentum and relative velocity vector of the
galaxy) over time is depicted in Figs.~\ref{fig:orbits_b} and
\ref{fig:orbits_c}.  All trajectories lie in the $xy$-plane, and the
model galaxies start close to the virial radius of the cluster.
Furthermore, in Fig.~\ref{fig:icm_prop} we show the ICM properties the
galaxy encounters while orbiting the cluster by plotting density,
relative velocity, corresponding ram pressure, and ICM temperature as a
function of time.

The set-up adopted here allows for more realistic scenarios as wind-tunnel
experiments.  While it is in principle possible to mimic temporal
changes of the relative velocity and density of the headwind, varying
the inclination angle of the galaxy correctly is difficult.
Furthermore, our new set-up also includes a full treatment of
gravitational effects from the interaction of the galaxy with the
 potential of the cluster.  This includes tidal truncation of the DM halo as well as tidal effects on the disk itself, which are
both not present in ordinary wind-tunnel simulations.  These effects
not only change the structure of the galaxy, but also influence the
conditions under which RPS takes place.
On the other hand, the smooth ICM distribution of our live 
galaxy-cluster models rather resembles a relaxed and virialised system. 
In future work, we will include dynamically perturbed clusters, especially studying the influence of 
shock fronts in the ICM on RPS and SFR of cluster galaxies 
\citep[e.g.][]{Pranger2014}.
Also, the investigation and disentanglement of tidal interaction in 
combination with RPS \citep[e.g.][]{Bischko2015} in a realistic cluster
set-up should be subject of future work.

\subsection{Numerical code and star formation model}
\label{sec:num_sfr}

All simulations presented in this work were performed with the
hydrodynamical moving-mesh code AREPO \citep{Springel2010}. Given a
set of mesh-generating points, the volume is discretised by a Voronoi
tessellation, yielding an unstructured grid.  Subsequently, fluxes
across each face of a cell are calculated using an unsplit Godunov
scheme with an exact Riemann solver in the form of the MUSCL-Hancock
finite-volume scheme \citep{VanLeer1984, VanLeer2006}, which is
second order in space and time (for a smooth gas distribution).

As the Voronoi mesh in AREPO is allowed to move with the gas flow and
the Riemann problem is solved in the frame of the (moving) face of a
cell, it can be shown that the scheme is manifestly
Galilean invariant, which is not the case for standard Eulerian codes
using static grids. This makes the truncation errors of AREPO
independent of the gas velocity. Effectively, the code combines
advantages of Eulerian and Lagrangian schemes such as smoothed
particle hydrodynamics (SPH) while avoiding some of their
disadvantages. In particular, the good treatment of shocks and contact
discontinuities of Eulerian schemes is preserved. On the other hand,
the scheme inherits the adaptivity and Galilean invariance of
SPH. Furthermore, self-gravity is treated as in SPH codes, resulting
in a continuous adjustment of the gravitational resolution in a
Lagrangian fashion.

In cosmological simulations, a very high dynamic range in the gas
density is present, ranging from
$\sim 10^{-30}\, \mathrm{g}\,\mathrm{cm}^{-3}$ in the ICM at the
virial radius to well above
$\sim 10^{-22}\, \mathrm{g}\,\mathrm{cm}^{-3}$ in the ISM in a
galaxy. To avoid numerical artifacts, in SPH an approximately equal
mass resolution needs to be used for particles that are in direct
contact \citep{Ott2003, Read2010}, making it difficult to resolve a
comparatively small galaxy accurately within a large cluster of
galaxies.  In AREPO this problem can be much better addressed because
there is no restriction on the relative mass content of neighbouring
cells, allowing steep gradients in resolution. It is then possible to
efficiently simulate a highly resolved galaxy moving around within a
poorly resolved background cluster by adaptively refining the region
the galaxy passes through.

We use a simple sub-resolution model for radiative cooling, SF,
and supernova feedback according to \citet{Springel2003}.
Radiative cooling is implemented for an optically thin gas of
primordial composition \citep{Katz1996}. This gives an effective
temperature floor of $\sim 10^4\, \mathrm{K}$ when metal-line cooling
is not included. However some gas can cool below this
temperature through adiabatic expansion. Star formation and
  feedback is modelled with a sub-resolution multi-phase description
  for the ISM on a cell-by-cell basis, since these processes cannot
  be resolved explicitely.  For each gas cell in the star-forming regime
  ($\rho>\rho_\mathrm{th}$), the amount of gas is split into cold
  clouds and a hot, supernova-heated ambient medium. Cold clouds are
  transformed into stars on a characteristic timescale $t_\star$, where a
  mass fraction $\beta$ is released immediately as supernovae. The
  released mass as well as evaporated gas from the cold cloud phase
  replenishes  gas in the hot phase again.  This leads to
  self-regulation and, assuming that the temperature of the hot medium
  evolves towards an equilibrium value governed by cooling and
  feedback heating from SF, the amount of gas contained in
  the hot and cold phase can be calculated. This then yields an
  effective equation of state that stabilises the ISM against too
  rapid collapse. We note that these processes are modelled
  separately for each gas cell and no mass exchange or energy release
  into neighbouring cells takes place. Also, the actual creation of
  star particles is  treated in a stochastic fashion such that
  the formed collisionless star particles all have roughly the same
  mass.

We follow \citet{Hess2012} in choosing the parameters of the
sub-resolution model. We set the SF timescale
to $t_\star = 1.5\,h^{-1}\mathrm{Gyr}$. For the supernova feedback model,
we use a cloud evaporation parameter of $A_0 = 10^4$ and a supernova
temperature of $T_\mathrm{SN} = 10^9\,{\rm K}$. Furthermore, adopting a 
\citet{Salpeter1955} IMF, the mass fraction of massive stars 
$(>8\,\mathrm{M}_\odot)$ blowing up instantaneously as supernovae is
$\beta=0.1$. The choice of those
parameters defines the density threshold for the onset of SF
$\rho_\mathrm{th} = 2.6 \, h^2 \mathrm{cm}^{-3}$ \citep[see][for a detailed 
discussion of these parameters]{Springel2003}. The resulting threshold is higher
than that obtained using the default choice of parameters in
\citet{Springel2003}, which were used~e.g.~in \citet{Steinhauser2012,
  Kronberger2008}. This is motivated by recent claims that higher
threshold values are needed to obtain more realistic spiral galaxies
\citep{Guedes2011}.

Furthermore, we switch on cooling and SF only for the
ISM of the galaxy but not for the cluster.  To this
end, we are using a `colouring' technique \citep[as
e.g.][]{Vogelsberger2012, Roediger2005} to trace the galactic gas
throughout the simulation. Each gas cell gets an additional scalar
value that  is initialised with the
gas mass of the cell in case of the cells of the galaxy.  This scalar value is then advected with the
calculated flux from the hydrodynamic solver. This technique allows us
to track the gas of the infalling galaxy  in the stripped wake also, in
addition to controlling cooling and SF. The latter is only
switched on if the galactic gas fraction in a cell exceeds
$f_\mathrm{ISM}=0.25$. This trick prevents the rest of the cluster from
radiatively cooling and forming stars at a high rate in the BCG, which would
quickly become numerically very expensive even though we are not
interested in the central cluster region itself. We checked that
varying this threshold in the range $f_\mathrm{ISM}=0.01 - 0.25$ has
almost no impact on the results.

\subsection{Refinement technique}
\label{sec:ref_tec}

As pointed out previously, we need to cover a high dynamic range in
density to account for the interaction of the 
ISM and the ICM. Ordinary grid codes usually
use an adaptive mesh refinement (AMR) technique
\citep[e.g.][]{O'Shea2004, Fryxell2000} to dynamically increase the
resolution where needed by creating a hierarchical subgrid in, for example
regions with a high density contrast. The resolution can be decreased
again by removing levels of the subgrid, for example in regions with a
smooth density distribution. Such an approach is also possible in
AREPO with its unstructured mesh, although the Lagrangian nature of
the code drastically reduces the required number of refinement and
derefinement operations when a roughly constant mass resolution (which
is often the case) is desired. To this end, single cells can be refined
easily by inserting a new mesh-generating point very close to the
original one, effectively splitting the cell into two and keeping the
rest of the Voronoi mesh unaltered \citep{Springel2010}. The
regularity of the mesh is restored in the following timesteps, as the
points are moved towards the geometric centre of the corresponding
cells.  In this work, we use the maximum face angle of a Voronoi cell
\citep{Vogelsberger2012} to decide whether a cell is too distorted and
consequently mesh regularisation motions are added to the ordinary
movement of the mesh-generating points.

The initial gaseous disks for our model galaxies are generated by
distributing gas cells of constant mass according to the desired density 
profile (see Sect.~\ref{sec:galaxy_cluster_models}). This leads to 
a higher spatial resolution in the dense, inner parts of the disk and
a lower spatial resolution in the outer parts. In our production runs, 
we are using a gas mass of $7.8\times10^{4}\,h^{-1}\mathrm{M}_\odot$,
corresponding to initially $210000$ gas cells in the disk of galaxy G1 
(see also Sect.~\ref{sec:app_res_study}) with a maximum and average spatial 
resolution of $10\,h^{-1}\,\mathrm{pc}$ and $160\,h^{-1}\,\mathrm{pc}$, 
respectively. Model galaxy G2 has the same mass 
resolution and hence fewer gas cells.
While the Lagrangian behaviour of the moving mesh adjusts the
resolution automatically to the clustering state of matter, we still
need a criterion for deciding whether a finer grid in some preferred
regions is needed, in our case, for the galaxy and  surrounding ICM.
For the galactic gas, we want to keep the mass resolution of all cells
almost constant at the value in the initial conditions that also defines 
the target mass for refinement to create star particles
of similar mass, which helps to keep two-body heating effects at a
minimum level \citep{Vogelsberger2012}.

However, if we impose the same
mass resolution for the ICM in the region around the virial radius
where the interaction with the galaxy starts, the corresponding ICM
cells would have a size comparable to the galaxy itself owing to the
low ICM densities there.  This can create heavily distorted cells and
an underestimation of the gas density in the outskirts of the galaxy.

However, in AREPO, there is no need to have the same mass or spatial
resolution for all gas cells, unlike in standard formulations of SPH
\citep{Ott2003}. One possible solution to this problem lies in
introducing a kind of background grid \citep{Springel2010}, which
simply means that there is an imposed upper limit for the maximum
volume of cells. In addition, we need to prevent that the denser parts
of the ICM that do not interact with the galaxy are all refined to our
high target mass resolution, otherwise the computational cost would be
completely dominated by following cluster material we are not really
interested in.

To address both of these issues, we apply the following three criteria
in sequence to decide whether a cell should be refined or
derefined:  Firstly, we use the refinement and derefinement criteria of
  \citet{Vogelsberger2012} to keep the mass resolution of the ISM
  close to the initial value for all gas cells with an ISM mass
  fraction larger than 1\%, which we identify using the same
  colouring technique as described in Sect.~\ref{sec:num_sfr}.  Such
  cells are refined or derefined if the mass exceeds two times or is
  lower than half the target mass of ISM gas cells.
Secondly, we require all gas cells to have a volume not larger
  than ten times the smallest volume of all its neighbouring cells
  \citep{Pakmor2013}, ensuring a gradual refinement of the ICM around
  the galaxy and its stripped gas.
The third criterion with lowest priority keeps gas cells, which have
  an ISM mass fraction that is less than 1\%, and that are not affected by the
  maximum volume criterion, at the initial mass resolution of the
  cluster.

Nevertheless, the volume criterion for refining cells in the
surrounding of the galaxy is not necessarily sufficient when the
galaxy moves at high speed through the cluster. Because a gas cell can
be refined only if it is sufficiently regular (meeting the above-mentioned criterion for its face angles), a certain time is needed
until all cells in a region can be transformed to a new (finer)
volume. To avoid potential problems from this lag, we refine cells
along the a priori known trajectory of the galaxy to the 
initial mean volume of gas cells in the galaxy itself, so that
the ICM gas cells already have a sufficiently small volume and the desired spatial
resolution when the galaxy arrives. In the case of our production runs, this 
initial mean volume corresponds to $52.72\,h^{-3}\,\mathrm{kpc}^{3}$ 
(see also Sect.~\ref{sec:app_res_study}).
Also, we make sure that the
gradual refinement mentioned above is already realised in the
initial conditions. This is achieved by simulating the cluster itself
for around $300\,h^{-1} \mathrm{Myr}$ with enabled refinement
criteria. Then we insert the galaxy in the relaxed galaxy cluster and
start the actual simulation. Tests of the correct functioning of these
special refinement techniques and a resolution study can be
found in Sect.~\ref{sec:app_res_study}.

\begin{figure}[tbp]
  \centering
  \subfigure[]{\includegraphics[width=\linewidth]{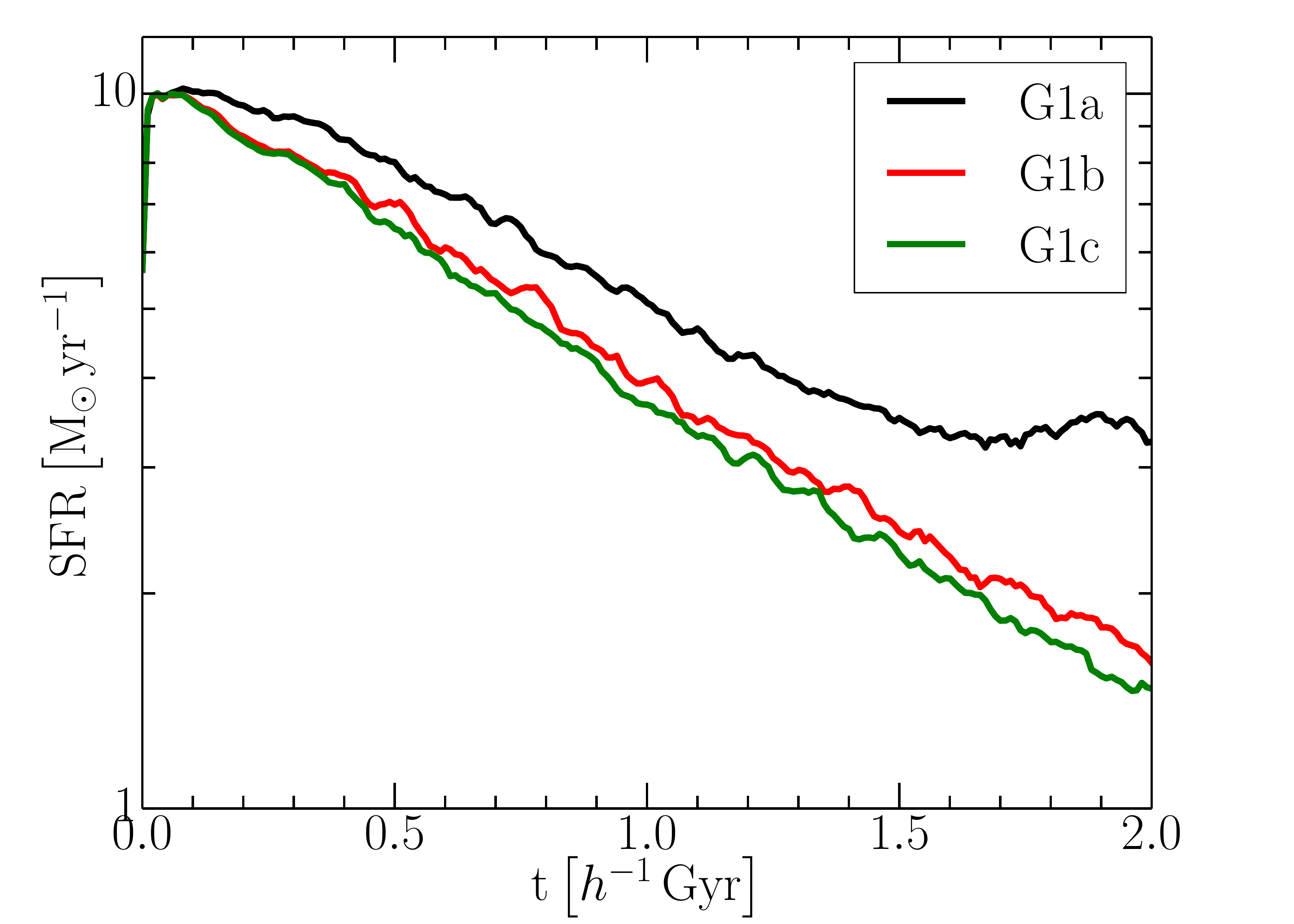} \label{fig:sfr_isolation_sfr}}
  \subfigure[]{\includegraphics[width=\linewidth]{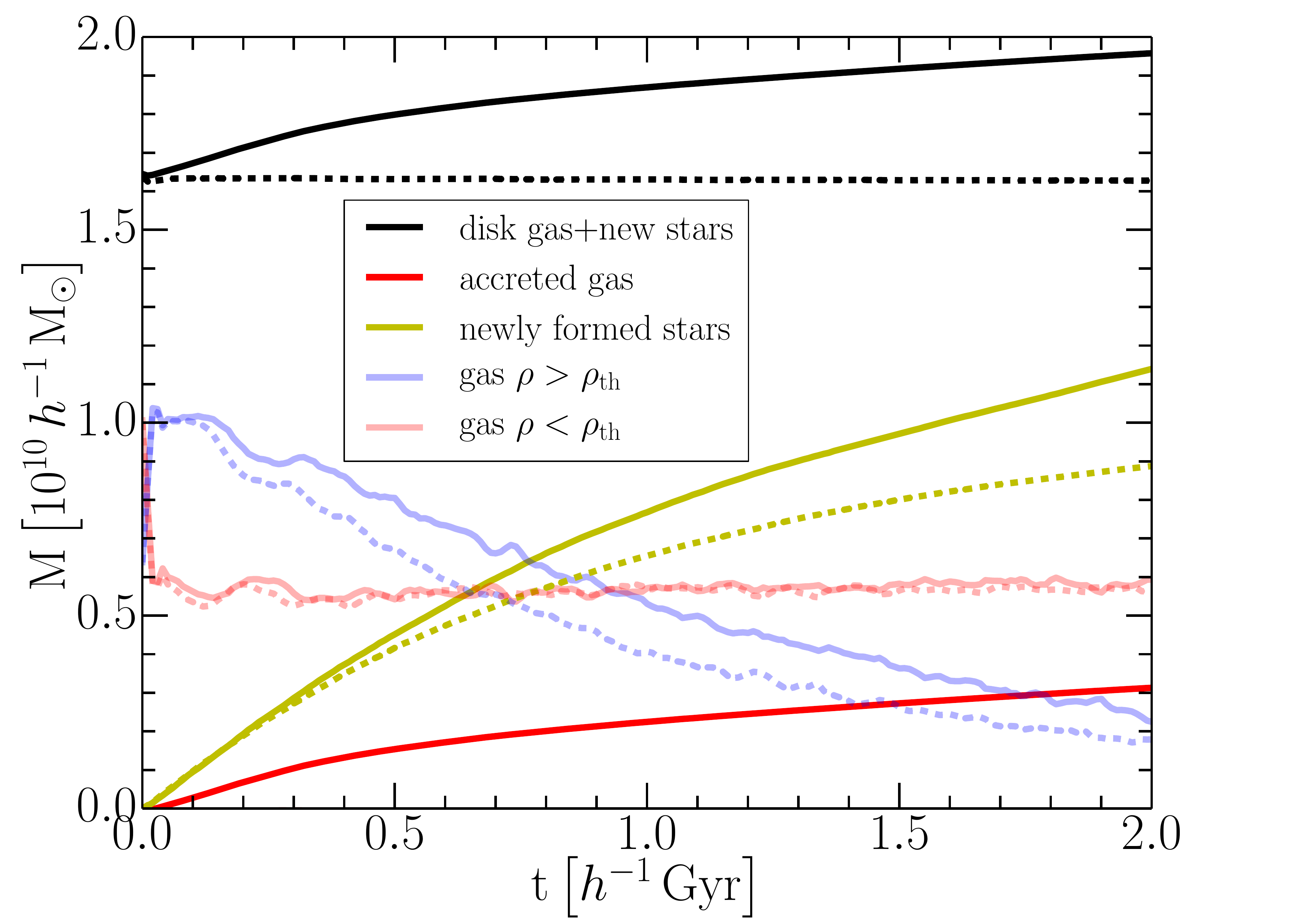} \label{fig:sfr_isolation_mass}}
  \caption{(a) Star formation rate of model galaxy G1 when evolved in
    isolation. The three variants shown correspond to different gas
    masses in the corona. Model G1a has the same baryonic mass in the
    halo as in the disk, G1b has half of that mass, and G1c has only 1\% of the
    disk mass in the gaseous halo. (b) Evolution of different baryonic
    components of galaxy model G1a (solid lines) and G1c (dashed lines) 
    evolved in isolation. The red line indicates
    accreted halo mass onto the disk. The mass components shown are
    newly formed stars, star-forming
    ($\rho > 2.6\,h^2\mathrm{cm}^{-3}$), and non-star-forming
    gas. The black line shows the total gas mass in the disk plus
    newly formed stars.}
  \label{fig:sfr_isolation}
\end{figure}

\section{Results} \label{sec:results}

\subsection{Galaxies in isolation}
\label{sec:results_isolation}

We begin by considering results for our model galaxy G1 when
evolved in isolation with different hot halo gas masses. This
investigates the evolution of field galaxies depending on the presence
of a more or less massive hot gas halo and provides the comparison
benchmarks for our simulations with RPS.  In
Fig.~\ref{fig:sfr_isolation_sfr}, the evolution of the SFR
of the three different galaxies is shown.  The model galaxy with the
most massive gas halo, which contains the same mass as the disk, shows an
enhanced SFR as the disk is refueled by accretion
from the hot halo; however, the other two models with only half, or
essentially no, baryonic mass in the halo show almost the same SF
activity.

In the bottom panel of Fig.~\ref{fig:sfr_isolation}, we show the
evolution of the baryonic mass in the disk with respect to simulation
time. We use a simple geometric criterion (a cylinder with 
$r=20\,h^{-1}\,\mathrm{kpc}$ and $h=3\,h^{-1}\,\mathrm{kpc}$, centred 
on the galaxy plane)
to decide whether gas cells and newly formed stellar particles belong to 
the disk or the halo of the galaxy and split the gas mass in the disk 
into star-forming (having a
density $> 2.6\,h^2\mathrm{cm}^{-3}$) and non-star-forming gas.  The
amount of low density, non-star-forming gas remains the same in all
of the three simulations.  Also, the mass of newly formed stars over time is
shown while the red line indicates mass that was accreted onto the
disk from the hot gas halo during the simulation. 
The higher mass of newly formed stars due to a higher SFR in the 
simulations, including a hot gas halo, matches the mass of gas accreted
after $2\,h^{-1}\,\mathrm{Gyr}$ of evolution.
We note, however, that our simulations do not include strong
feedback processes that could expel significant ISM mass or even unbind
hot gas halo \citep[the influence of feedback is discussed e.g.
  in][]{Nelson2015}.  Furthermore, we note that we follow the common
assumption that there is no angular momentum transport from the dark
matter to the hot gas halo, hence we initialise the hot halo with the
same specific angular momentum as the DM halo. \citet{Moster2011}
carried out simulations of galaxies in isolation using GADGET-2
\citep{Springel2005}, varying the total angular momentum of the hot
gas halo. They found that by assigning  the DM halo angular
momentum four times to the hot gas halo yields the best results compared to
observations of stellar mass and disk scale length.  However, also applying
 the angular momentum four times to our model galaxy G1 does not
lead to very different results.  Although this lowers the SF
slightly, the most important parameter is still the hot halo
mass. Presumably, the influence of the angular momentum is also small
because our simulations start with a relative high gas mass in the
disk with respect to the hot gas halo, compared to the \citet{Moster2011} studies.

\begin{figure*}[htbp]
  \centering 
  \includegraphics[width=0.92\textwidth]{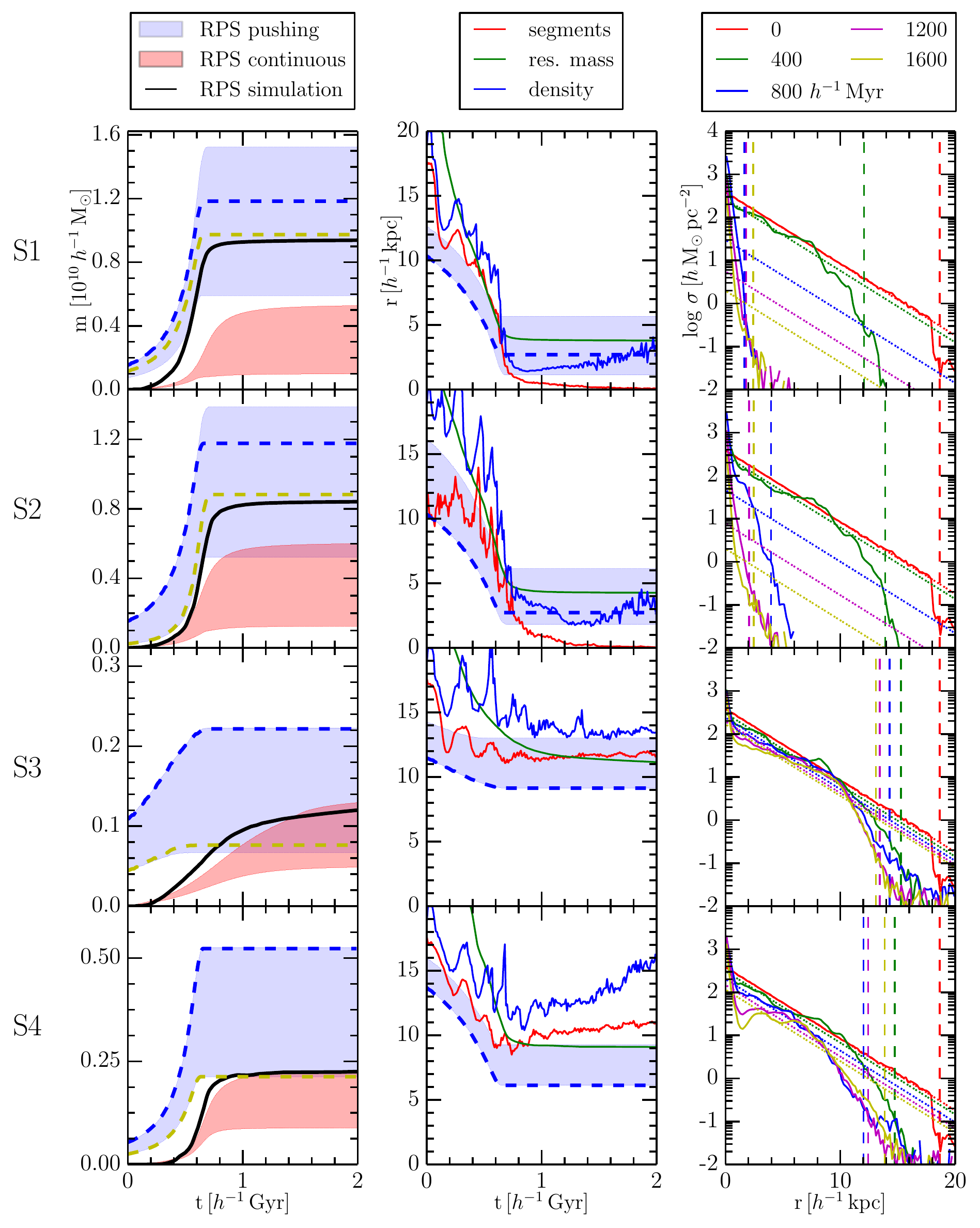}
  \caption{Ram-pressure stripping of galaxy G1a for runs S1-S4 in the
    corresponding clusters. The panels of the column on the
    right-hand side show gas surface density profiles of the galaxy
    (solid lines) and exponential surface density fits based on the 
    gas mass in the galaxy at a particular timestep (dotted lines) 
    as coded by the colours. Vertical dashed lines indicate the radius, 
    at which the actual surface density drops more than one dex 
    below the exponential fit.
    The panels of the middle column give the
    evolution of the stripping radius in the simulations, which are determined with
    three different measurement methods (defined in Sect.~\ref{sec:rps}). 
    The blue shaded area encloses the
    minimum and maximum stripping radius, calculated by different 
    theoretical predictions based on the standard \citet{Gunn1972} criterion
    (blue dashed line), considering initial or current and total disk mass, 
    the inclination of the galaxy as well as Rankine-Hugoniot post-shock 
    conditions. The black lines in the panels of the left
    column show the effective mass lost via stripping in the
    simulations. The edges of the blue shaded region indicate
    the minimum and maximum mass loss, considering the stripping radius 
    calculated with the different models. The red shaded area 
    depicts the corresponding extrema of mass loss due to a continuous stripping
    assumption. The yellow dashed lines indicate the model that best fits the 
    corresponding simulation.}
  \label{fig:stripping_p1h1}
\end{figure*}

\begin{figure*}[htbp]
  \centering 
  \includegraphics[width=0.92\textwidth]{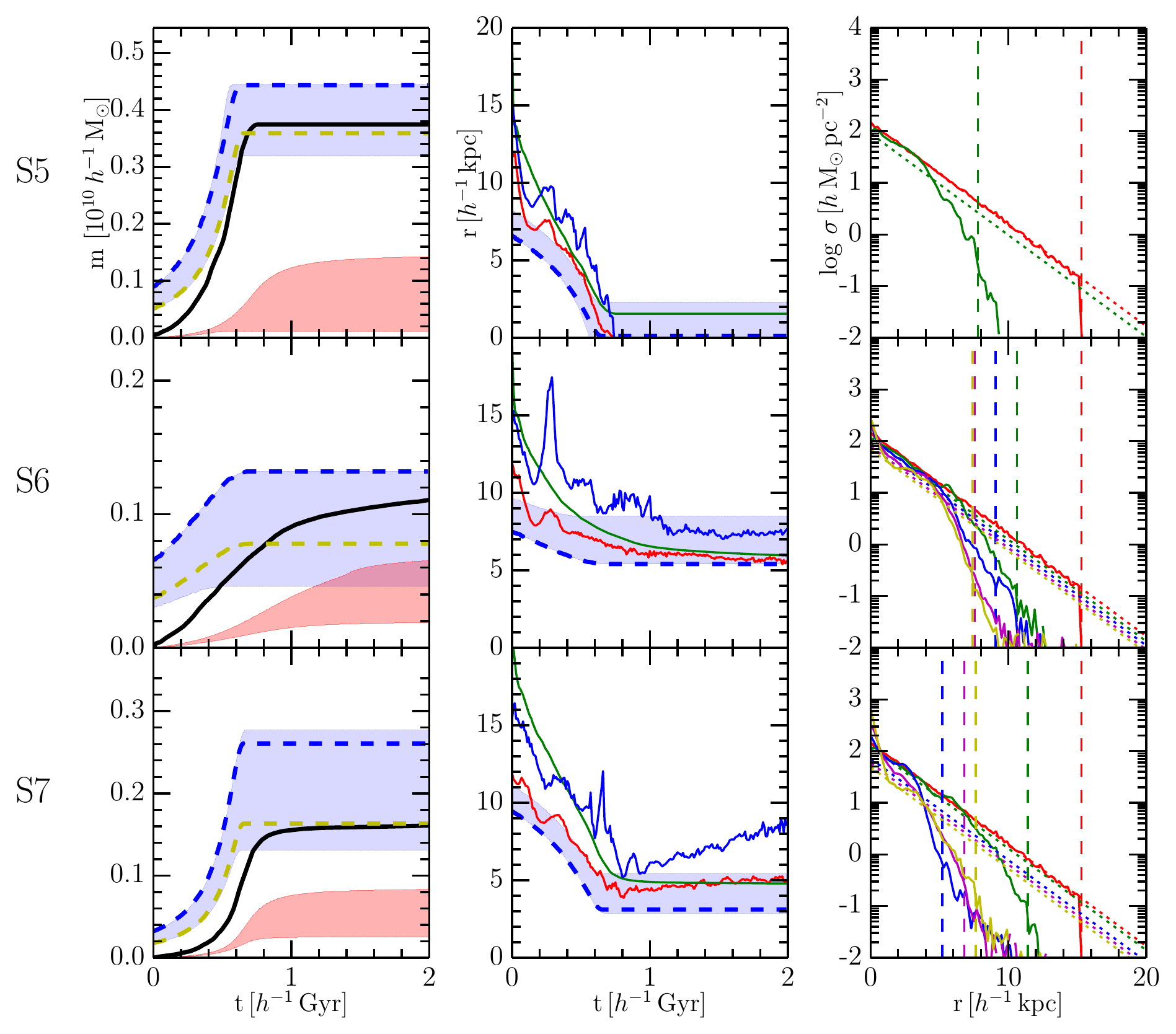}
  \caption{Same as Fig.~\ref{fig:stripping_p1h1} but for galaxy
    G2. }
  \label{fig:stripping_p2h3}
\end{figure*}

\subsection{Ram-pressure stripping}
\label{sec:rps}

Stripping can be broadly subdivided into two classes: ram-pressure
pushing \citep{Gunn1972}, and gas dragging from the outskirts of 
a disk of a galaxy from Kelvin-Helmholtz instabilities
\citep{Nulsen1982}.  Ram-pressure pushing can truncate the gas disk of
a galaxy out to the so-called stripping radius, where the pressure of
the impinging ICM wind
\begin{equation}
 P_\mathrm{ram} = \rho_\mathrm{ICM}v_\mathrm{rel}^2 
\end{equation}
\noindent
exceeds the gravitational restoring force of the disk, 
\begin{equation}
P_\mathrm{ram} \geq 2\pi\mathrm{G}\sigma_\star\sigma_\mathrm{gas}.
\end{equation}

Assuming a double exponential profile for gas and stars
in the disk, the stripping radius can be
calculated as
\begin{equation}
  r_\mathrm{strip} = \frac{r_0}{2}\,\ln \left( \frac{G M_\star M_\mathrm{gas}}{\rho_\mathrm{ICM}v_\mathrm{rel}^2\,2\pi r_0^4} \right),
  \label{eq:stripping_radius}
\end{equation}
where $r_0$, $M_\mathrm{gas}$ and $M_\star$ is the initial disk
scale length, and gas and stellar masses, respectively.  The
density of the impinging ICM wind is given by $\rho_\mathrm{ICM}$, and
$v_\mathrm{rel}$ denotes the velocity of the galaxy relative to the
ICM.  As this model is very often used in SAMs
\citep[e.g.][]{DeLucia2004, Guo2011, Benson2012}, 
below we test how well it agrees with our
simulations.  This formulation does not take a possible
inclination of the disk with respect to the velocity vector of the
galaxy into account.

In previous work, it has been found that the above relation is quite
accurate for face-on galaxies \citep[e.g.][]{Kronberger2008,
  Roediger2007}, and that the inclination angle does not seem to
have a large impact on the amount of stripped gas as long as the
inclination is not close to edge-on \citep[e.g.][]{Steinhauser2012,
  Jachym2009, Roediger2006a}. Nevertheless, there are modified
theoretical descriptions that take the inclination of the galaxy into
account \citep[e.g.][]{Lanzoni2005} by adding a
$\cos^2\left( i \right)$ factor to the RP term.  Other
variations consider the total mass of the galactic disk when
calculating its potential, hence computing a more accurate estimate of
the restoring force \citep[e.g.][]{Font2008}.

Also, the above model is intended for a single stripping
event at constant RP, and assumes that the surface density
profile of the disk within the stripping radius is not altered after
the stripping event. However, in reality, RPS is
not an instantaneous process with the stripping radius and hence the
amount of stripped gas settling down only after $\sim 10 -
100\,\mathrm{Myr}$ in typical wind-tunnel experiments.  Furthermore,
during cluster or subhalo infall, the density and velocity change
continuously.  Also, galaxies can move at a higher velocity than the
local sound speed, leading to the formation of a bow shock in the ICM that
is impacted.  Calculating the velocity and density of the ICM behind
the bow shock with the Rankine-Hugoniot conditions \citep{Shu1992},
one finds that the RP behind the bow shock is less effective
than the classical approach suggests when the shock is ignored.

\begin{figure*}[htbp]
  \centering
  \subfigure[]{\includegraphics[width=0.5\linewidth]{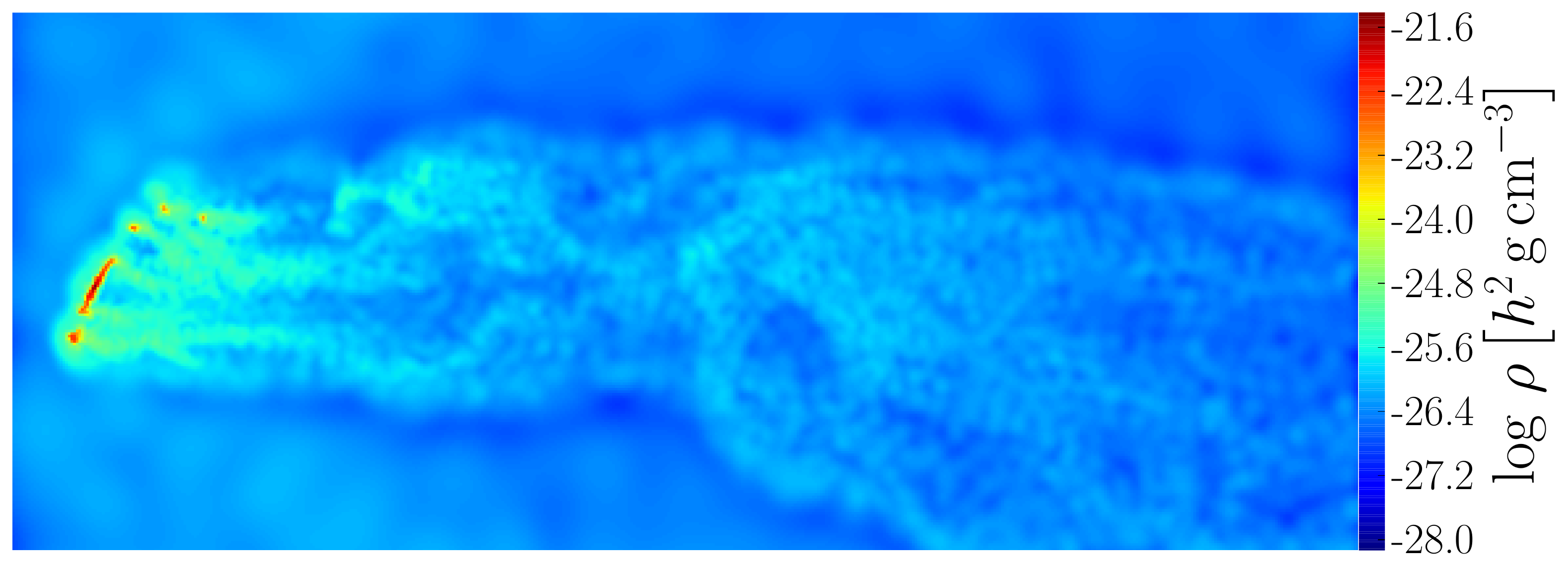}}
  \subfigure[]{\includegraphics[width=0.5\linewidth]{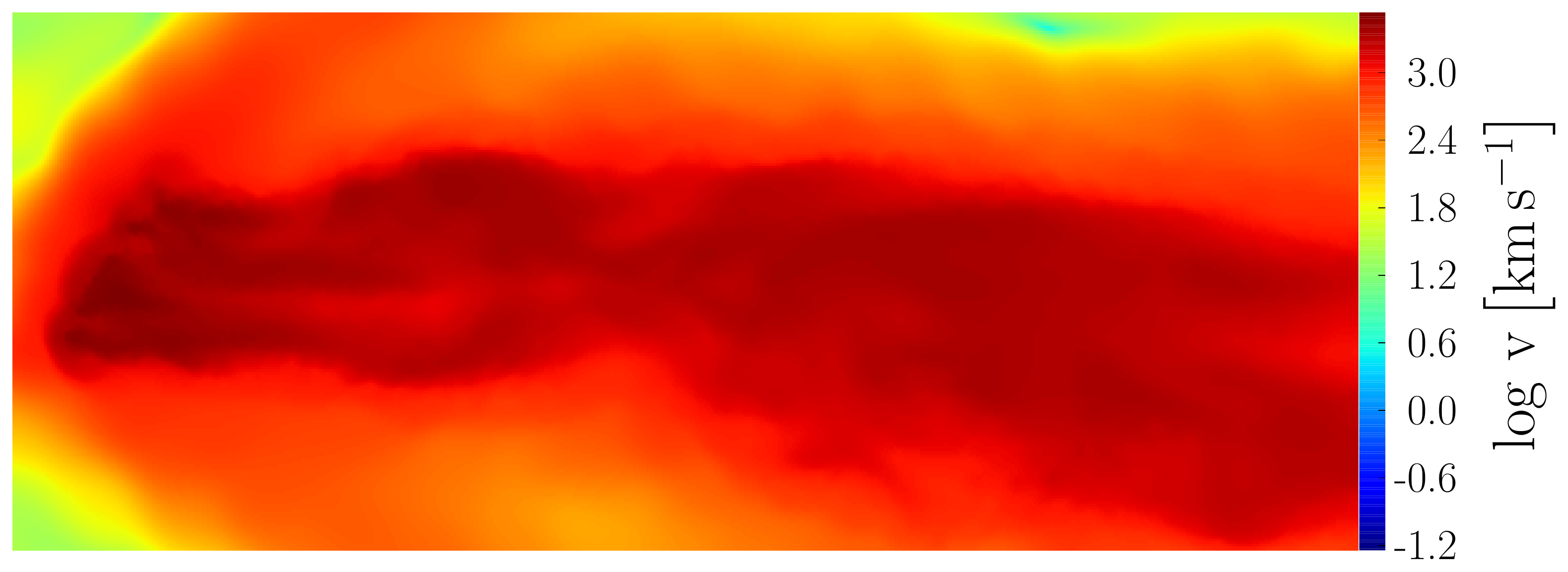}}
  \subfigure[]{\includegraphics[width=0.5\linewidth]{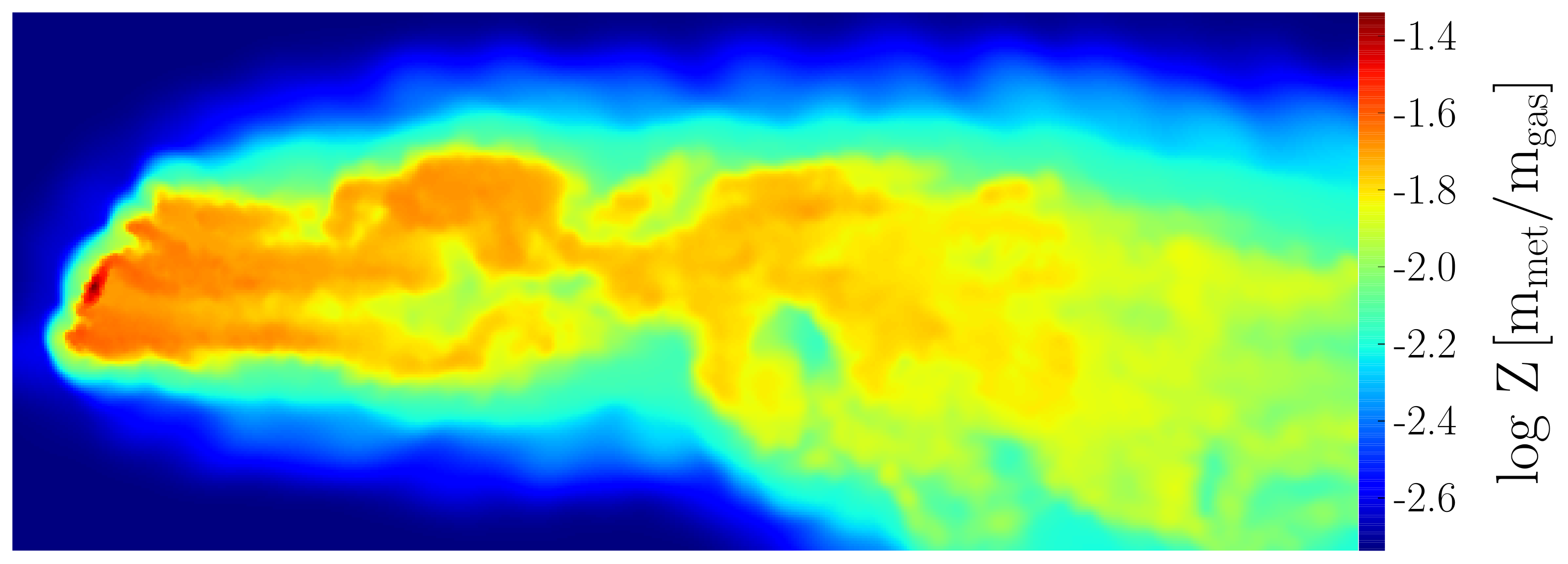}\label{fig:ortho_gas_metals_metals}}
  \subfigure[]{\includegraphics[width=0.5\linewidth]{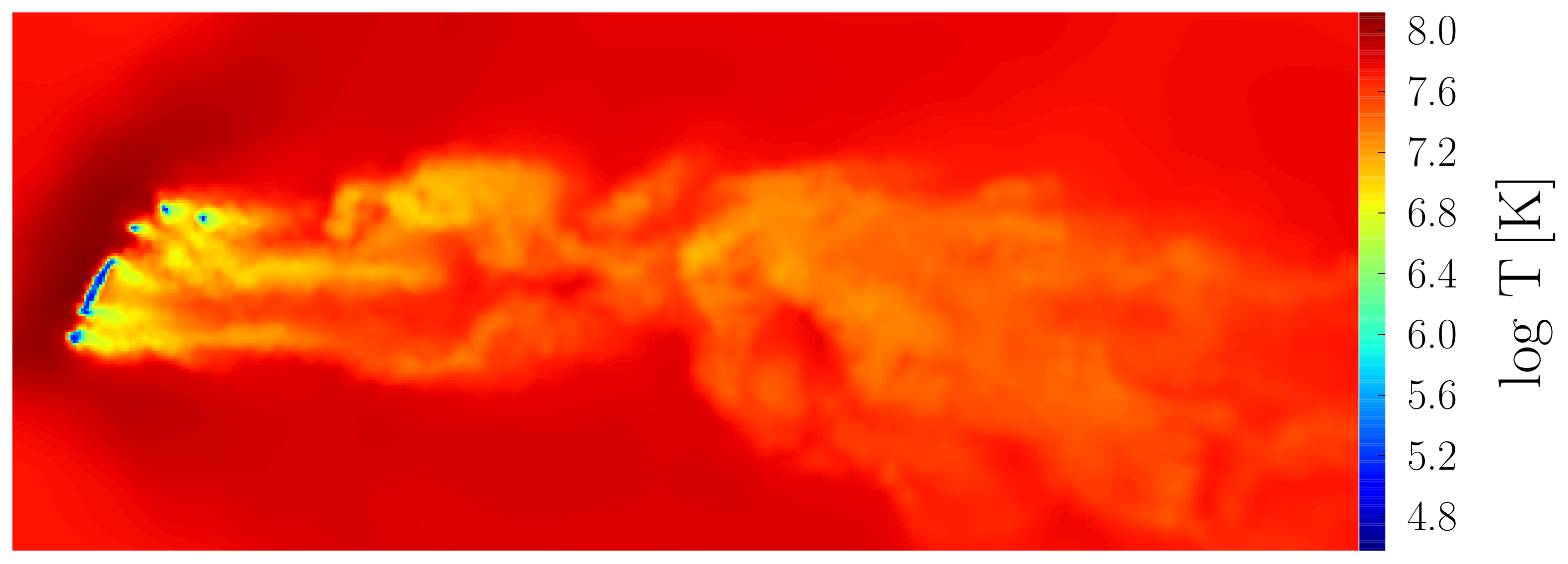}}
  \subfigure[]{\includegraphics[width=0.5\linewidth]{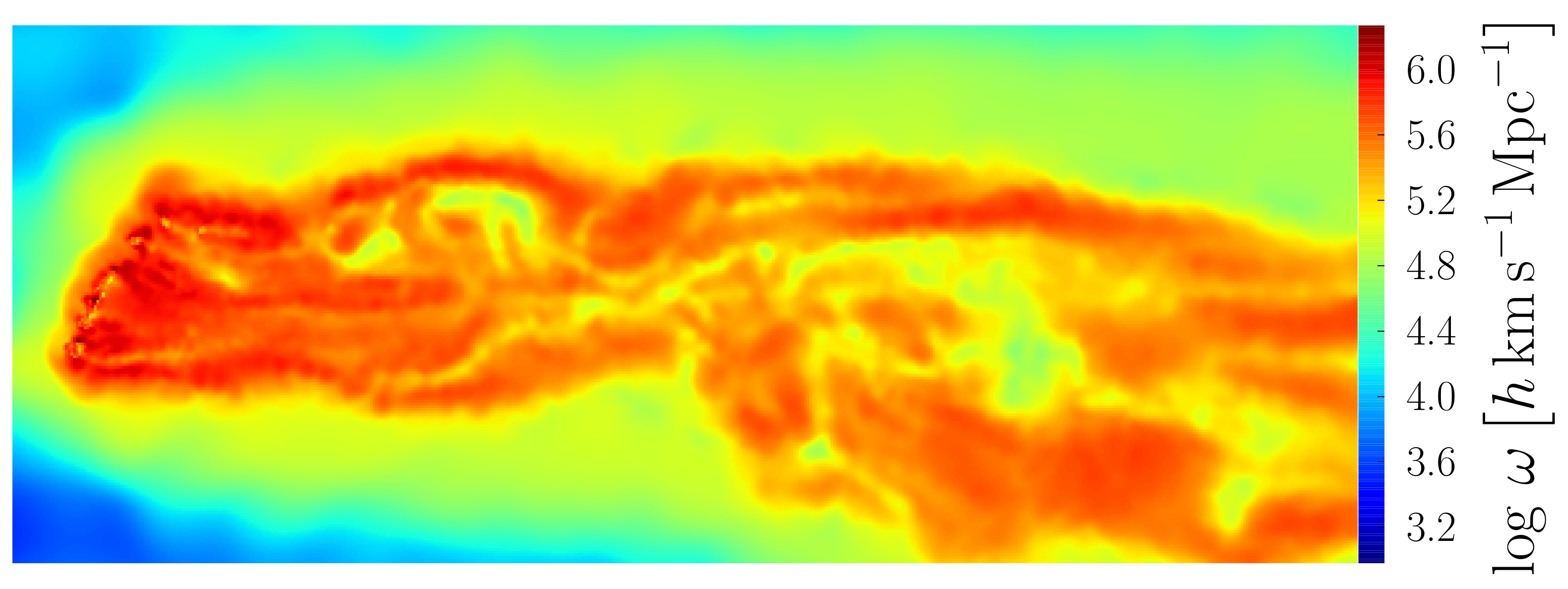}} 
   \caption{Slices of thickness $0.2\,h^{-1}\mathrm{kpc}$ and
     extension $100\,h^{-1} \mathrm{kpc} \times40\,h^{-1}\mathrm{kpc}$
     through the galaxy and its wake in run S1, after
     $600\,h^{-1}\mathrm{Myr}$ of evolution. Quantities are mapped
     using an SPH kernel. The different panels show: (a) gas density,
     (b) velocity magnitude, (c) metallicity, (d) temperature, and (e)
     vorticity magnitude.}
  \label{fig:ortho_gas_metals}
\end{figure*}

In the following, we hence compare the stripping radius and gas
loss in our simulations with all possible combinations of the
aforementioned variations of the original \citet{Gunn1972} criterion,
using $\rho_\mathrm{ICM}$ and $v_\mathrm{rel}^2$ and also the
total remaining gas and stellar mass of the galaxy from every snapshot
of the simulations.  We calculate the gas loss due to Kelvin-Helmholtz
instabilities, which are also referred to as continuous stripping, following
\citet{Nulsen1982} and \citet{Roediger2007}, as
\begin{equation}
\dot{{M}}_\mathrm{KH} = 0.5 \, r_\mathrm{strip}^2\, \pi \,\rho_\mathrm{ICM}\,v_\mathrm{rel}
\label{eq:kh_strip}
,\end{equation}
using the calculated theoretical stripping radius for all
distinct variations.  

In Figs.~\ref{fig:stripping_p1h1} and \ref{fig:stripping_p2h3} we show
the results for galaxies G1a and G2 respectively, for
different environments.  In the column on the right-hand side, the
surface density profiles of the galaxy (solid lines) are shown after 0,
400, 800, 1200, and 1600 $h^{-1}\mathrm{Myr}$ of evolution in the
corresponding cluster; this is compared to the theoretical, exponential
density profile the galaxy would have when the scale radius is kept
constant at the initial value and the normalisation is updated to the
remaining gas mass in the disk at a particular timestep, as
coded by the colours (dotted lines). The vertical dashed lines indicate
the radius, where the surface density of the galaxy falls more than one dex
below the theoretical, exponential profile mentioned above. This is useful
for calculating the stripping radius, as described in the next paragraph, and 
also allows us to investigate a possible redistribution of gas to the outer
parts of the disk.

In the middle column, the stripping radius versus simulation time is
shown.  Determining the exact radius in an automated way is not a
trivial task; even investigating every snapshot visually often does
not lead to clearly defined radii.  Therefore, we plot three different
stripping radii assigned with different methods.  
In the first case (blue solid line), we determine the stripping
radius at the point where the actual surface density profile in the simulation
drops more than one dex below an exponential surface density profile 
(\ref{eq:surf_dens_disk}), using the remaining gas mass in the 
disk of a particular timestep and the initial disk scale length (see previous 
paragraph and compare the dotted to the solid lines at different timesteps 
in the panel on the right-hand side).
Secondly, following the approach by
\citet{Roediger2006a}, we divide the disk into 12 segments and
determine the radius for each.  As the mass in the cells should be
equal and close to the target mass of the ISM gas (see
Sect.~\ref{sec:ref_tec}), we divide the segments into radial bins and
define the stripping radius as the bin with less than ten gas cells.
Then, the red line indicates the stripping radius as the mean value of all
segments.  This approach has been found to yield the most robust
measure for all simulations performed in this work.  Finally, we also
consider the still gravitationally bound gas mass, considering also
gas transformed into stars. For each snapshot we calculate the radius that
the galaxy would have if the initial surface density profile were not altered (green line).
As can be seen in the plots, this method
does not consider the delay between the gas that is pushed out of the
disk and the gas that is not gravitationally bound to the galaxy anymore.
However, considering this lag, and in cases of constant RP (as in
cluster B), it is a reasonably good measure and agrees well with the
other methods.  Finally, we sketch the results of theoretical
predictions described at the beginning of this section, comprising 
the standard \citet{Gunn1972} criterion and variations including the 
total mass and inclination angle of the disk as well as considering
Rankine-Hugoniot post-shock conditions. The blue dashed line indicates
the stripping radius calculated with the standard criterion 
(\ref{eq:stripping_radius}). The blue shaded area delimits the range of
stripping radii from all theoretical predictions mentioned above, with 
the edges corresponding to minimum and maximum values at any time.

In the column on the left-hand side, the actual gas mass stripped from
the disk is plotted (black solid line).  We consider that all gas not bound
gravitationally to the galaxy anymore (gas cells that have a positive
total energy
$E_\mathrm{tot} = E_\mathrm{pot} +
E_\mathrm{therm} + E_\mathrm{kin}$)
are stripped off the disk.  Usually, gas is being pushed out of
the disk and stays gravitationally bound until the gas is sufficiently
decelerated.  To calculate the stripping radius and the surface
density profile, we therefore only consider the gas in a cylinder
around the galaxy.  We note that gas is lost also due to SF.
Furthermore, we calculate the mass loss considering
the theoretical predictions of the stripping radius from the different 
models mentioned above. Again, the edges of the blue shaded area depict 
the minimum and maximum of the mass loss calculated from different 
models. The blue dashed line once more indicates  the mass loss using the 
standard \citet{Gunn1972} criterion. The red shaded area delimits 
minimum and maximum mass loss due to continuous stripping, using 
the stripping radius from the different models,  denoted above  
in (\ref{eq:kh_strip}) and applying the current RP at any time.
The yellow dashed line depicts the best model for the corresponding
simulation, which is described later in this section.

The results in Fig.~\ref{fig:stripping_p1h1} show different behaviour
for the RPS of a Milky-Way sized galaxy
G1a in different environments.  The first extreme case of
RPS in cluster A (see also Sect.~\ref{sec:galaxy_cluster_models})
strips the gas disk of the galaxy almost completely during pericentre
passage, as expected.  Considering also the SF activity,
only 12\% of the gas is left in the very centre of the disk.
Similarly, when the galaxy starts with a different inclination angle
(run S2), around 15\% of the gas is left
after pericentre passage.  This indicates that the inclination does
not have a major influence on the total amount of stripped gas.
However, the inclination of the galaxy in both runs differs most at
the beginning of the simulation when RP is still low (face-on
vs. edge-on), whereas at the peak of RP the inclination becomes quite
similar (see Fig.~\ref{fig:orbits_c}).  Furthermore, the
stripping radius for both inclinations is also similar, dropping from
$18\,h^{-1}\mathrm{kpc}$ to around $1-2\,h^{-1}\mathrm{kpc}$ after
$600\,h^{-1}\mathrm{Myr}$.  Ram-pressure pushing lasts until the
galaxy reaches the peak of RP at pericentre passage. Afterwards, 
continuous stripping proceeds at a very low rate as a result of the small 
remaining gas disk, which is depleted by forming new stars.

In the smaller cluster B, the trajectory was chosen to have a slowly
changing, medium RP acting on the galaxy (see
Fig.~\ref{fig:icm_prop}).  As a consequence, the galaxy keeps losing
gas throughout the whole simulation mainly by continuous stripping,
but at a much smaller rate than in cluster A.  In cluster/group C
on the other hand, the trajectory again leads  more towards the centre and has an overall less pronounced RP but a higher peak at pericentre
passage.  In this case, gas is lost mainly from
$400 - 700\,h^{-1}\mathrm{Myr}$ when the galaxy experiences high RP. At
the same time the stripping radius is reduced to $10\,h^{-1}\mathrm{kpc}$.
Before and afterwards, the galaxy is mainly affected by continuous
stripping, which proceeds however at a very low rate of
$<1\,\mathrm{M}_\odot\,\mathrm{yr}^{-1}$.

In simulations with a high peak of RP (S1, S2 and S4), the radius of the 
gas disk is again increasing when the galaxy leaves the central 
region of a cluster (red and blue lines in the middle column of 
Fig.~\ref{fig:stripping_p1h1}). The strength of the effect however depends 
on the method used for determining the radius (see description 
above for the different methods used). Using a fixed cut (red line), 
the effect is small, showing a maximum expansion of 
$2\,h^{-1}\mathrm{kpc}$ in run S4. However, comparing the actual surface density
with an exponential profile, considering initial disk scale length and remaining
gas mass in the particular timestep, shows an increase of up to $5\,h^{-1}\mathrm{kpc}$.
Hence, the strong increase, seen in this case, is mainly due to a shallower exponential 
profile the actual surface density of the galaxy is compared to.
Furthermore, in such cases, the surface density can be described well with the mentioned
exponential profile at radii sufficiently below the stripping radius. It is as if an 
`overcompressed' disk expands slightly again.

\begin{figure*}[htbp]
  \centering 
  \subfigure[]{\includegraphics[width=0.49\textwidth]{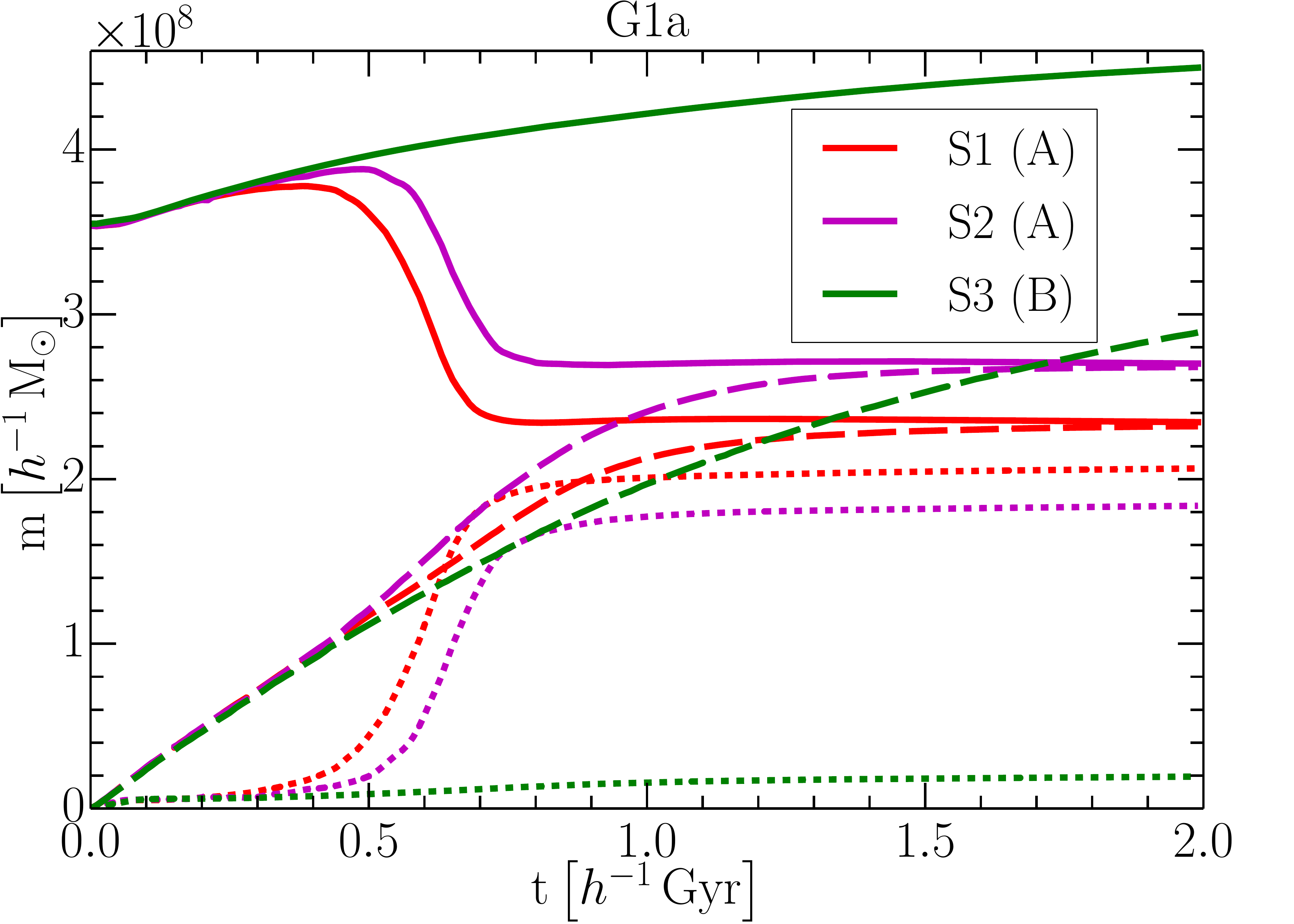}\label{fig:metals_strip_a}}
  \subfigure[]{\includegraphics[width=0.49\textwidth]{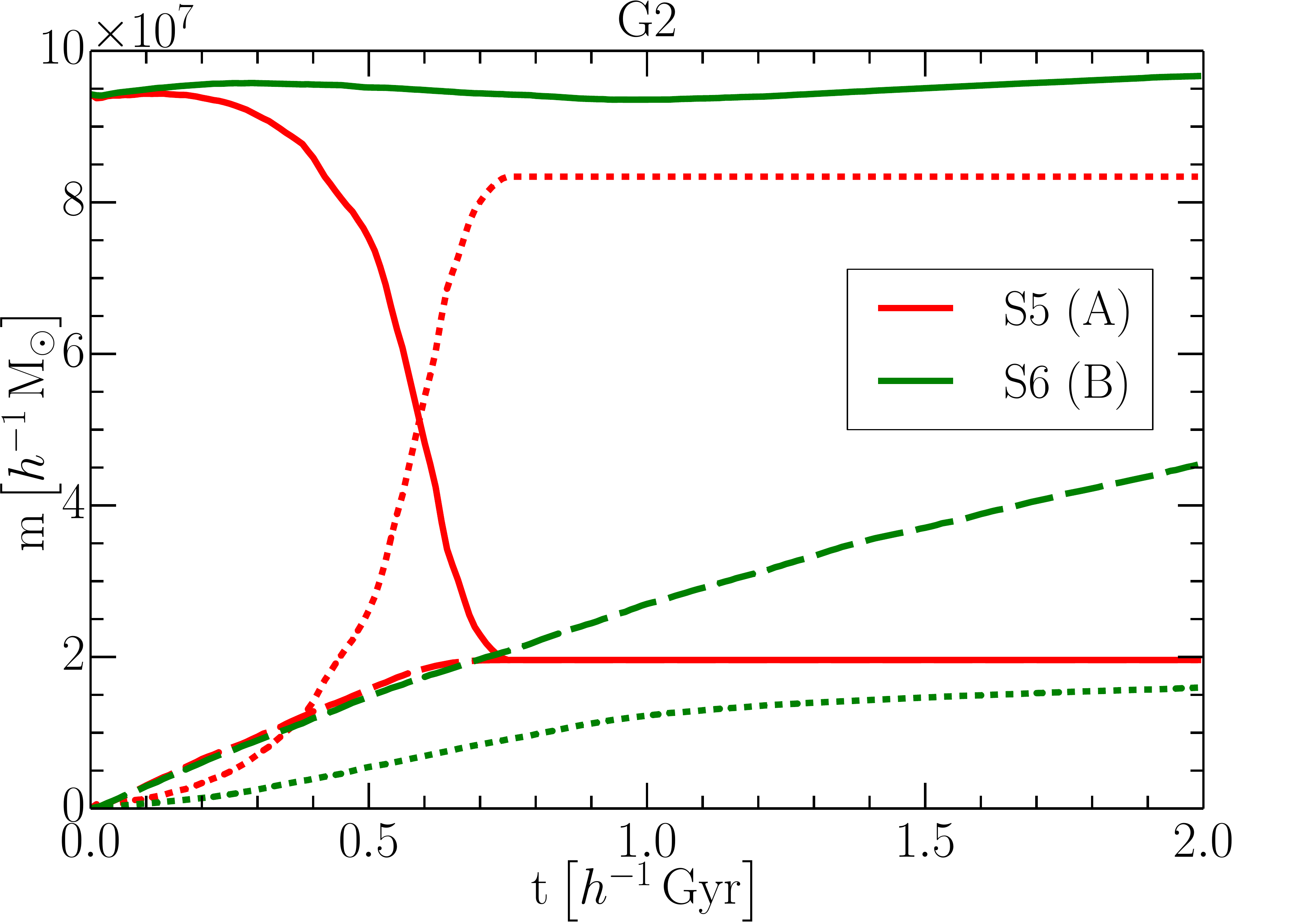}\label{fig:metals_strip_b}}
  \subfigure[]{\includegraphics[width=\linewidth]{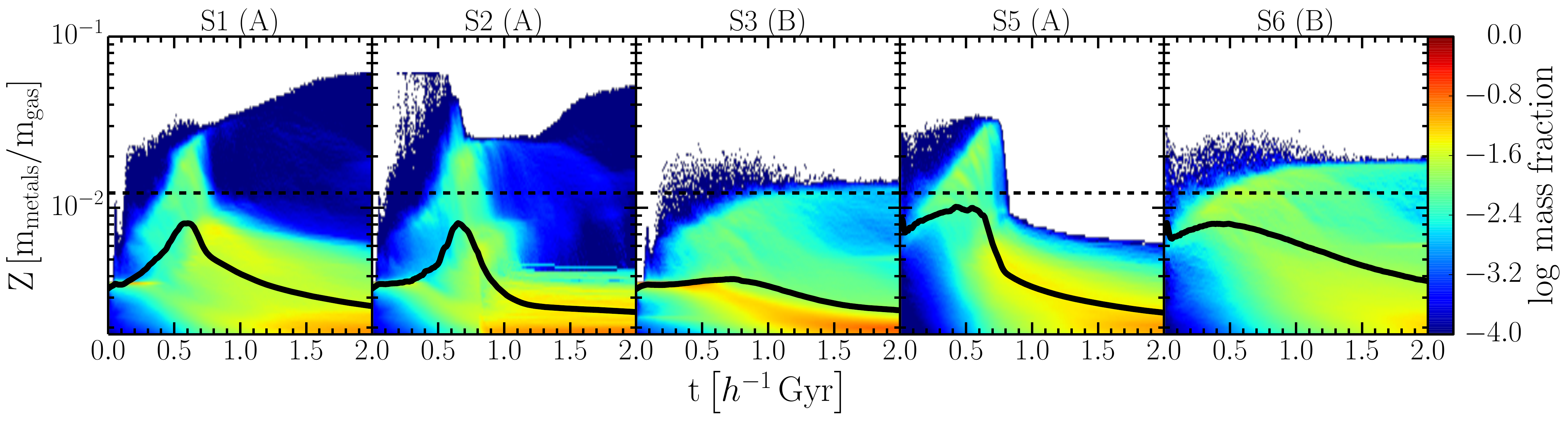}\label{fig:metals_strip_c}}
  \caption{Panels \subref{fig:metals_strip_a} and
    \subref{fig:metals_strip_b} show the evolution of metal mass in
    both model galaxies as a function of simulation time for
    interactions with clusters~A or B, respectively. The solid lines
    indicate the metal mass in both gas and newly formed stars that 
    are gravitationally bound to the galaxy, 
    whereas the dashed lines give the metal mass 
    locked up only in newly formed stars. The dotted lines indicate the metal 
    mass in the stripped gas wake. Panel \subref{fig:metals_strip_c} shows
    metallicity histograms of stripped gas (considering gas cells not
    gravitationally bound to the galaxy anymore and with ISM mass
    fraction > 1\%) as a function of time.  The colour encodes the mass 
    fraction in each metallicity bin of the total gas mass stripped for 
    each timestep. The black solid line shows
    the mean metallicity value of the gas cells in the histogram,
    while the dashed line indicates $Z_\odot$. 
    }
  \label{fig:metals_strip}
\end{figure*}

Very similar results are found for model galaxy G2 and are presented in
Fig.~\ref{fig:stripping_p2h3}.  This smaller galaxy gets stripped
completely when passing the central region of cluster A.  Passing
through clusters B and C, the stripping radius drops from the initial
$15$ to $6$ and $4-5\,h^{-1}\mathrm{kpc}$, respectively.
Similar to galaxy G1, in cluster C the redistribution of
the gas in the disk can be observed after pericentre passage when the 
strength of RP is decreasing again. However, the effect is limited
to the outer parts of the disk. The surface density at smaller radii, 
not affected by RPS, shows no change compared to the initial profile.

\begin{figure*}[]
  \centering
  \subfigure[]{\includegraphics[width=0.49\textwidth]{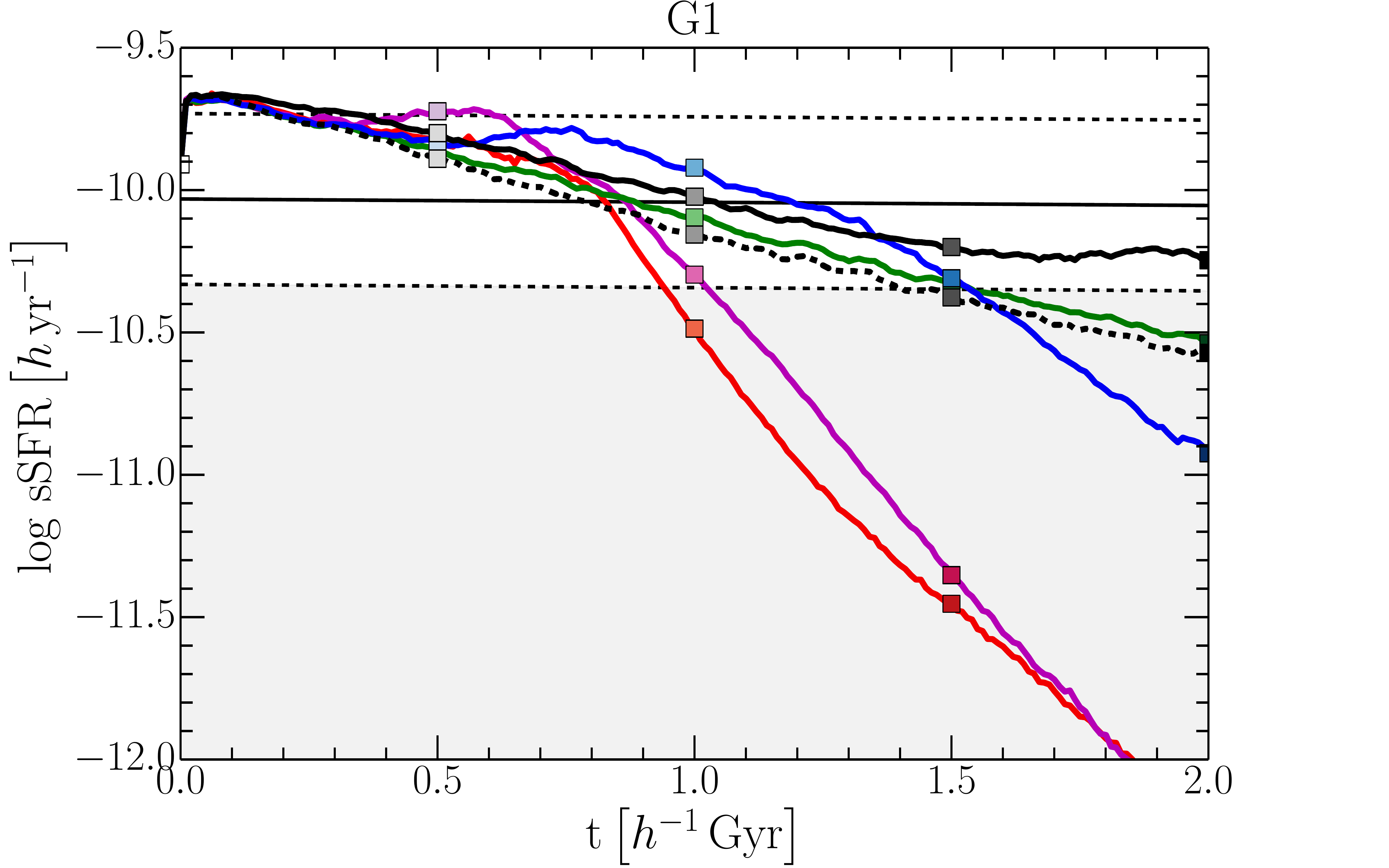}\label{fig:sfr_main_1}}
  \subfigure[]{\includegraphics[width=0.49\textwidth]{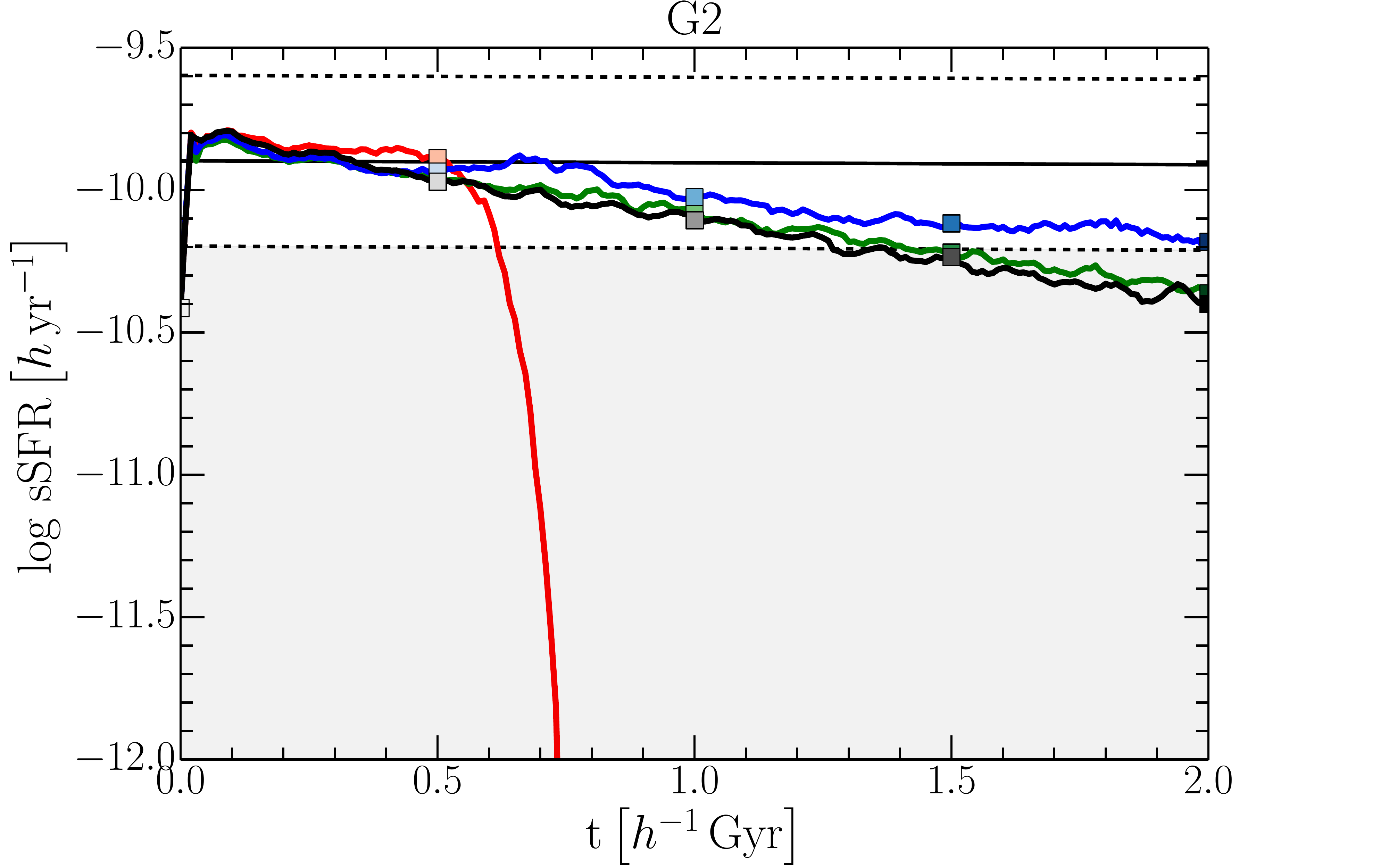}\label{fig:sfr_main_2}}
  
  \subfigure[]{\includegraphics[width=0.49\textwidth]{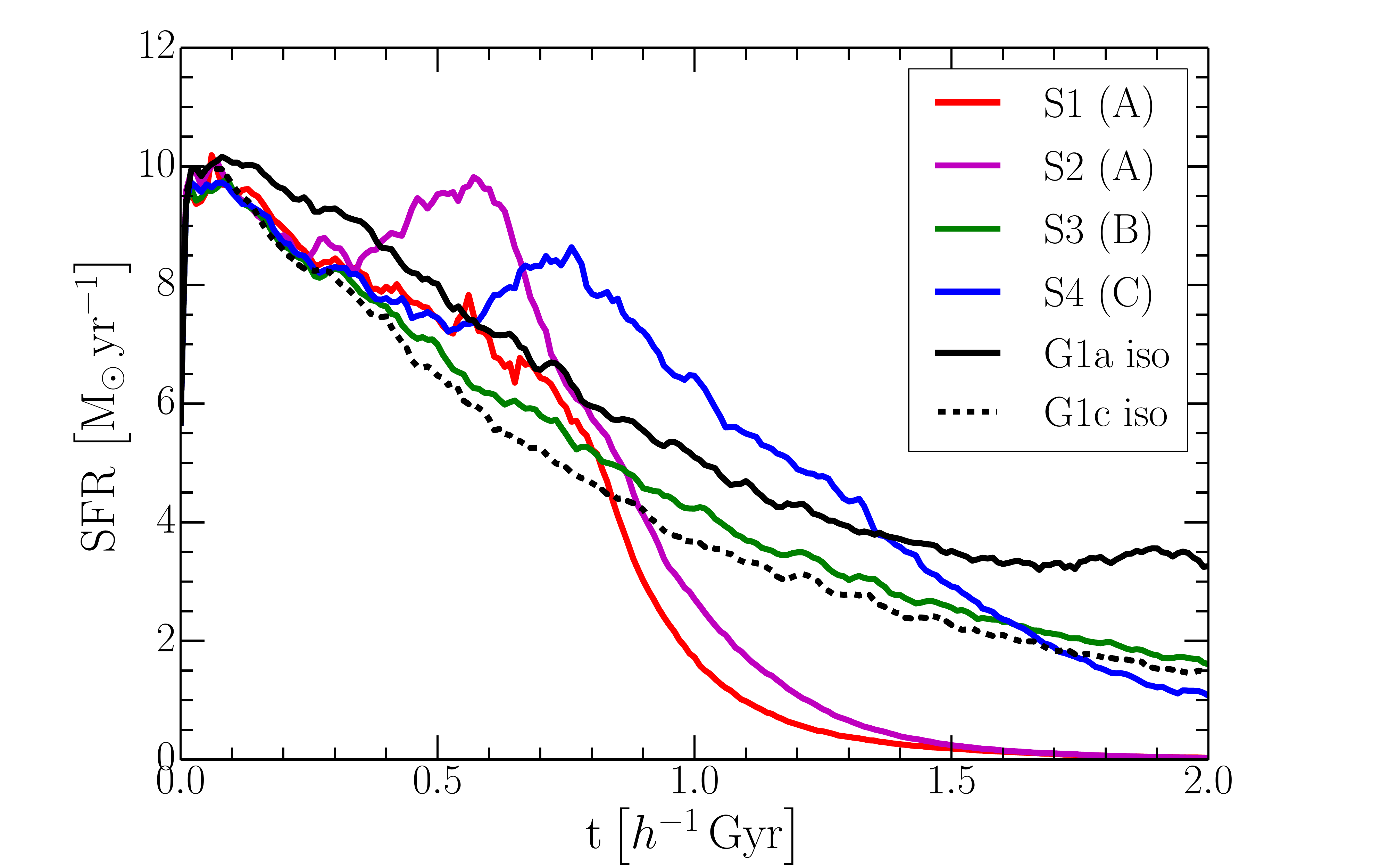}\label{fig:sfr_1}}
  \subfigure[]{\includegraphics[width=0.49\textwidth]{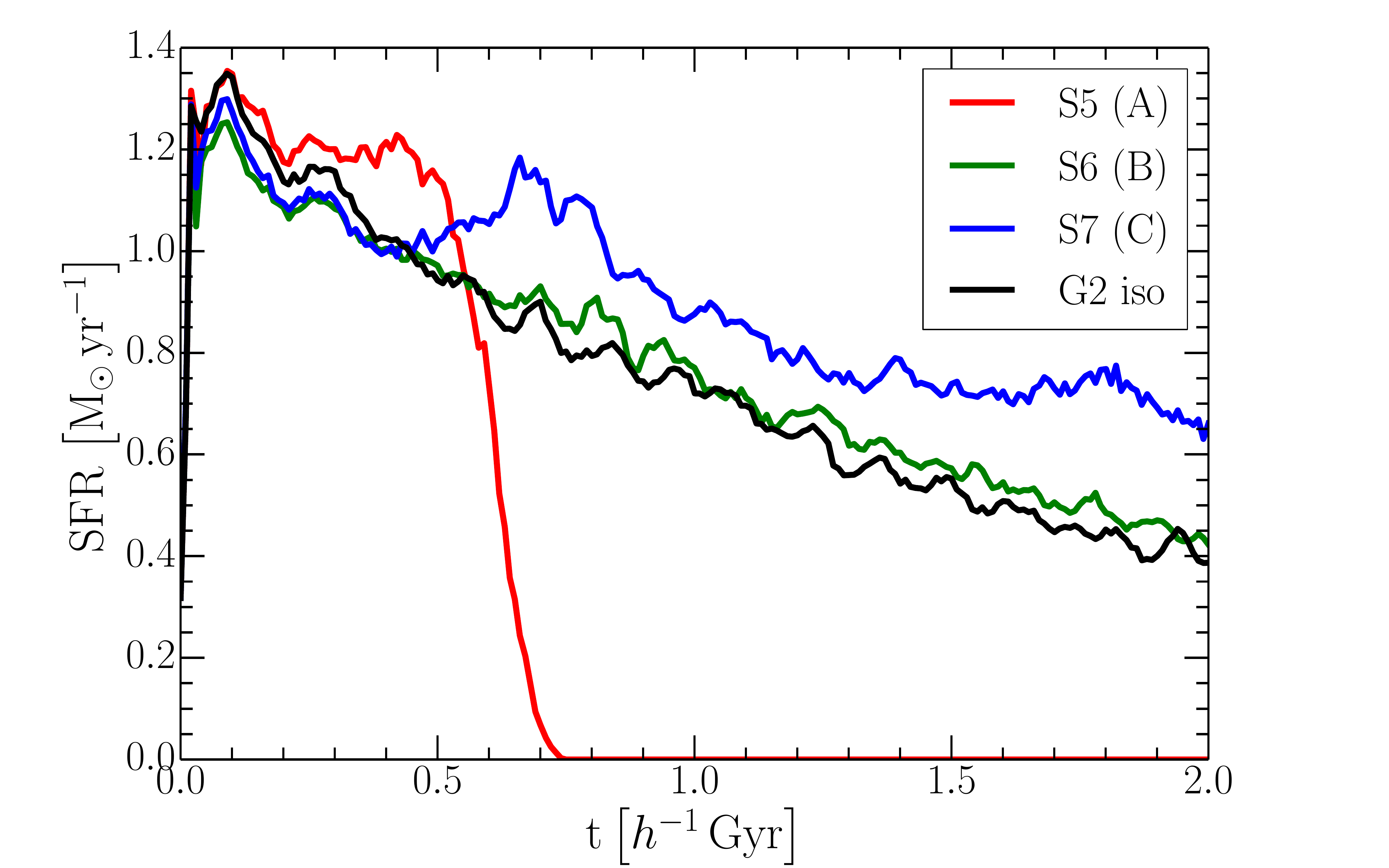}\label{fig:sfr_2}}
  
  \caption{(a,b) Evolution of the $\mathrm{sSFR}\equiv \mathrm{SFR}\,h\,\mathrm{M}_\star^{-1}$
    of model galaxies G1a, G1c, and G2 in isolation and cluster environments. 
    The value of the main sequence of SF for the particular galaxy (black solid line) and 
    0.3 dex scatter (dashed black lines), taken from \citet{Schawinski2014},
    are shown. The grey shaded area indicates the region where they find 'green-valley' galaxies.
    (c,d) SFR of the model galaxies.}
  \label{fig:sfr}
\end{figure*}

The stripping radii calculated with theoretical RPS models
described above can differ by as much as $5\,h^{-1}\mathrm{kpc}$ with
corresponding consequences for the estimated gas mass that is
stripped as well, especially when it comes to the inner parts of the disk.
The best model fitting our simulations depends on environment, but also
on the mass of the galaxy itself. For the more massive galaxy, G1,
the theoretical model, which takes into account the appearance of supersonic
velocities and the current gas and stellar mass in every
snapshot, yields results closest to
runs S2 and S3, if one also considers the mass lost after pericentre
passage from continuous stripping in the simulations. With different initial
inclination of the galaxy (run S1), the standard model and when
RP is scaled with the effective inclination of the galaxy also provides
reasonable results. In run S4, with a high peak of RP, using the
initial disk mass gives the best results.  For galaxy G2, the
model that best fits all simulations includes Rankine-Hugoniot
post-shock conditions and uses the initial disk mass, hence assumes
a constant surface density profile to calculate the stripping
radius. However, in all of the cases and for both model galaxies,
considering the total mass in the disk yields much better results
than the standard criterion. All of the mentioned models are depicted
in the panels on the right-hand side in
Figs.~\ref{fig:stripping_p1h1} and \ref{fig:stripping_p2h3} as
yellow dashed lines.
Furthermore, SF needs to be taken into account, as a
sizable fraction of gas in the disk is transformed into stars on the
timescale of one cluster centre passage 
(see also Sect.~\ref{sec:samcomparison}).

In Fig.~\ref{fig:ortho_gas_metals}, we show $100\times40\,h^{-1}\mathrm{kpc}$
slices in the $yz$-plane that are centred on the disk of the galaxy of run S1 at
time $600\,h^{-1}\mathrm{Myr}$ (just at pericentre passage), which are
produced using adaptive kernel smoothing by first computing SPH-like
smoothing lengths for the mesh-generating points and then distributing
the corresponding quantity onto a regular grid using the SPH kernel.
Although in the chosen snapshot the highest RP of all simulations acts
on the galaxy, losing mass at the highest rate, gas is still diluted
and heated fast as soon it is stripped.  Around
$50\,h^{-1}\mathrm{kpc}$ behind the disk, most of the gas has heated
up and is already diluted to the background density of the ICM.  On
the other hand, mixing with the ICM proceeds at a slower rate as can
be seen in the lower panel: the metallicity shows a slower decline in
the stripped wake.

The stripping of the hot gas halo proceeds very fast in all of the
simulations.  After $50 - 150\,h^{-1}\mathrm{Myr}$, 80-90\% of the gas
mass in the hot halo is being stripped.  Only a small fraction that
either resides very close to the galaxy or is in the slipstream of the
disk stays bound to the galaxy.  In SAMs \citep[e.g.][]{Guo2011},
stripping of the hot gas halo is very often modelled with two
processes: by either assuming full stripping at a radius where the RP
exceeds the gas pressure of the halo or by invoking tidal stripping in
which the gas mass is reduced from the hot halo in direct proportion
to the DM halo mass during infall.  Looking at our
simulations, only the first process comes into play, as the stripping
advances quickly, already in the outskirts of the cluster, whereas the
DM halo loses mass significantly only after pericentre
passage. We also note that \citet{McCarthy2008} have shown that
the truncation radius due to tidal stripping is only smaller compared 
to the RPS radius if the galaxy exceeds 10\% of the
 cluster mass, which is not the case in any of our simulations.

\subsubsection{Metal enrichment}
\label{sec:metal_enrichment}

Ram-pressure stripping is an important process \citep{Domainko2006}
contributing to the metal enrichment of the ICM in galaxy clusters
\citep[e.g.][]{Schindler2005, Schindler2007, Werner2008, Zhang2011}.
Here, we investigate the enrichment of the ICM by means of RPS from a
single galaxy.  The treatment of SF we use
\citep{Springel2003} tracks only the total metal mass and assumes solar
composition.  This is done by means of a scalar value that is advected
with the calculated mass flux derived from the hydrodynamic solver,
which is similar to the colouring technique employed in the refinement. The
metal yield is set to $p=0.0122$ \citep{Asplund2005} and is used
throughout all of the simulations.

As we start our simulations with a gas fraction of 35\% in the disk,
the initial metal content is important for the metal enrichment of the 
ICM due to RPS. Observations show a huge diversity in
radial metallicity profiles of disk galaxies
\citep[e.g.][]{Zaritsky1994, Bresolin2012}.  However, as in
\citet{Holler2014}, we decided to model the metallicity gradient in
our model galaxies using mean values of observations from
\citet{Ferguson1998}, 
\begin{equation} 
  \log \left(O/H\right) = -2.8 -
  0.65\, r/\mathrm{R}_{25} ,
\end{equation}
approximating $\mathrm{R}_{25} \simeq 3.5 r_0$, and using solar
values from \citet{Asplund2005} again to translate from oxygen abundance to
total metallicity values.  The total metal mass in the gas disk is
given by the mass of the old stellar disk, according to the model
presented in Sect.~\ref{sec:colour_evolution}.

\begin{figure}[htb]
  \centering
  \subfigure[]{\includegraphics[width = \linewidth]{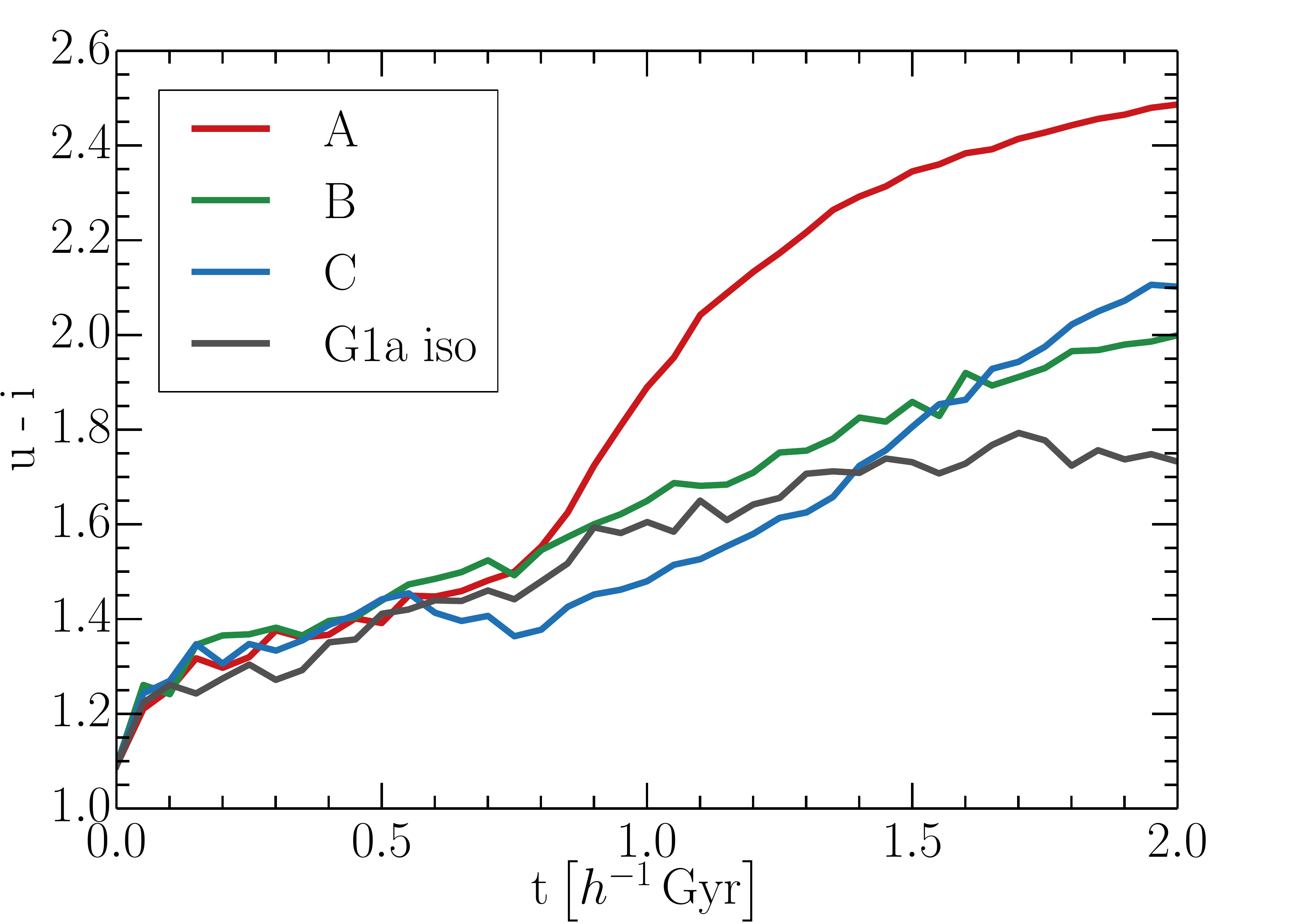}}
  \subfigure[]{\includegraphics[width = \linewidth]{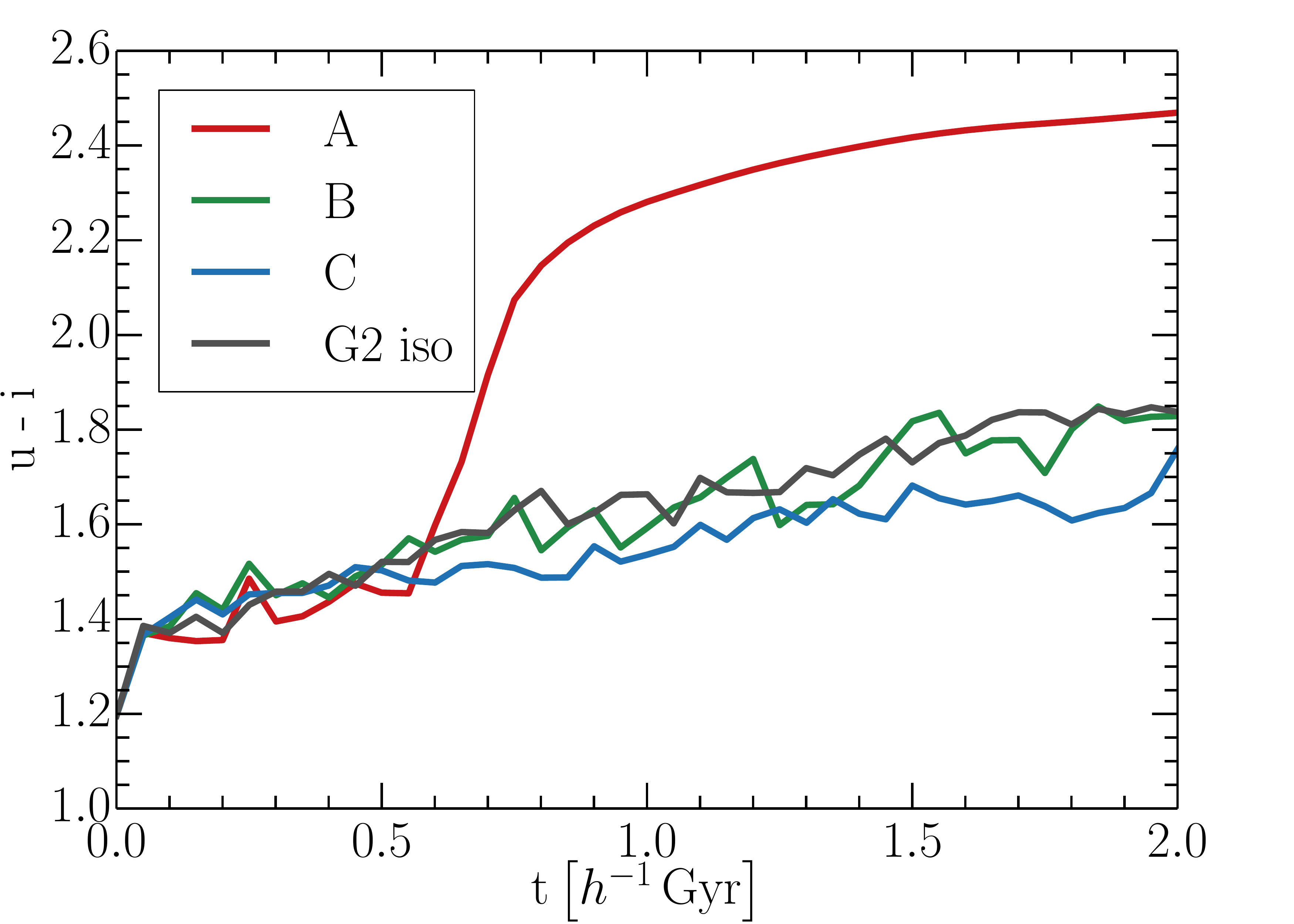}}
  \caption{Colour evolution (in the u-i index of the SDSS) of galaxies
    G1 (top panel) and G2 (bottom panel) as a function of time while
    interacting with different cluster environments. The slow
    evolution towards redder colours in the interactions with
    clusters~B and C is actually very similar to the evolution of the
    galaxies in isolation, showing that the gas consumption timescale
    is not significantly affected by RPS in these
    models. This is different in the deep interaction with cluster A,
    where a rapid shutdown of SF is achieved.}
  \label{fig:u_i_time}
\end{figure}

\begin{figure}[htb]
  \centering
  \subfigure[]{\includegraphics[width = \linewidth]{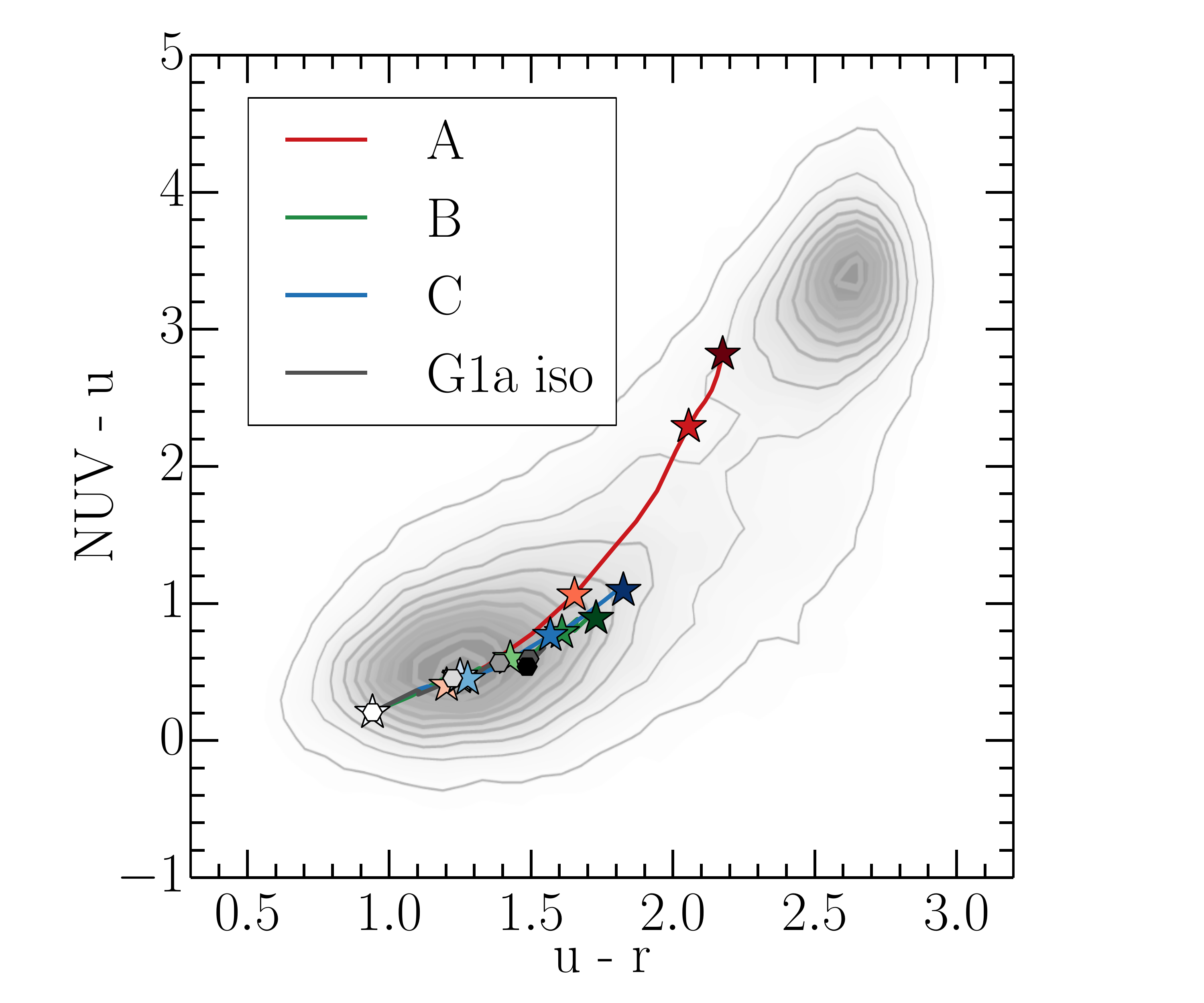}}
  \subfigure[]{\includegraphics[width = \linewidth]{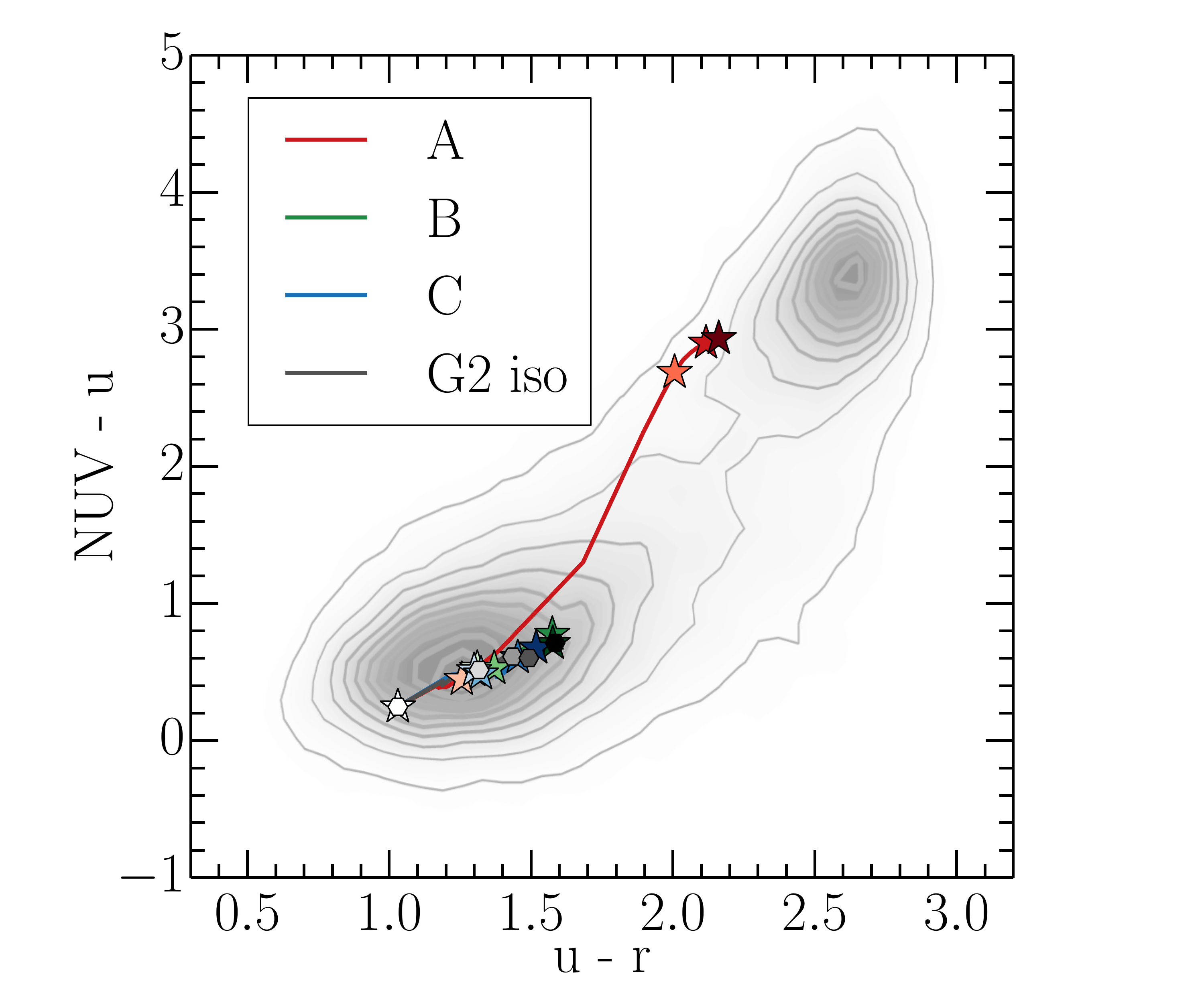}}
  \caption{Colour-colour diagrams showing the evolution of galaxies in
    NUV-u vs. u-r space as they interact with different clusters. The
    top panel is for model galaxy G1, and the bottom panel for G2. The
    colour is indicating different environments as shown in the
    legend.  Stars mark simulation times with their brightness
    encoding 0, 0.5, 1.5 and $2\,h^{-1}{\rm Gyr}$ (the brightness
    corresponds to the one used for the big squares in
    Figs.~\ref{fig:sfr_main_1} and \ref{fig:sfr_main_2}). The grey shaded
    contours represent the densities of the observed early- (upper
    right concentration) and late-type (lower left peak) galaxy
    populations, as determined by \citet{Schawinski2014}.}
  \label{fig:nuv_colour}
\end{figure}

In Figs.~\ref{fig:metals_strip_a} and \ref{fig:metals_strip_b}, we
plot the total mass of metals in gas and stars still gravitationally bound to
the galaxy (solid lines) and metals locked up in newly formed
stars (dashed lines).  Furthermore, the metal mass in stripped gas 
(dotted lines) for galaxies G1a and G2 is depicted, respectively. Here
we do not include the metals in the old stellar disk, as the mass
there does not change throughout the simulation and no stars are
stripped.  As expected, as SF is quenched after pericentre
passage in cluster~A in both galaxies, fewer metals are present in
total than in cluster~B, although the effect is very small.  The more
massive galaxy can enrich the ICM with around
$2 \times 10^{8}\,h^{-1}\mathrm{M}_\odot$, which is almost half of its
initial metal content in the ISM.  The smaller galaxy
loses $0.8 \times 10^{8}\,h^{-1}\mathrm{M}_\odot$, which
accounts for about 80\% of its initial metal mass.  
In cluster~B, where only weak but continuous stripping takes place,
only a few metals ($<5\%$ for G1a, $<20\%$ for G2) are transported to the
ICM because only the outer parts of the galaxy with a lower metallicity
are stripped.  
Hence, with respect to their gas mass and metal content in the disk, 
the smaller galaxy G2 is enriching the ICM with a relative higher amount 
of metals. We note that the gas of G2 is more 
easily stripped because of the shallower gravitational potential when
the same RP is exerted.
  
The inclination of the galaxy has an effect on the metal enrichment
of the ICM and the amount of stripped gas.
The purple line corresponds to run S2 with the galaxy starting
edge-on. During pericentre passage, more metals are locked up
in newly formed stars, as SF is enhanced with respect to run S1 
(see Sect.~\ref{sec:sfr}).
In total, less gas is stripped in
S2 with respect to S1 (see Sect.~\ref{sec:rps}), which translates to a
relatively smaller amount of metals transported to the ICM.
  
The differences however are more evident in Fig.~\ref{fig:metals_strip_c}.
We plot the metallicity $Z$ of stripped gas in a histogram ranging
from $0.15\,Z_\odot$ (our initial value adopted for ICM gas) to
$5\,Z_\odot$ for each snapshot throughout the simulation.  Thereby, we
consider all gas cells with at least 1\% of its mass originating in
the galaxy itself.  The colour encodes the mass fraction of
  stripped material in different metallicity bins.
The black solid line indicates the mean metallicity of the gas
considered in the histogram.  In cluster~A, the mean metallicity in
stripped material for both simulations with the galaxy starting
edge-on and face-on are not very different, reaching a peak of almost
$Z_\odot$ at pericentre passage.  Afterwards, when no gas is stripped
anymore and owing to the mixing with the ICM, the mean metallicity
decreases again.  However, more metal-rich gas from the centre of the
galaxy is stripped edge-on in the first $500\,h^{-1}\mathrm{Myr}$ of
the simulation, which is diluted quickly afterwards.  Stripping is going
on for the larger galaxy until the end of the simulations.  In both
cases, progressively more of the enriched material from the centre is
stripped, but the total amount nevertheless stays  small.  
The stripped material of the smaller galaxy on the other hand
has a higher metallicity right from the beginning, as the metal-rich
inner parts of the disk are stripped earlier. However, mixing of ISM and
ICM in the wake proceeds in the same way for both galaxies (runs S1, S2, and S5,
respectively), hence considering that they have a similar ISM metallicity
their mean metallicities in the wake also end up  
the same at the end of the simulations when all the gas has been
stripped.

In cluster~B, with a slow but continuous stripping rate,  the gas
wake of the smaller galaxy has a higher metallicity again, reaching up to
$1.6\,Z_\odot$ with the mean value reaching as high as $0.6\,Z_\odot$.
For both galaxies, the supply of stripped, enriched material balances
mixing with the ICM.  Mixing and hence a drop of the metallicity can
also be seen in Fig.~\ref{fig:ortho_gas_metals_metals},
where a slice of the metallicity in the stripped wake of run S1 after
$600\,h^{-1}\mathrm{Myr}$ is shown.  The metallicity of stripped gas
right behind the galaxy drops by a factor of 2 - 3 already after
around $100\,h^{-1}\mathrm{kpc}$ behind the disk while the gas is
mixed with the ICM.

\begin{figure*}[htb]
  \centering
  \includegraphics[width=\linewidth]{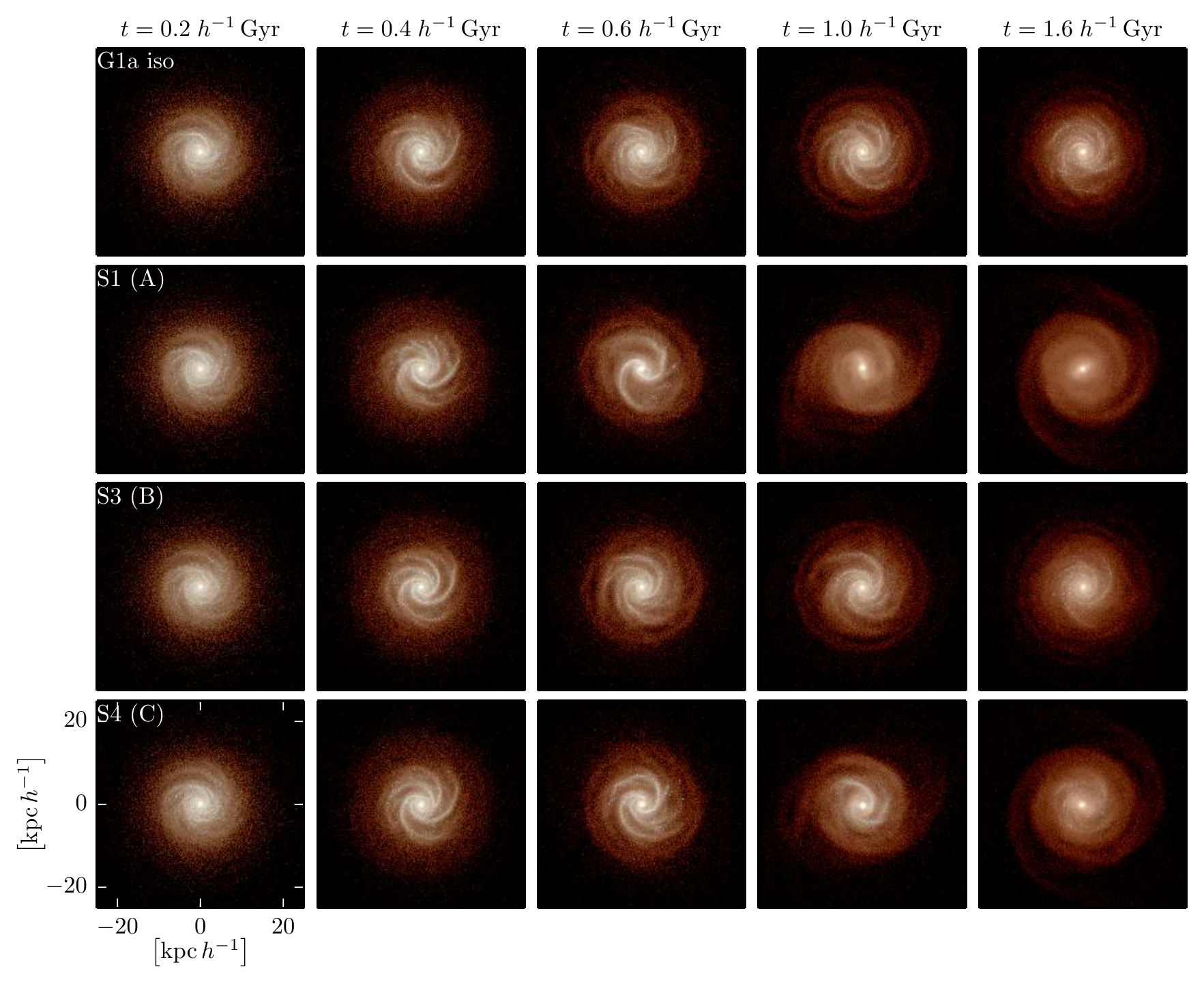}
  \caption{Mock colour images of galaxy G1a in different
      environments.  The evolution in isolation and runs S1, S3 and S4
      are depicted in the different rows.  In each column, the state
      at a selected time (indicated on top) is shown for each
      simulation.}
  \label{fig:colour_img_A}
\end{figure*}

\subsection{Star formation}
\label{sec:sfr}

In the following section, we investigate the SFR
in the simulations with a focus on the dependence of galaxy mass and
environment on quenching. 

As the hot gas halo of the more massive galaxy G1a is 
stripped right at the beginning of the simulation, in the first
$500\,h^{-1}\mathrm{Myr}$, the SFR in all of the clusters drops to the values
of the isolated galaxy without hot gas halo (G1c); this is also because
in this phase, not much gas in the disk is lost due to RP 
(see Fig.~\ref{fig:sfr_1}). 
In cluster~B, the SFR of the galaxy follows the same profile as 
G1c in isolation throughout the whole simulation. The weak RP in this case,
which only leads to continuous stripping in the outskirts, does not
affect the SFR significantly.  In contrast, in cluster~C, at the short
but high peak of RP, the SFR is enhanced by at most
$2\,\mathrm{M}_\odot \, \mathrm{yr}^{-1}$ as the disk gets
compressed.  Afterwards, the SFR only slowly declines and finally
drops below the SFR in isolation from a lack of gas in the disk.  
A similar scenario happens in cluster~A: first SFR is enhanced and then
drops below isolation levels after big parts of the gas disk are stripped and 
finally  depleted by residual SF. 
However, this process proceeds much faster than in cluster C, since a higher 
amount of gas is stripped during pericentre passage because of the stronger RP.
In run S1, only a slight enhancement of the SFR
to isolation levels (G1a) can be observed. As a result of a different inclination
and consequently delayed gas stripping in run S2, more gas remains in the 
galaxy at pericentre passage when the gas disk is being compressed most.
In that case, the SFR can be enhanced by up to
$3\,\mathrm{M}_\odot \, \mathrm{yr}^{-1}$, but drops in the same way
afterwards.
However, almost $10^9\,h^{-1}\mathrm{M}_\odot$ more
stellar mass can be formed in total by the end of the simulation.  In
cluster~C, the cumulative amount of new stars is largest, and very
similar to the amount formed in isolation.

The less massive galaxy G2 shows basically the same behaviour as G1 in 
the distinct cluster environments (see Fig.~\ref{fig:sfr_2}).
In cluster~B, similar to galaxy G1, SFR is slightly enhanced compared 
to simulations of the respective galaxy in isolation, including the least massive 
hot gas halo (G1c and G2, see Table~\ref{tab:galaxy}). In cluster~A, again an initial 
enhancement of the SFR can be observed before most parts of the gas disk are 
stripped. However, the momentary enhancement of around 20\% is smaller than for 
galaxy G1, which shows an increase of 30-60\%, depending on the initial inclination
of the galaxy. Finally, the biggest difference between the two model galaxies
can be found in cluster~C. Despite the similar enhancement of the SFR at
pericentre passage, compared to G1 the SFR of G2 is higher than in isolation until 
the end of the simulation, although relatively more gas is stripped in G2.

The specific SFR
($\mathrm{sSFR} \equiv \mathrm{SFR}\,h \mathrm{M}_\star^{-1}$) is
shown in Figs.~\ref{fig:sfr_main_1} and \ref{fig:sfr_main_2}.
The value of the main sequence of SF and 0.3 dex scatter for the
  particular stellar mass of both model galaxies, taken from
  \citet{Schawinski2014}, are represented with the horizontal black
  solid and dashed lines, respectively.  The grey shaded area below
  the main sequence indicates the region where 'green-valley'
  galaxies, a transition population between star-forming and passively
  evolving quenched galaxies, can be found.
The more massive galaxy starts its evolution 0.3 dex above the
main sequence of \citet{Schawinski2014}.  Throughout the simulation in
isolation and in cluster~B, the galaxy stays within the 0.3 dex
scatter around the main sequence.  In cluster~C, the galaxy only   falls
below the main sequence at the end of the simulation.  In the case of
extreme RP (cluster~A), the galaxy leaves the main sequence only after
$1\,h^{-1}\mathrm{Gyr}$ of simulated time, a few hundred
$h^{-1}\mathrm{Myr}$ after pericentre passage when the peak of RP is
experienced.  The difference in run S2 for a different inclination of
the galaxy is again small, except for the more pronounced
enhancement after around $600\,h^{-1}\mathrm{Myr}$. 
Afterwards, the galaxy still needs half of
the simulation time to pass through the green valley region in both
runs S1 and S2.  The smaller galaxy also stays on the main sequence
of SF in all environments, except in cluster~A.  As the SFR suddenly
drops to zero when all the gas is stripped during pericentre passage,
the galaxy leaves the main sequence quickly and passes through the
green valley in less than a hundred $h^{-1}\mathrm{Myr}$.

In summary, a weak, continuous RP (B) removes the gas supply from
the hot halo but does not further influence the SFR. On the other hand,
a strong, short RP event (C) , although removing double the gas mass
from the disk through stripping, can compress the disk in such
a way that the SFR is enhanced such that the same amount of stars is
formed by the end of the simulation as in the isolated case.  
A galaxy can lose all of its gas and drastically quench its SF only in
extreme cases of RPS (A). As expected, as less massive galaxies are
more prone to lose their gas due to RPS, they are also quenched more
rapidly.

\subsection{Colour evolution}
\label{sec:colour_evolution}

In this section, we study the colour evolution of the model galaxies
evolving in isolation and undergoing RPS.  We are
particularly interested in how fast a galaxy becomes red when
infalling to a galaxy cluster. Furthermore, studying the colour
evolution of the galaxy, we can compare with observations and
investigate the morphology by generating mock colour images.

As the stellar particles of our model galaxies have masses in the
range of $\sim 10^4 - 10^5\,\mathrm{M}_\odot$, they can be considered
simple stellar populations \citep[SSP, 
e.g.][]{Querejeta2015, Murante2015}.
We are using the flexible
stellar population synthesis code \texttt{FSPS} \citep{Conroy2009,
  Conroy2010} to calculate SSPs out of the stellar particles of the
simulation.  The output of \texttt{FSPS} includes observed mock
spectra of an SSP given its metallicity and age.  By convolving with a
desired filter function, the luminosity and hence magnitude of the SSP
in the desired band can be obtained.  In this work, we use the
SDSS bands u, g, r, i, and z as well as Buser B, U, V, and IR K filters.
We then calculate the colour for a stellar particle (SSP) via a
2D interpolation of the \texttt{FSPS} tables for different stellar age
and metallicity.

Newly formed stars generated during the simulation inherit the
metallicity of the gas cell they were formed of.  Also, the age of the
SSP for those particles is known as the time of generation is
stored. For such particles, colour calculation is straightforward, but
it is more difficult for the old stellar disk and old bulge particles
(type 2 and 3) already present in the initial conditions (see
Sect.~\ref{sec:galaxy_cluster_models}) because they do not contain age
and metallicity information.  Therefore, for our model galaxies we
assume a $\tau$-model for SF prior to the start of the
actual simulations.  We use Eq. 1 of \citet{Feldmann2015} to
calculate the SF history of our model galaxy,
\begin{equation}
  \mathrm{SFR}(t) = \mathrm{SFR}(t_\star)\left[ (1-c) \exp\left(\frac{t-t_\star}{\tau}\right) + c \right],
\end{equation}
 and assign age and metallicity to the old stellar disk and bulge
 particles.  In this model, the SFR
 of the galaxy peaks at time $t_\star$. For our model galaxies this occurs at the
 beginning or shortly after the start of the simulation. The value  $\tau$ is
 the SF timescale. The value of $\tau$ is positive for an
 exponentially growing SFR, which we assume to be the case for our
 model galaxies for times prior to the start of the simulation. Hence,
 $\mathrm{SFR}(t_\star)$ is the peak SFR, which we get
 directly from our simulations.  The constant $c$ is an offset for the
 exponential growth or decline of the SFR, modelling
 different phases of galaxy formation like gas accretion or starvation
 phase \citep[for more details see][]{Feldmann2015}. We neglect these
 different phases of galaxy formation and calculate the parameters of
 this model for our simulated galaxies in the following way.  Given the
 mass of the old stellar disk and bulge, the integral of the SF
 history is set. Assuming that the overall age of our model galaxies at the
 beginning of our simulations is 7 Gyr and given the peak star
 formation rate $\mathrm{SFR}\left(t_\star\right)$ from the
 simulation, all parameters can be set.  We use $c=0$ for all of our model galaxies.
The stellar mass produced in each
time bin can be calculated by dividing the 7 Gyr of galaxy formation into time bins given the
SFR by the above described model.  Randomly, a number of stellar particles are selected,
first from the bulge \citep[which is usually considered older than the
disk, e.g.][]{Athanassoula2005} and then from the old stellar disk
according to the mass in the time bin. The time of this
bin is then assigned to those particles.  To compute metallicities for
those particles (see also Sect.~\ref{sec:metal_enrichment}), we proceed 
as in \citet{Springel2003}.  Given the produced stellar mass,
the metallicity of surrounding ISM can be calculated immediately for
each time bin.  The metallicity obtained in this way is assigned to the
stellar particles the same way as their age.  Values for age and
metallicity for the old stellar and bulge particles are assigned once
and used to calculate galaxy colours for all snapshots of the
simulation to correctly model morphological evolution of the galaxy.

In Fig.~\ref{fig:u_i_time} we present the SDSS u-i colour evolution
of our different runs.  The values agree well with SDSS data from
\citet{Weinmann2010} for both of our model galaxies.  In isolation,
both galaxies have practically the same colour.  Also in clusters~B and
C, the colour is very similar to the isolated case, although the SFR is
quite different, tending slowly towards a redder colour. As soon as SF is quenched, the colour turns red quite quickly   in
cluster~A alone; this is
the case especially for the less massive galaxy on timescales of a few hundred
Myr.  Afterwards, reddening proceeds again at a smaller pace,
comparable to galaxies that still form stars.

\begin{table*}[tb]
  \centering
  \caption{Different runs carried out for the resolution study.}
  \begin{tabular}{l|rrrrr}
    \hline \hline
     & & ICM target & \# stellar particles / & \# DM particles / & refinement \\
    run & \# gas cells & volume $\left[h^{-3}\,\mathrm{kpc}^3\right]$ & softening $\left[h^{-1}\,\mathrm{kpc}\right]$ & softening $\left[h^{-1}\,\mathrm{kpc}\right]$ & method \\
    \hline
    R1 & 21000 & 383.80 & 44000 / 0.130 & 60000 / 0.280 & ad-hoc \\
    R2 & 21000 & 383.80 & 44000 / 0.130 & 60000 / 0.280 & full \\
    R3 & 210000 & 52.72 & 440000 / 0.060 & 600000 / 0.130 & ad-hoc \\
    R4 & 210000 & 52.72 & 440000 / 0.060 & 600000 / 0.130 & full \\
    R5 & 210000 & 20.00 & 440000 / 0.060 & 600000 / 0.130 & ad-hoc\\
    R6 & 2100000 & 5.24 & 4400000 / 0.026 & 6000000 / 0.072 & ad-hoc\\
    \hline \hline
  \end{tabular}
  \vspace*{0.2cm}
  \tablefoot{All of the runs correspond to the set-up of run S1 (see Sect.~\ref{sec:merger_orbits}). 
    The softening is scaled with the cubic root of the mass resolution of cells/particles.}
  \label{tab:resolution_runs}
\end{table*}

When SF is quenched, immediate differences can be seen at
short wavelengths.  Therefore, in Fig.~\ref{fig:nuv_colour}, we show
a colour-colour diagram considering NUV-u versus u-r colours according
to \citet{Schawinski2014}.  By investigating the different paths of
the model galaxies, one can infer quenching timescales as well.  The
grey shaded contours in these plots indicate the early- and
late-type galaxy populations that we reproduced from \citet{Schawinski2014}.
The colour of the lines corresponds to different paths for the
distinct simulations with the shading indicating simulation time as
in Figs.~\ref{fig:sfr_main_1} and \ref{fig:sfr_main_2}.  Every
$500\,h^{-1}\mathrm{Myr}$ is indicated by a star for galaxies in a
cluster environment and by a circle for galaxies in isolation.
In all of the scenarios, except for the extreme RP case, the model galaxies
stay within the late-type population.   The path for
both galaxies connects to the early-type population in cluster~A alone.  Whilst the
more massive galaxy needs almost a $h^{-1}\mathrm{Gyr}$ to pass the region between the
two galaxy populations, referred to as green valley, the transition
for the smaller galaxy is much faster, taking a few hundred
$h^{-1}\mathrm{Myr,}$ because SF is quenched immediately.

Finally, we produce spatially resolved mock observations of the model
galaxies such as those in \citet{Lupton2004},  however, we use a logarithmic
stretch for the colours and mapping IR K to red, Buser B to green, and
Buser U to the blue image channel following \citet{Marinacci2014}.
First, to generate an image, particle colours need to be mapped to a
grid.  To this extent, we use the same approach as for
Fig. \ref{fig:ortho_gas_metals}, distributing colour information of the
particles with an SPH kernel to a regular grid.  Afterwards, a
projection of the different bands is mapped to RGB image channels to
generate a 2D image.  The evolution of galaxy G1a in isolation
and the three model clusters (runs S2, S3, and S4) is shown in 
Fig.~\ref{fig:colour_img_A}, respectively.
Clearly, in isolation, the spiral structure of the galaxy is retained
throughout the whole simulation.  Also, more blue light is coming from
the spiral arms where SF is also going on at the end of the
simulation.  The morphology of the galaxy is almost the same in
cluster~A for the first $\sim500\,h^{-1}\,\mathrm{Myr}$.  At pericentre
passage, morphology gets distorted as a result of tidal interaction with the
gravitational potential of the galaxy cluster.  In both cases, a bar in the
centre of the galaxy is visible and, apparently, in the cluster the
spiral structure is lost.  With moderate RP in run S3, the spiral structure 
is retained the same way as in the isolated case.  However, the galaxy is 
redder in the last snapshot presented at $t=1.6\,h^{-1}\mathrm{Gyr}$, as the
SFR drops below the isolated case.  On the other hand, in run S4, morphology
gets distorted, especially in the outer parts, showing the typical patterns 
of tidal interaction, although the overall colour of the galaxy is the same
as in run S3.

\subsection{Comparison to semi-analytic models}
\label{sec:samcomparison}

\begin{figure}[htb]
   \centering
   \subfigure[]{\includegraphics[width=\linewidth]{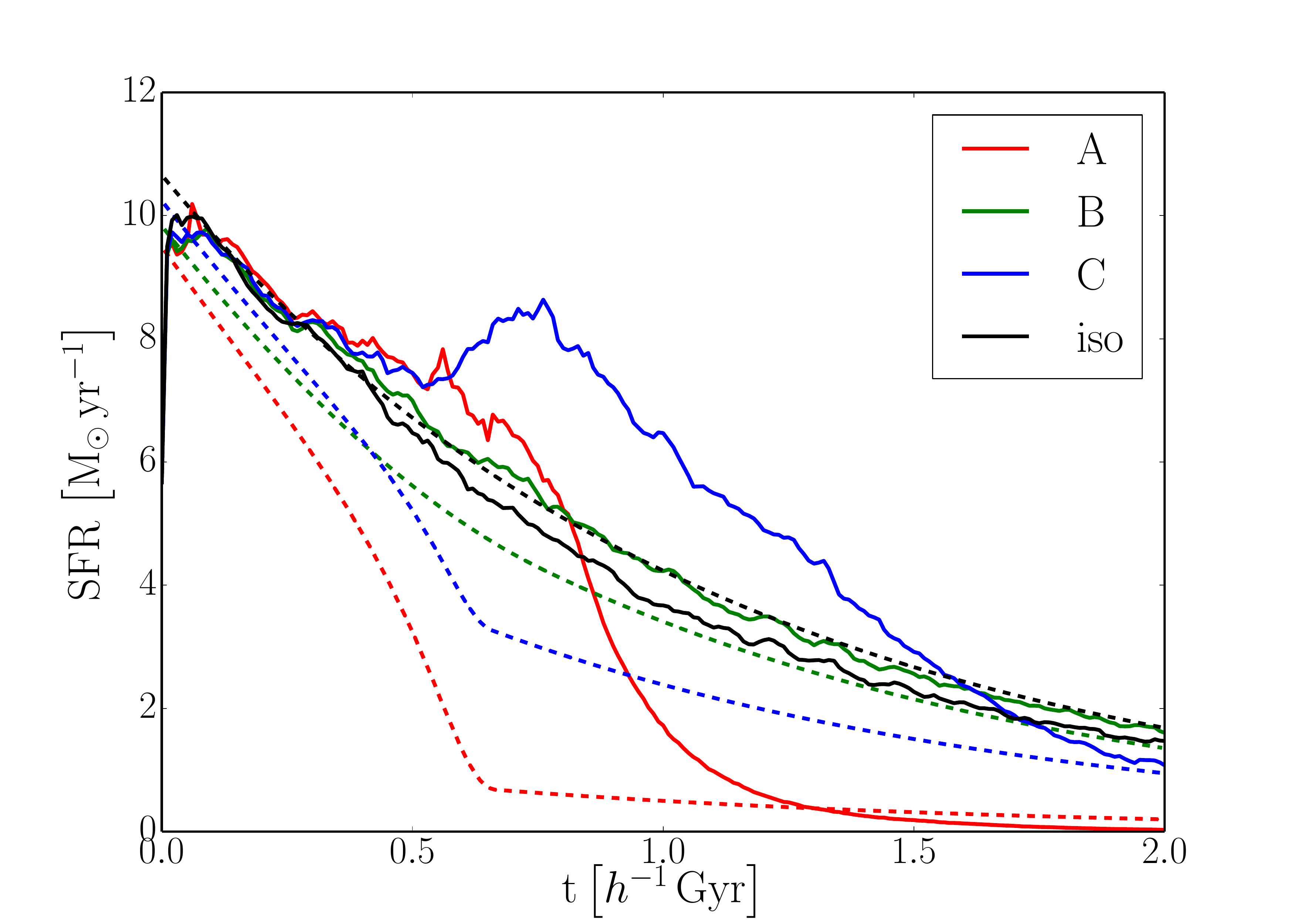}\label{fig:sam_comparison_sfr}}
   \subfigure[]{\includegraphics[width=\linewidth]{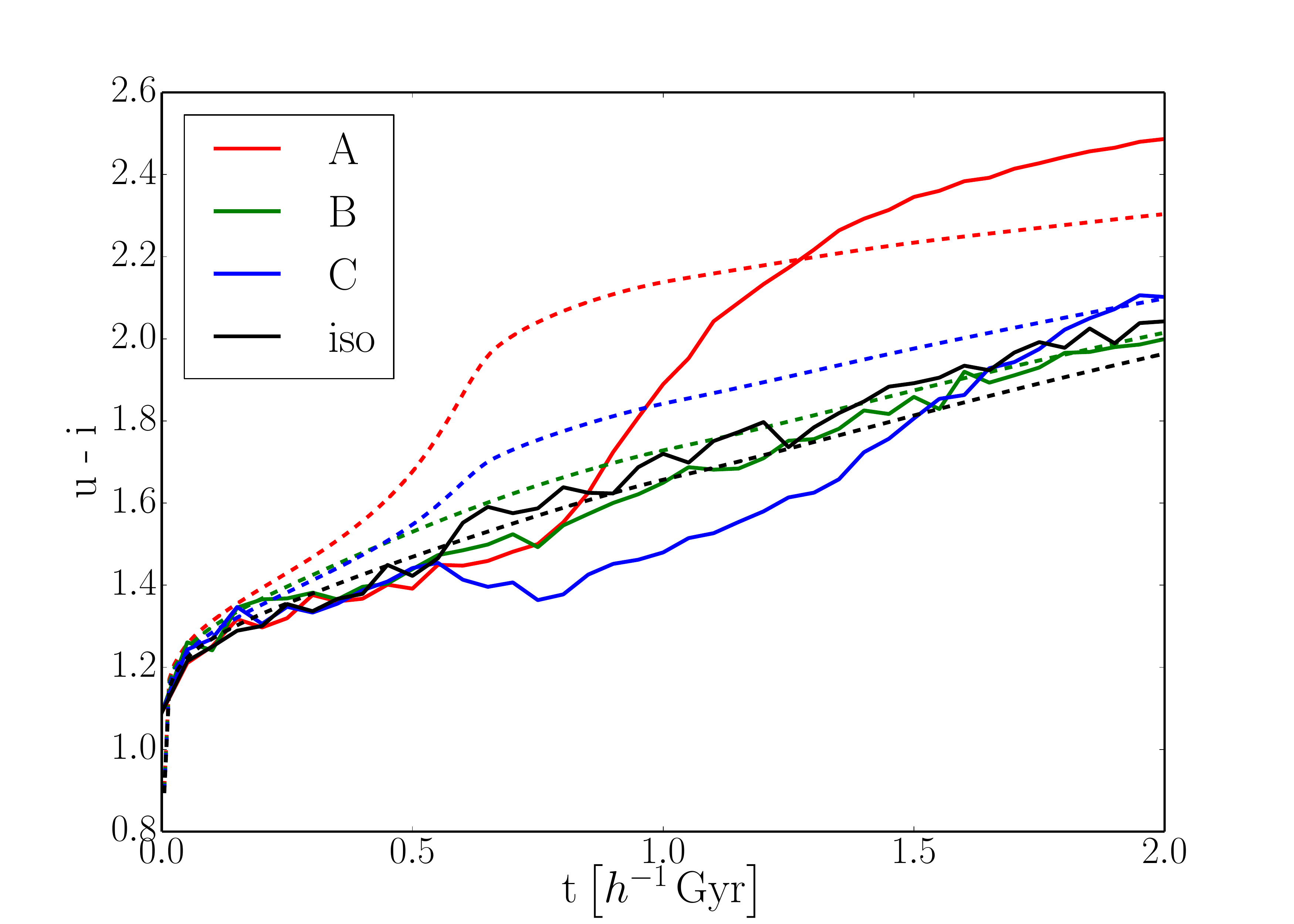}\label{fig:sam_comparison_color}}
   \caption{Comparison of the simulations with SAM expectations. (a)
     The SFR of galaxy G1a in cluster A, B, and C (runs
     S1, S3, and S4, respectively) and galaxy G1c is depicted
     with solid lines.  The dashed lines correspond to the SFR of
     galaxy G1c with standard prescriptions for RPS \citep{Gunn1972}
     and SF \citep{Cora2006}. In (b), the same comparison is shown
   for the time evolution of the galaxy colours.}
   \label{fig:sam_comparison}
\end{figure}

In Sect.~\ref{sec:rps} we have seen that in all of the scenarios the hot gas
halo is stripped almost instantaneously when RP sets in. Hence,
although there are different implementations for stripping the hot gas
halo (see also Sect.~\ref{sec:rps}) in SAMs, we only consider
stripping of the gasesous disk of galaxies.  To this intent, using the
parameters of model galaxy G1c without hot gas halo, we implement the
standard stripping model for the disk, according to
\citet{Gunn1972}. Besides gas stripping, we include SF
\citep{Cora2006, Lagos2008} but no models for AGN feedback.  In order
to determine the colour evolution, assuming instantaneous recycling
approximation \citep[e.g.][]{Benson2012}, we calculate the metal
evolution according to \citet{Springel2003}. In each timestep, we
assume that a SSP is being formed, calculating its mass from the
current SFR and assigning current gas metallicity. An initial
evolution of the stellar population of the galaxy is carried out as described in
Sect.~\ref{sec:colour_evolution}.

In Fig.~\ref{fig:sam_comparison}, we compare the SFR (a) and colour
evolution (b) of our simulations of model galaxy G1 in three different
environments (runs S1, S3 and S4) with the evolution of the
corresponding galaxy in a SAM, as described above. Thus, extracting
the ICM properties along the orbit of the respective simulation, we
calculate the theoretical stripping radius and stripped gas mass
according to Eq.~\ref{eq:stripping_radius} in 
Sect. \ref{sec:rps}.  Then, the SFR is calculated according to
Eq. 3 in \citet{Cora2006},
\begin{equation}
  \frac{\mathrm{d}\,M_\star}{\mathrm{d}\,t} = \frac{\alpha\,M_\mathrm{cold}}{t_\mathrm{dyn}},
  \label{eq:sfr_sam}
\end{equation}
where cold gas mass  is $M_\mathrm{cold}$ and the dynamical time of the
galaxy is $t_\mathrm{dyn}$.  For the dimensionless parameter 
$\alpha=\alpha_0\left(v_\mathrm{vir}/220\,\mathrm{km}\,\mathrm{s}^{-1}\right)^n$,
we chose $\alpha_0 = 0.18$ and $n = 2.2$ such that the SFR and total
stellar mass produced after $2\,h^{-1}\mathrm{Gyr}$ in isolation
resemble values of model galaxy G1c.  Initially, for the cold gas mass
$M_\mathrm{cold}$, we use the mass in the star-forming regime of
model galaxy G1c.

\begin{figure*}[tb]
  \centering
  \includegraphics[width = \linewidth]{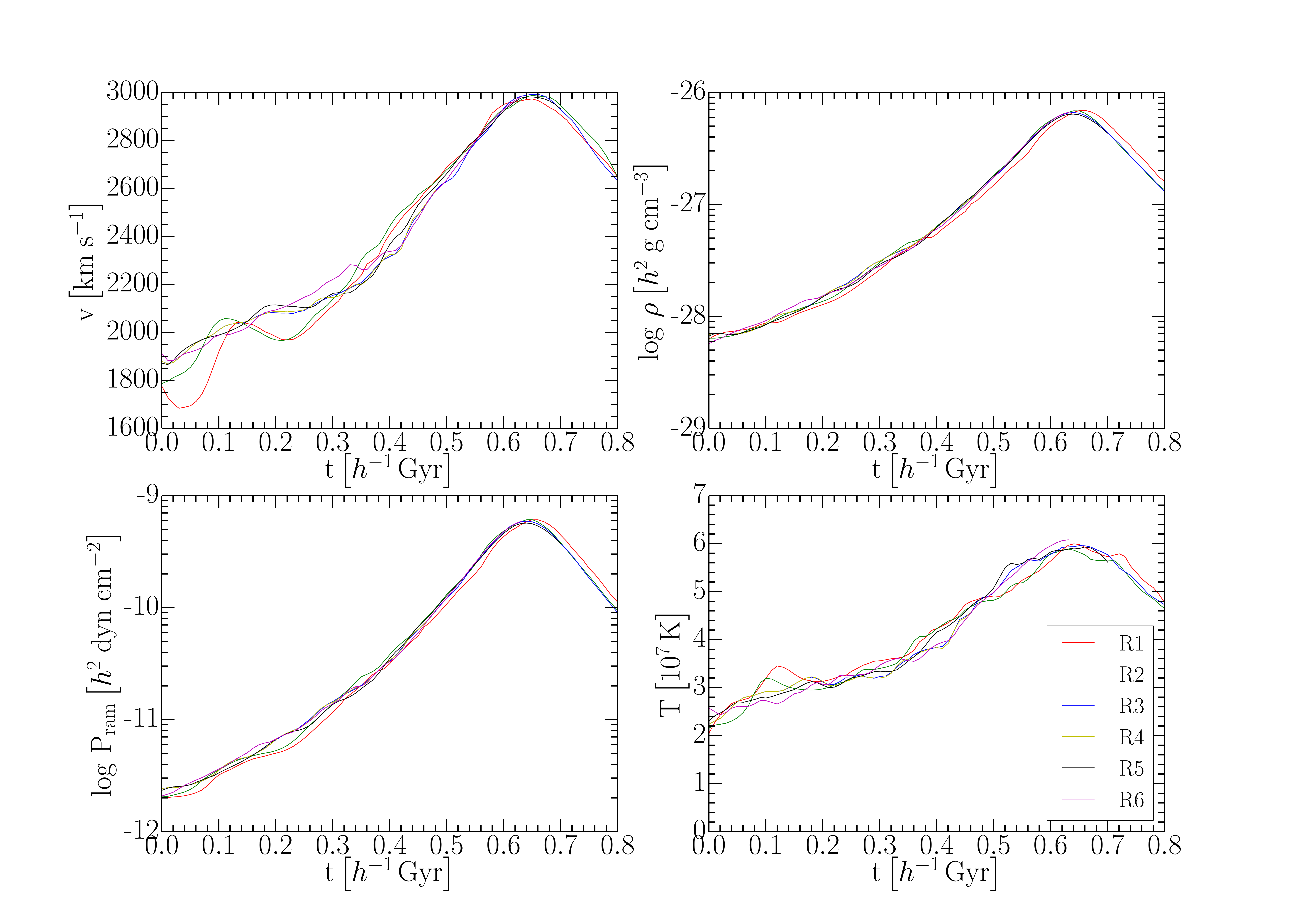} 
  \caption{Convergence study of the time evolution of the ICM
    properties measured in the surrounding and $45\,h^{-1}\mathrm{kpc}$ in
    front of the galaxy.}
  \label{fig:res_icmprop}
\end{figure*}

Adopting the prescription for SFR of a SAM (\ref{eq:sfr_sam}) combined 
with the RPS model (\ref{eq:stripping_radius}) as described above, leads to 
the following results. Since the SFR is directly related
to the available cold gas mass $M_\mathrm{cold}$ in the disk,
in cluster~A the SFR drops below
$1\,\mathrm{M}_\odot\,\mathrm{yr}^{-1}$ at pericentre when the
stripping radius attains its minimum.  The same behaviour can be
observed in cluster~C. After pericentre passage, the remaining gas is
consumed as new stars are formed, since a growing theoretical
stripping radius does not provoke further stripping nor is the disk
refueled by gas. In cluster~B, the weak RP leads to a slow but
constant mass loss  that leads, however, to a lower SFR than in
isolation.  Comparing those results with our simulations, mainly two
differences are evident. First, even a weak RP can enhance the SFR in
the disk, leading to a higher SFR in cluster~B than in isolation,
which is not reproduced by the SAM, where the SFR is actually lower in
the cluster environment.  The difference is even more striking in
cluster~A and C.  As a consequence, the simulation produces almost
double the amount of stars in cluster~A and C at the end of the
$2\,h^{-1}\mathrm{Gyr}$ of evolution than the SAM. In cluster~B, the
stellar mass only differs by $26\%,$ however.  Furthermore, as RPS
takes some time to strip the gas from the galaxy (see
Sect. \ref{sec:rps}), SFR also does not drop immediately as it is the
case in SAMs. So although the drop of the SFR in cluster~A is very
similar as in the simulation, it is delayed by a few hundred Myr.
Despite those differences, at the end of the simulation the SFRs
agree well in all of the cases.

The corresponding colour evolution is shown in
Fig.~\ref{fig:sam_comparison_color}.  Obviously, the differences are
largest when the current SFR of simulations and SAM differ most at
pericentre passage. The trend of higher SFR in the simulations
compared to the SAM expectation is again reflected in the u-i colour
as simulated galaxies are generally bluer. However, at the end of the
simulation, again the colours coincide well between simulation and
SAM, except in cluster~A, where the simulated galaxy is much redder
than expected from the low level of SF predicted by the
SAM.

\section{Resolution tests}
\label{sec:app_res_study}

Given the large dynamic range in density that needs to be covered, it
is important to examine how robust our results are when the numerical
resolution is changed.  The initial conditions of \citet{Springel2005a}
for the model galaxy evolving in isolation and in combination with our
SF recipe have already been subjected to a resolution study
as well as a comparison between AREPO, VPH \citep{Hess2010}, and SPH
(GADGET-2) in \citet{Hess2012}.  These studies found that the SFR 
grows slightly  with increasing resolution, but tends to
converge at high resolution.  Simplified wind-tunnel \citep{Hess2012}
and cosmological simulations \citep{Vogelsberger2012} show that gas
stripping is more efficient in AREPO than in SPH.  Furthermore,
stripping becomes less efficient with increasing resolution in AREPO.

Mixing and fluid instabilities (especially Kelvin-Helmholtz
instabilities) are important processes in RPS
simulations.  \citet{Springel2010} showed that comparatively such fluid
instabilities are modelled accurately in AREPO thanks to
its Galilean-invariance even for large bulk flows, which is the case
for galaxies falling into a galaxy cluster.  Also, overmixing due to
advection errors in bulk flows should be reduced.  Yet, limited
numerical resolution can qualitatively change the outcome of
simulations that are sensitive to the development of fluid
instabilities, hence we carry out a numerical resolution study of our
RPS simulations to quantify the uncertainty
related to this.

For our resolution study, we use the trajectory and model cluster~A of
our fiducial case, run S1 (see Table~\ref{tab:orbits} in Sect.~\ref{sec:merger_orbits}).  
This rather extreme case fits best to test
our simulation set-up because the galaxy experiences a very high
relative velocity and encounters high ICM densities on its way through
the cluster, which is best suited to test the proper functioning of
the refinement of the ICM gas surrounding the galaxy. 

In our resolution study, we carry out simulations of the same model
with different resolutions for both ISM and ICM as well as different
refinement approaches. The different runs are listed in
Table~\ref{tab:resolution_runs}.  First, as we are mainly interested
in the evolution of the model galaxy, we change its mass resolution
by factors of ten and adapt the resolution of the refined ICM in front
of the galaxy accordingly.  The corresponding target volume of ICM gas
cells is given in Table~\ref{tab:resolution_runs} as well.  The target
volume is only valid for gas cells in the 'tunnel' along the trajectory
of the galaxy through the cluster and is defined as the mean
volume of the gas cells within the model galaxy.  As a result of the gradual
refinement around ISM gas cells, the resolution of ICM gas cells
interacting with the galaxy is defined by the resolution of the ISM as
long as the refinement of gas cells in AREPO (and rather the
subsequent mesh regularisation) is fast enough.  To test the
behaviour related to a different target volume of the ICM, in run R5 we
adopt our fiducial resolution for the galaxy but increase the ICM
resolution around the galaxy, applying a target volume of
$20\,h^{-3}\mathrm{kpc}^3$.  Finally we made two additional runs, R2
and R4, where the ICM gas cells are refined to the target volume
along the trajectory of the galaxy throughout the whole simulation.  These
simulations intend to test the influence of the mesh
regularisation in the surroundings of the galaxy when ICM cells
are being refined only shortly before the galaxy passes by.

\begin{figure}[tb]
  \centering
  \includegraphics[width = \linewidth]{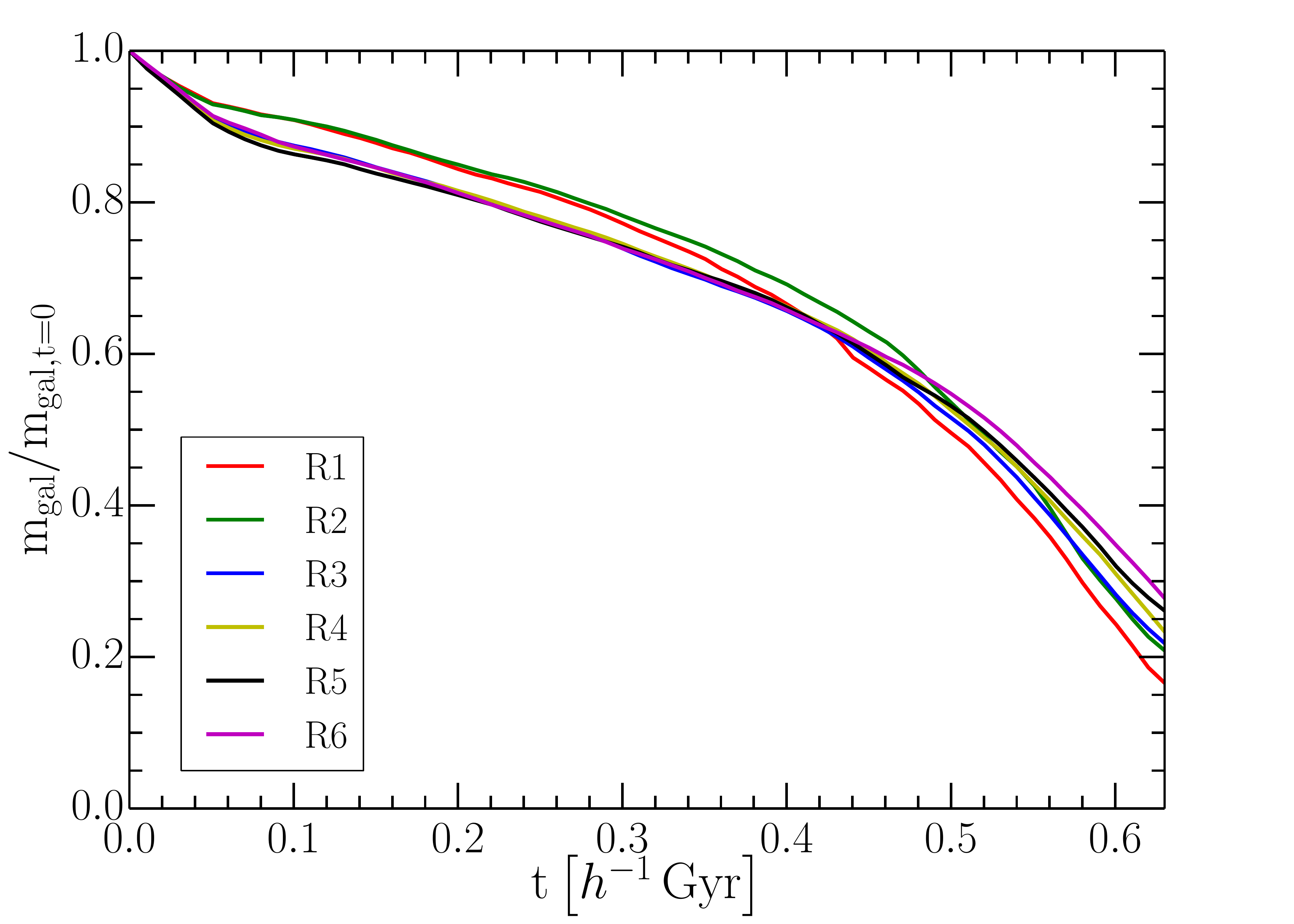}
  \caption{Fraction of remaining ISM bound to the galaxy compared
    to the initial value is shown. In the runs with lower resolution,
    less gas is stripped at the beginning in the outskirts of the
    galaxy cluster, but more gas is stripped at pericentre
    passage.}
  \label{fig:res_strippedMass}
\end{figure}

\begin{figure}[htb]
  \centering
  \includegraphics[width = \linewidth]{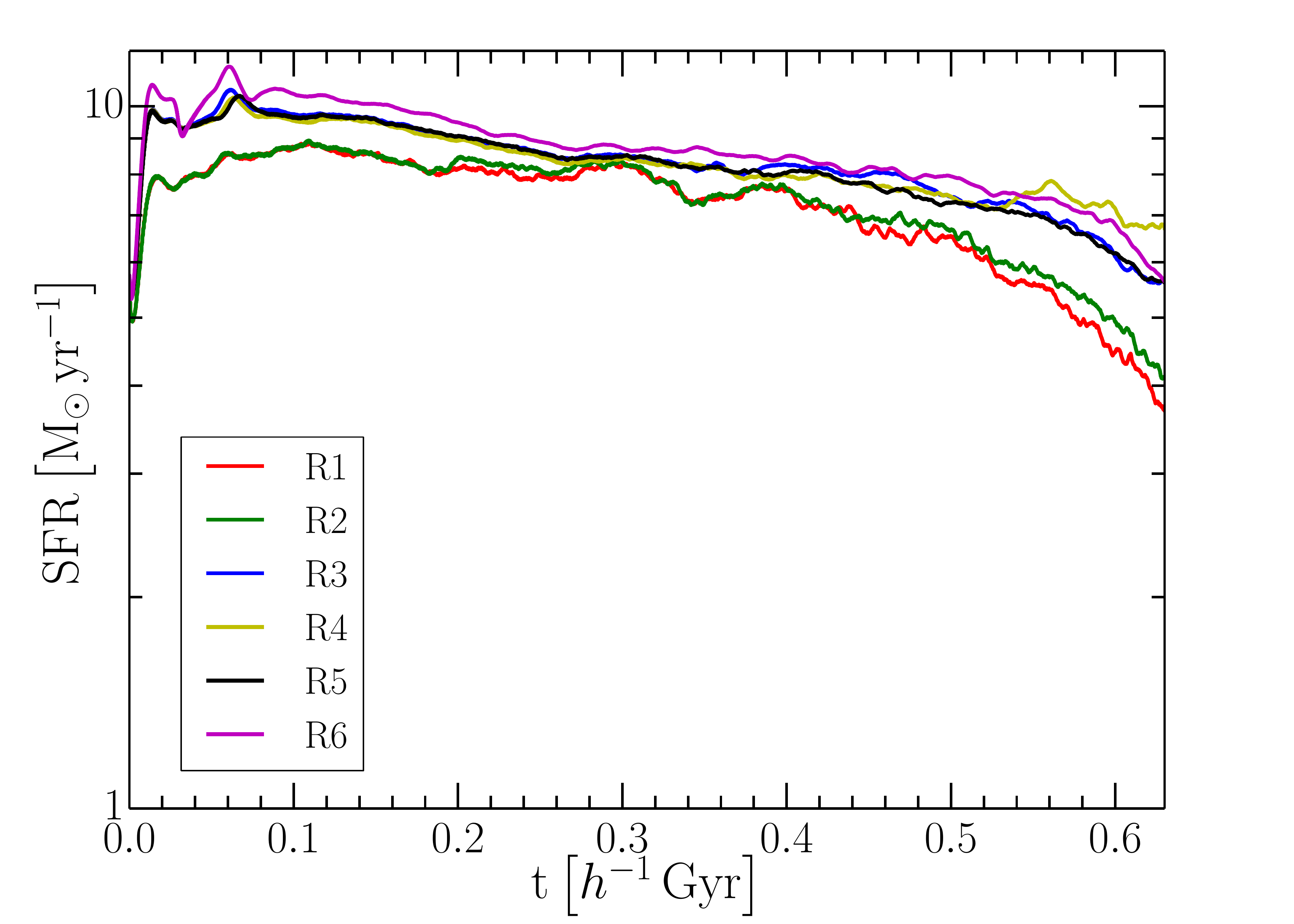}
  \caption{Star formation rate of runs with different resolution are
    shown for the interaction of galaxy G1a with cluster A. Higher
    resolution leads to higher SFR.  The SFR
    clearly converges with increasing resolution.}
  \label{fig:res_sfr}
\end{figure}

\begin{figure*}[htbp]
  \centering 
  \subfigure{\includegraphics[width=0.32\textwidth]{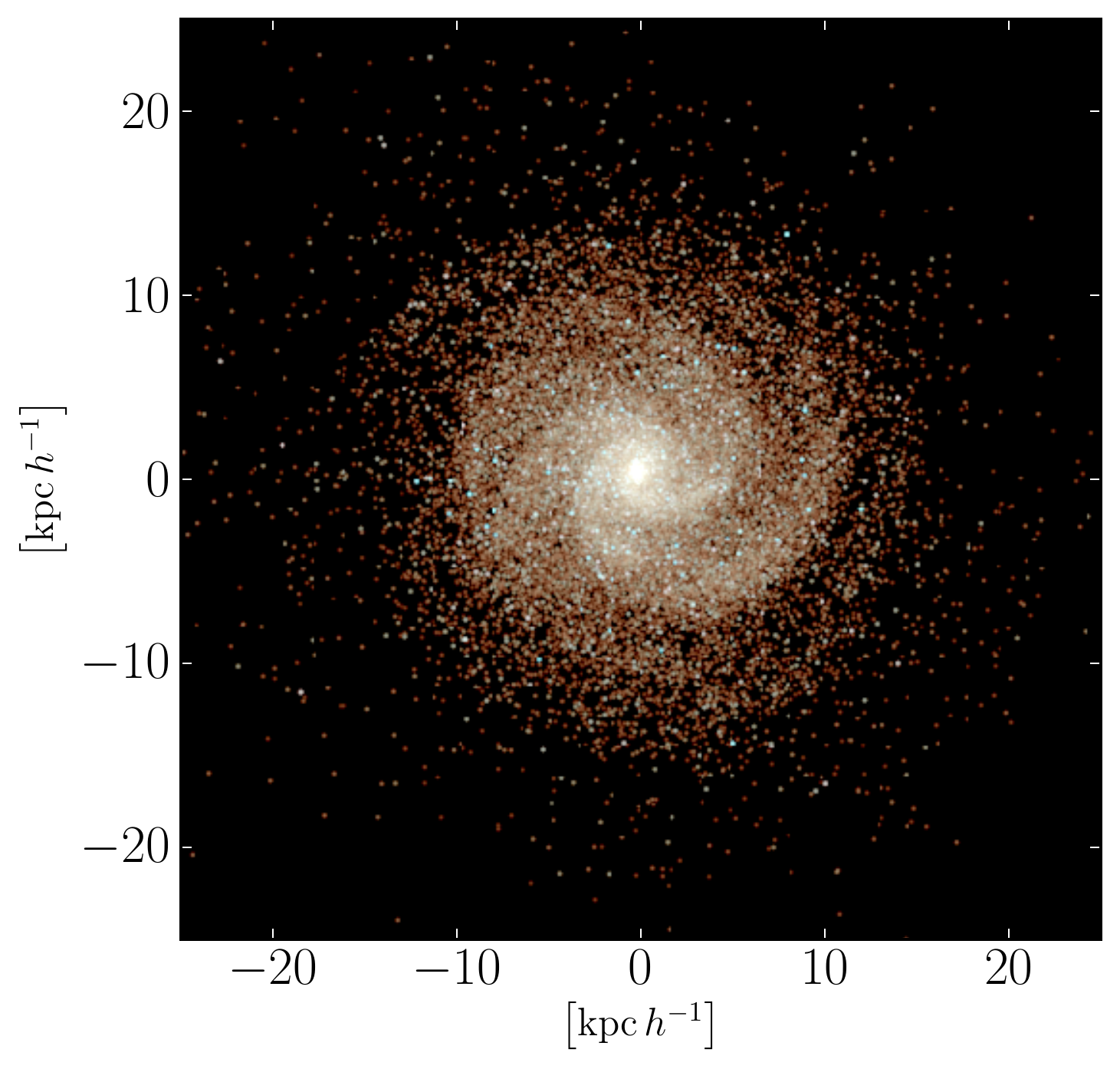}\label{fig:colour_lo}}
  \subfigure{\includegraphics[width=0.32\textwidth]{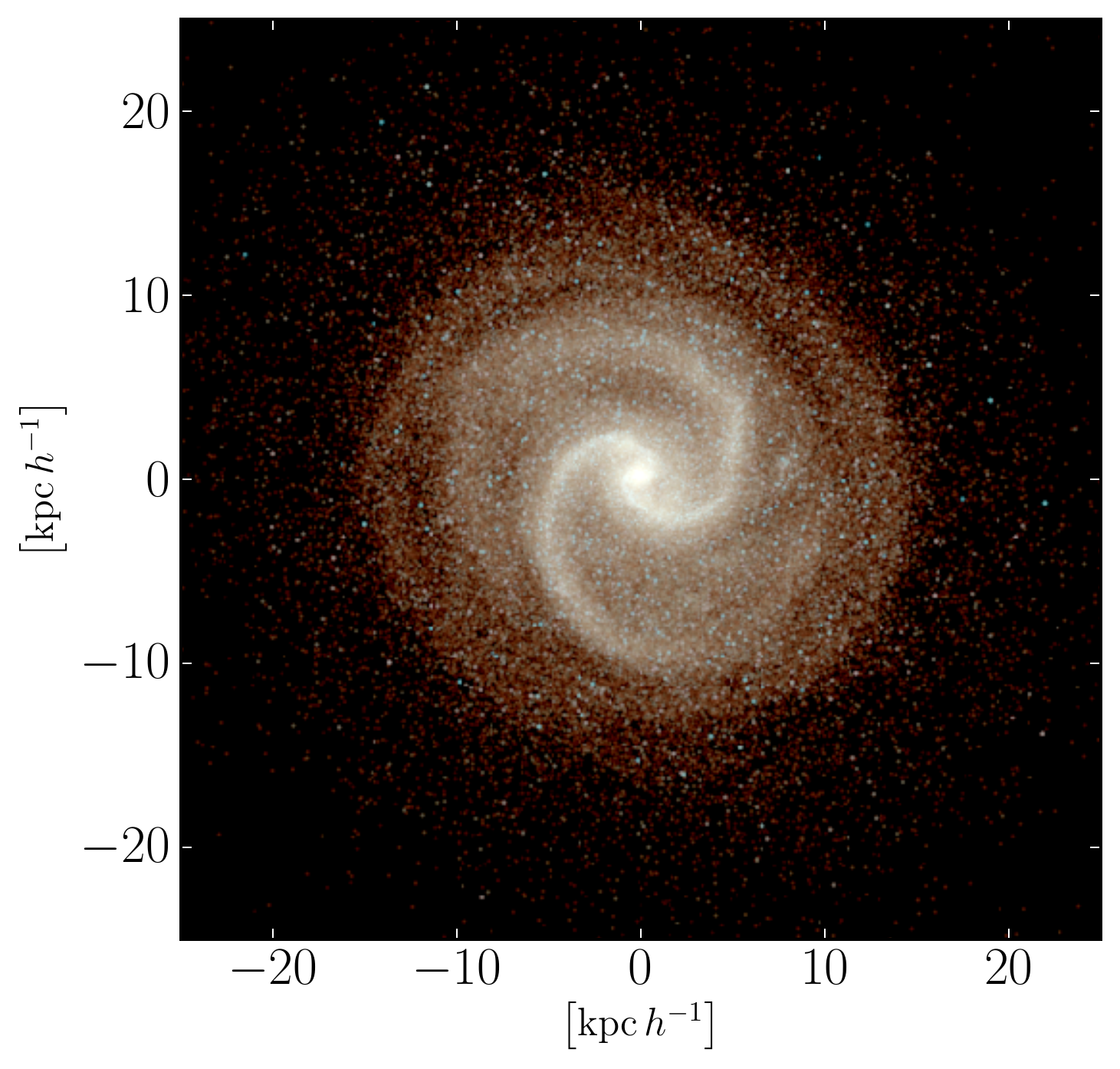}\label{fig:colour_mid}}
  \subfigure{\includegraphics[width=0.32\textwidth]{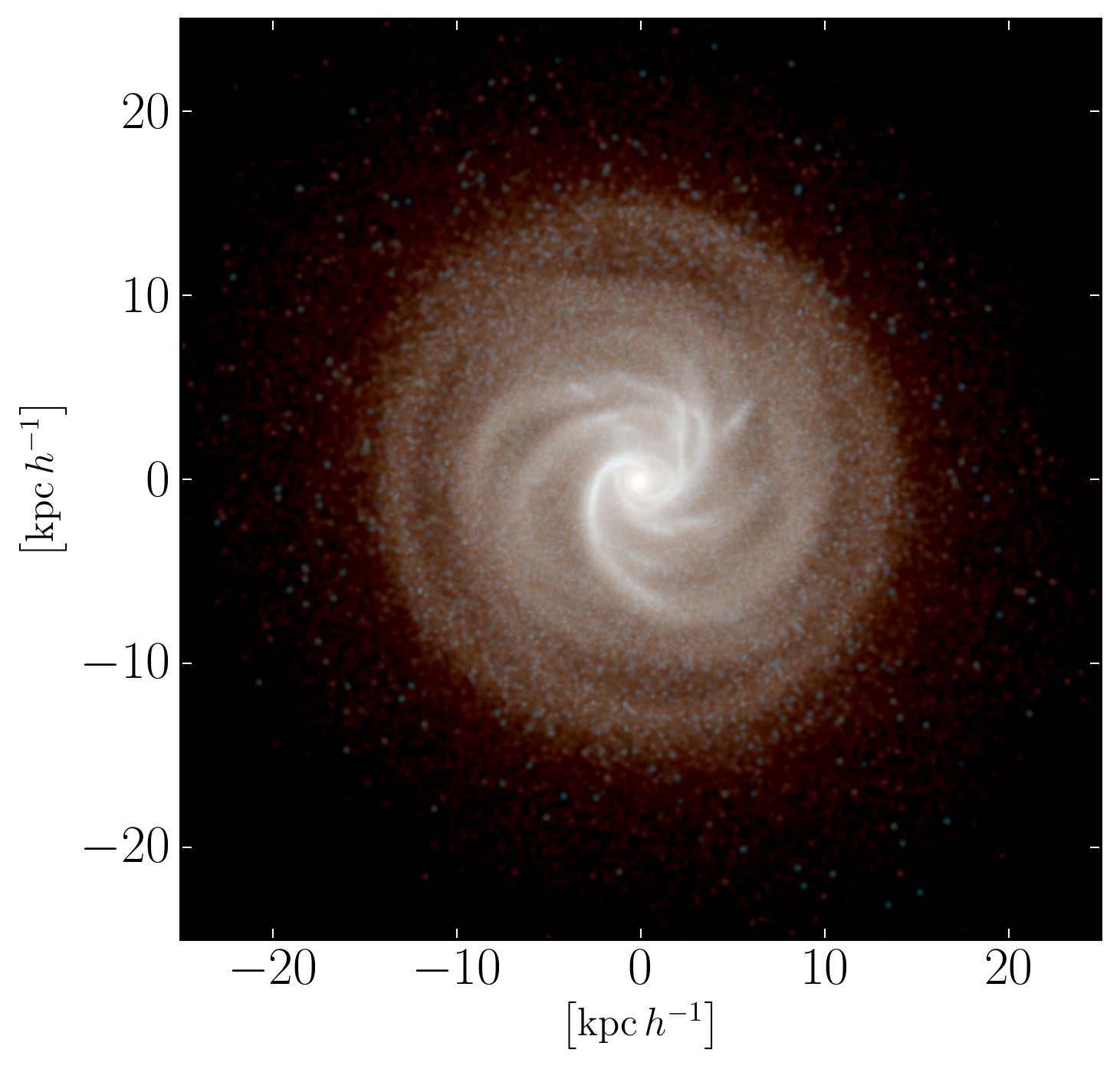}\label{fig:colour_hi}}
  \caption{Convergence of mock images of the galaxy G1 after
    $600\,h^{-1}\mathrm{Myr}$ of evolution in the galaxy cluster for
    run R1, R3, and R6 (from left to right), respectively.}
  \label{fig:res_colour}
\end{figure*}

First, we check the ICM properties along the trajectory
of the galaxy within the galaxy cluster. In Fig.~\ref{fig:res_icmprop} we show
relative velocity, density, corresponding ram pressure, and temperature
of the ICM in front of the galaxy.  The low resolution runs R1-R2 show
a substantial scatter in velocity and temperature at the beginning of
the simulation.  The initial ICM resolution at that quite low density
is very low in those runs and hence gas cells cannot be refined fast
enough.  Furthermore, the peak of ram pressure is shifted in time by
about $20\,h^{-1}\mathrm{Myr}$.  On the other hand, the medium
(R3-R6) and high resolution (R7) runs are in very good agreement.

The amount of stripped gas for different resolutions is shown in
Fig.~\ref{fig:res_strippedMass}.  As in the wind-tunnel experiments of
\citet{Hess2012}, stripping becomes less efficient with increasing
resolution. Reassuringly, the amount of stripped gas is however in
good agreement especially for the medium and high resolution run.
Even at pericentre passage, the amount of stripped gas differs by only
a few percent.  Interestingly, the low resolution runs show less gas
stripping in the outer parts of the cluster at low ICM densities.
Afterwards, however, as expected more gas is stripped as the ICM
density gets higher.

The SFR for the distinct runs is shown in
Fig.~\ref{fig:res_sfr}.  As expected from simulations of model
galaxies in isolation \citep{Springel2003, Springel2005, Hess2012},
SF is enhanced and converges with increasing resolution
 in our new simulation set-up as well.  The target volume of the ICM has no
influence on the SFR of the galaxy.  On the other
hand, at pericentre passage, when the galaxy has a very high relative
velocity, the SFR is lower when the gas cells in front of the galaxy
are refined to the target volume immediately before the galaxy passes
by. This effect is evident in both low and medium resolution runs
(R1-R2 and R3-R5).

Related to SFR and morphology of the galaxy with different
resolutions, in Fig.~\ref{fig:res_colour} we provide mock colour images
of the galaxy for run R1, R3, and R6 after $600\,h^{-1}\mathrm{Myr}$ of
evolution in the galaxy cluster.  As the SFR is enhanced with higher
resolution, the galaxy appears bluer in R6 but redder in R1.
Furthermore, structural features, such as spiral arms, are much more
pronounced at higher resolution.

These tests hence show that the new refinement strategy, which is
crucial to carry out high resolution RPS
simulations in realistic live models of galaxy clusters, is valid and
working well.  There is no influence on the amount of stripped gas
when the refinement is applied. In particular, there is no relevant
difference in the mixing of the gas with the ICM, and SF
in the wake is not prevented either.  The higher amount of gas in the
star-forming regime is mostly present in the disk itself, as less gas
is dragged from the outskirts of the disk because of the higher resolved
cells present there.  Furthermore, it is very important that the gas
in front of the galaxies is refined sufficiently in advance,
which is an issue especially when the galaxy moves fast.  This is
ensured in all of our production runs.

\section{Conclusions} \label{secconclusions}

In this work, we introduced a new set-up for realistic ram-pressure
stripping simulations that employs the moving-mesh code AREPO combined
with a special refinement strategy to cope with the high dynamic range
in density encountered in this problem and the need to restrict the
high resolution to the path of the galaxy for reasons of computational
efficiency. In addition, we prevented radiative cooling in the
background cluster and the computational cost this would entail
through a suitable marker field.  With our approach, the interaction
of a live galaxy cluster with an infalling galaxy can be followed at
high resolution.

We constructed several different galaxy models, including variations
of the amount of gas in the halos of the galaxies, and collided these 
models with galaxy clusters of different mass. We then analysed the
evolution of SFR, galaxy colours, and bound baryonic 
masses to the galaxies in comparison to control
simulations carried out in isolation. Through extensive convergence
tests we verified that all our conclusions are numerically robust.

In general, we find that the cold disks of our model galaxies are
remarkably resilient against RPS, unless the
pericentre distance of their orbit is very small and the encounter is
'deep'. Typically, the galaxies continue to form stars with only
slightly modified rates as a result of the environmental interaction,
and in some cases, they even show an enhanced SFR.
These modifications primarily result from a stripping of the hot
gaseous halos of the galaxies and, in some cases, from a compression of
the ISM of the disks. We confirm that the inclination angle has only a
moderate influence on the overall strength of the stripping
process. The colour evolution of our galaxies, augmented with sensible
assumptions about their initial age, is in good agreement with
observational data. 

In particular, quenching timescales do not differ significantly from
isolation $(\tau_\mathrm{quench} > 1\,h^{-1}\,\mathrm{Gyr})$.
Inefficient stripping is balanced by increased SF in the
RP compressed disk. Hence, the influence of RPS
on galaxy quenching is mainly due to the missing gas supply
from the gas halo. Only extreme, supposedly rare cases of RPS
in the centre of massive clusters are able to completely
strip the gas disk and quench SF on short timescales of
a few hundred Myr.

Furthermore, as predicted by findings of wind-tunnel simulations,
  the turbulent wake of the stripped galaxies is narrower in our
  simulations, which include radiative cooling for the ISM, compared
  to previous studies of RPS in live galaxy clusters,
  which only implemented adiabatic gas physics.
The stripped gas is heated up and diluted quickly to background
temperature and density and is thus free of SF.  The
metallicity of stripped gas can reach near solar values but is
diluted fast because of high relative velocities. After cluster passage,
in all of the scenarios metallicity in stripped gas is mostly identical,
although the stripped metal mass can differ by a factor of eight.

When compared to commonly employed semi-analytic prescriptions for RPS, we find that our simulations in general show less
stripping than inferred in SAMs with the exact size of the mismatch
depending on the details of the employed SAM. It thus
seems clear that the widely reported difficulty of SAMs to account for
relatively blue galaxies in dense environments may largely be a result
of an inaccurate treatment of RPS. The work
presented here should be helpful in re-calibrating the corresponding
model prescriptions and putting them on a quantitatively more accurate
physical basis, such that one can continue to use SAMs as a shortcut
for the expensive but more reliable full hydrodynamical simulations.

\begin{acknowledgements}

DS thanks Federico Marinacci, Markus Haider, Matteo Bianconi, Andreas Bauer, Kevin Schaal, and R\"udiger Pakmor for
many valuable discussions. DS acknowledges support of the Austrian Federal Ministry of Science, Research and Economy 
through the Uni-Infrastrukturprogramm of the Focal Point Scientific Computing at the University of Innsbruck, 
the doctoral school Computational Interdisciplinary Modelling FWF DK plus (W1227), the research grant
from the office of the vice rector for research of the University of Innsbruck, 
and the award for research stay abroad of the International Relations Office of the University
of Innsbruck.
VS acknowledges support through the European Research Council
through ERC-StG grant EXAGAL-308037 and the Klaus Tschira
Foundation.
Furthermore, we would like to acknowledge the valuable comments of the anonymous referee.
\end{acknowledgements}

\bibliographystyle{aa}
\bibliography{library_ads}

\begin{thebibliography}{106}
\expandafter\ifx\csname natexlab\endcsname\relax\def\natexlab#1{#1}\fi

\bibitem[{{Abramson} \& {Kenney}(2014)}]{Abramson2014}
{Abramson}, A. \& {Kenney}, J.~D.~P. 2014, \aj, 147, 63

\bibitem[{{Asplund} {et~al.}(2005){Asplund}, {Grevesse}, \&
  {Sauval}}]{Asplund2005}
{Asplund}, M., {Grevesse}, N., \& {Sauval}, A.~J. 2005, in Astronomical Society
  of the Pacific Conference Series, Vol. 336, Cosmic Abundances as Records of
  Stellar Evolution and Nucleosynthesis, ed. T.~G. {Barnes}, III \& F.~N.
  {Bash}, 25

\bibitem[{{Athanassoula}(2005)}]{Athanassoula2005}
{Athanassoula}, E. 2005, \mnras, 358, 1477

\bibitem[{{Bekki}(2009)}]{Bekki2009}
{Bekki}, K. 2009, \mnras, 399, 2221

\bibitem[{{Benson}(2012)}]{Benson2012}
{Benson}, A.~J. 2012, \na, 17, 175

\bibitem[{{Bischko} {et~al.}(2015){Bischko}, {Steinhauser}, \&
  {Schindler}}]{Bischko2015}
{Bischko}, J.~C., {Steinhauser}, D., \& {Schindler}, S. 2015, \aap, 576, A76

\bibitem[{{Boissier} {et~al.}(2012){Boissier}, {Boselli}, {Duc}, {Cortese},
  {van Driel}, {Heinis}, {Voyer}, {Cucciati}, {Ferrarese}, {C{\^o}t{\'e}},
  {Cuillandre}, {Gwyn}, \& {Mei}}]{Boissier2012}
{Boissier}, S., {Boselli}, A., {Duc}, P.-A., {et~al.} 2012, \aap, 545, A142

\bibitem[{{Boselli} {et~al.}(2009){Boselli}, {Boissier}, {Cortese}, \&
  {Gavazzi}}]{Boselli2009}
{Boselli}, A., {Boissier}, S., {Cortese}, L., \& {Gavazzi}, G. 2009,
  Astronomische Nachrichten, 330, 904

\bibitem[{{Boselli} \& {Gavazzi}(2006)}]{Boselli2006}
{Boselli}, A. \& {Gavazzi}, G. 2006, \pasp, 118, 517

\bibitem[{{Boselli} {et~al.}(1997){Boselli}, {Gavazzi}, {Lequeux}, {Buat},
  {Casoli}, {Dickey}, \& {Donas}}]{Boselli1997}
{Boselli}, A., {Gavazzi}, G., {Lequeux}, J., {et~al.} 1997, \aap, 327, 522

\bibitem[{{Bresolin} {et~al.}(2012){Bresolin}, {Kennicutt}, \&
  {Ryan-Weber}}]{Bresolin2012}
{Bresolin}, F., {Kennicutt}, R.~C., \& {Ryan-Weber}, E. 2012, \apj, 750, 122

\bibitem[{{Chung} {et~al.}(2009){Chung}, {van Gorkom}, {Kenney}, {Crowl}, \&
  {Vollmer}}]{Chung2009}
{Chung}, A., {van Gorkom}, J.~H., {Kenney}, J.~D.~P., {Crowl}, H., \&
  {Vollmer}, B. 2009, \aj, 138, 1741

\bibitem[{{Conroy} \& {Gunn}(2010)}]{Conroy2010}
{Conroy}, C. \& {Gunn}, J.~E. 2010, \apj, 712, 833

\bibitem[{{Conroy} {et~al.}(2009){Conroy}, {Gunn}, \& {White}}]{Conroy2009}
{Conroy}, C., {Gunn}, J.~E., \& {White}, M. 2009, \apj, 699, 486

\bibitem[{{Cora}(2006)}]{Cora2006}
{Cora}, S.~A. 2006, \mnras, 368, 1540

\bibitem[{{Cortese} {et~al.}(2007){Cortese}, {Marcillac}, {Richard},
  {Bravo-Alfaro}, {Kneib}, {Rieke}, {Covone}, {Egami}, {Rigby}, {Czoske}, \&
  {Davies}}]{Cortese2007}
{Cortese}, L., {Marcillac}, D., {Richard}, J., {et~al.} 2007, \mnras, 376, 157

\bibitem[{{De Lucia} {et~al.}(2004){De Lucia}, {Kauffmann}, \&
  {White}}]{DeLucia2004}
{De Lucia}, G., {Kauffmann}, G., \& {White}, S.~D.~M. 2004, \mnras, 349, 1101

\bibitem[{{Domainko} {et~al.}(2006){Domainko}, {Mair}, {Kapferer}, {van
  Kampen}, {Kronberger}, {Schindler}, {Kimeswenger}, {Ruffert}, \&
  {Mangete}}]{Domainko2006}
{Domainko}, W., {Mair}, M., {Kapferer}, W., {et~al.} 2006, \aap, 452, 795

\bibitem[{{Duc} {et~al.}(2000){Duc}, {Brinks}, {Springel}, {Pichardo},
  {Weilbacher}, \& {Mirabel}}]{Duc2000}
{Duc}, P.-A., {Brinks}, E., {Springel}, V., {et~al.} 2000, \aj, 120, 1238

\bibitem[{{Ebeling} {et~al.}(2014){Ebeling}, {Stephenson}, \&
  {Edge}}]{Ebeling2014}
{Ebeling}, H., {Stephenson}, L.~N., \& {Edge}, A.~C. 2014, \apjl, 781, L40

\bibitem[{{Eke} {et~al.}(1998){Eke}, {Navarro}, \& {Frenk}}]{Eke1998}
{Eke}, V.~R., {Navarro}, J.~F., \& {Frenk}, C.~S. 1998, \apj, 503, 569

\bibitem[{{Feldmann} \& {Mayer}(2015)}]{Feldmann2015}
{Feldmann}, R. \& {Mayer}, L. 2015, \mnras, 446, 1939

\bibitem[{{Ferguson} {et~al.}(1998){Ferguson}, {Gallagher}, \&
  {Wyse}}]{Ferguson1998}
{Ferguson}, A.~M.~N., {Gallagher}, J.~S., \& {Wyse}, R.~F.~G. 1998, \aj, 116,
  673

\bibitem[{{Font} {et~al.}(2008){Font}, {Bower}, {McCarthy}, {Benson}, {Frenk},
  {Helly}, {Lacey}, {Baugh}, \& {Cole}}]{Font2008}
{Font}, A.~S., {Bower}, R.~G., {McCarthy}, I.~G., {et~al.} 2008, \mnras, 389,
  1619

\bibitem[{{Fryxell} {et~al.}(2000){Fryxell}, {Olson}, {Ricker}, {Timmes},
  {Zingale}, {Lamb}, {MacNeice}, {Rosner}, {Truran}, \& {Tufo}}]{Fryxell2000}
{Fryxell}, B., {Olson}, K., {Ricker}, P., {et~al.} 2000, \apjs, 131, 273

\bibitem[{{Fumagalli} {et~al.}(2014){Fumagalli}, {Fossati}, {Hau}, {Gavazzi},
  {Bower}, {Sun}, \& {Boselli}}]{Fumagalli2014}
{Fumagalli}, M., {Fossati}, M., {Hau}, G.~K.~T., {et~al.} 2014, \mnras, 445,
  4335

\bibitem[{{Guedes} {et~al.}(2011){Guedes}, {Callegari}, {Madau}, \&
  {Mayer}}]{Guedes2011}
{Guedes}, J., {Callegari}, S., {Madau}, P., \& {Mayer}, L. 2011, \apj, 742, 76

\bibitem[{{Gunn} \& {Gott}(1972)}]{Gunn1972}
{Gunn}, J.~E. \& {Gott}, III, J.~R. 1972, \apj, 176, 1

\bibitem[{{Guo} {et~al.}(2013){Guo}, {Cole}, {Eke}, {Frenk}, \&
  {Helly}}]{Guo2013}
{Guo}, Q., {Cole}, S., {Eke}, V., {Frenk}, C., \& {Helly}, J. 2013, \mnras,
  434, 1838

\bibitem[{{Guo} {et~al.}(2011){Guo}, {White}, {Boylan-Kolchin}, {De Lucia},
  {Kauffmann}, {Lemson}, {Li}, {Springel}, \& {Weinmann}}]{Guo2011}
{Guo}, Q., {White}, S., {Boylan-Kolchin}, M., {et~al.} 2011, \mnras, 413, 101

\bibitem[{{Hernquist}(1990)}]{Hernquist1990}
{Hernquist}, L. 1990, \apj, 356, 359

\bibitem[{{He{\ss}} \& {Springel}(2010)}]{Hess2010}
{He{\ss}}, S. \& {Springel}, V. 2010, \mnras, 406, 2289

\bibitem[{{He{\ss}} \& {Springel}(2012)}]{Hess2012}
{He{\ss}}, S. \& {Springel}, V. 2012, \mnras, 426, 3112

\bibitem[{{Hester} {et~al.}(2010){Hester}, {Seibert}, {Neill}, {Wyder}, {Gil de
  Paz}, {Madore}, {Martin}, {Schiminovich}, \& {Rich}}]{Hester2010}
{Hester}, J.~A., {Seibert}, M., {Neill}, J.~D., {et~al.} 2010, \apjl, 716, L14

\bibitem[{{H{\"o}ller} {et~al.}(2014){H{\"o}ller}, {St{\"o}ckl}, {Benson},
  {Haider}, {Steinhauser}, {Lovisari}, \& {Pranger}}]{Holler2014}
{H{\"o}ller}, H., {St{\"o}ckl}, J., {Benson}, A., {et~al.} 2014, \aap, 569, A31

\bibitem[{{J{\'a}chym} {et~al.}(2009){J{\'a}chym}, {K{\"o}ppen}, {Palou{\v s}},
  \& {Combes}}]{Jachym2009}
{J{\'a}chym}, P., {K{\"o}ppen}, J., {Palou{\v s}}, J., \& {Combes}, F. 2009,
  \aap, 500, 693

\bibitem[{{J{\'a}chym} {et~al.}(2007){J{\'a}chym}, {Palou{\v s}}, {K{\"o}ppen},
  \& {Combes}}]{Jachym2007}
{J{\'a}chym}, P., {Palou{\v s}}, J., {K{\"o}ppen}, J., \& {Combes}, F. 2007,
  \aap, 472, 5

\bibitem[{{Kapferer} {et~al.}(2009){Kapferer}, {Sluka}, {Schindler}, {Ferrari},
  \& {Ziegler}}]{Kapferer2009}
{Kapferer}, W., {Sluka}, C., {Schindler}, S., {Ferrari}, C., \& {Ziegler}, B.
  2009, \aap, 499, 87

\bibitem[{{Katz} {et~al.}(1996){Katz}, {Weinberg}, \& {Hernquist}}]{Katz1996}
{Katz}, N., {Weinberg}, D.~H., \& {Hernquist}, L. 1996, \apjs, 105, 19

\bibitem[{{Kauffmann} {et~al.}(2013){Kauffmann}, {Li}, {Zhang}, \&
  {Weinmann}}]{Kauffmann2013}
{Kauffmann}, G., {Li}, C., {Zhang}, W., \& {Weinmann}, S. 2013, \mnras, 430,
  1447

\bibitem[{{Kenney} {et~al.}(2014){Kenney}, {Geha}, {J{\'a}chym}, {Crowl},
  {Dague}, {Chung}, {van Gorkom}, \& {Vollmer}}]{Kenney2014}
{Kenney}, J.~D.~P., {Geha}, M., {J{\'a}chym}, P., {et~al.} 2014, \apj, 780, 119

\bibitem[{{Kennicutt}(1998)}]{Kennicutt1998}
{Kennicutt}, Jr., R.~C. 1998, \apj, 498, 541

\bibitem[{{Kimm} {et~al.}(2009){Kimm}, {Somerville}, {Yi}, {van den Bosch},
  {Salim}, {Fontanot}, {Monaco}, {Mo}, {Pasquali}, {Rich}, \&
  {Yang}}]{Kimm2009}
{Kimm}, T., {Somerville}, R.~S., {Yi}, S.~K., {et~al.} 2009, \mnras, 394, 1131

\bibitem[{{Kronberger} {et~al.}(2008){Kronberger}, {Kapferer}, {Ferrari},
  {Unterguggenberger}, \& {Schindler}}]{Kronberger2008}
{Kronberger}, T., {Kapferer}, W., {Ferrari}, C., {Unterguggenberger}, S., \&
  {Schindler}, S. 2008, \aap, 481, 337

\bibitem[{{Lagos} {et~al.}(2008){Lagos}, {Cora}, \& {Padilla}}]{Lagos2008}
{Lagos}, C.~D.~P., {Cora}, S.~A., \& {Padilla}, N.~D. 2008, \mnras, 388, 587

\bibitem[{{Lanzoni} {et~al.}(2005){Lanzoni}, {Guiderdoni}, {Mamon},
  {Devriendt}, \& {Hatton}}]{Lanzoni2005}
{Lanzoni}, B., {Guiderdoni}, B., {Mamon}, G.~A., {Devriendt}, J., \& {Hatton},
  S. 2005, \mnras, 361, 369

\bibitem[{{Larson} {et~al.}(1980){Larson}, {Tinsley}, \&
  {Caldwell}}]{Larson1980}
{Larson}, R.~B., {Tinsley}, B.~M., \& {Caldwell}, C.~N. 1980, \apj, 237, 692

\bibitem[{{Lupton} {et~al.}(2004){Lupton}, {Blanton}, {Fekete}, {Hogg},
  {O'Mullane}, {Szalay}, \& {Wherry}}]{Lupton2004}
{Lupton}, R., {Blanton}, M.~R., {Fekete}, G., {et~al.} 2004, \pasp, 116, 133

\bibitem[{{Marinacci} {et~al.}(2014){Marinacci}, {Pakmor}, \&
  {Springel}}]{Marinacci2014}
{Marinacci}, F., {Pakmor}, R., \& {Springel}, V. 2014, \mnras, 437, 1750

\bibitem[{{McCarthy} {et~al.}(2008){McCarthy}, {Frenk}, {Font}, {Lacey},
  {Bower}, {Mitchell}, {Balogh}, \& {Theuns}}]{McCarthy2008}
{McCarthy}, I.~G., {Frenk}, C.~S., {Font}, A.~S., {et~al.} 2008, \mnras, 383,
  593

\bibitem[{{McPartland} {et~al.}(2016){McPartland}, {Ebeling}, {Roediger}, \&
  {Blumenthal}}]{McPartland2016}
{McPartland}, C., {Ebeling}, H., {Roediger}, E., \& {Blumenthal}, K. 2016,
  \mnras, 455, 2994

\bibitem[{{Merluzzi} {et~al.}(2013){Merluzzi}, {Busarello}, {Dopita}, {Haines},
  {Steinhauser}, {Mercurio}, {Rifatto}, {Smith}, \& {Schindler}}]{Merluzzi2013}
{Merluzzi}, P., {Busarello}, G., {Dopita}, M.~A., {et~al.} 2013, \mnras, 429,
  1747

\bibitem[{{Mo} {et~al.}(1998){Mo}, {Mao}, \& {White}}]{Mo1998}
{Mo}, H.~J., {Mao}, S., \& {White}, S.~D.~M. 1998, \mnras, 295, 319

\bibitem[{{Moster} {et~al.}(2011){Moster}, {Macci{\`o}}, {Somerville}, {Naab},
  \& {Cox}}]{Moster2011}
{Moster}, B.~P., {Macci{\`o}}, A.~V., {Somerville}, R.~S., {Naab}, T., \&
  {Cox}, T.~J. 2011, \mnras, 415, 3750

\bibitem[{{Murante} {et~al.}(2015){Murante}, {Monaco}, {Borgani}, {Tornatore},
  {Dolag}, \& {Goz}}]{Murante2015}
{Murante}, G., {Monaco}, P., {Borgani}, S., {et~al.} 2015, \mnras, 447, 178

\bibitem[{{Navarro} {et~al.}(1996){Navarro}, {Frenk}, \& {White}}]{Navarro1996}
{Navarro}, J.~F., {Frenk}, C.~S., \& {White}, S.~D.~M. 1996, \apj, 462, 563

\bibitem[{{Nelson} {et~al.}(2015){Nelson}, {Genel}, {Vogelsberger}, {Springel},
  {Sijacki}, {Torrey}, \& {Hernquist}}]{Nelson2015}
{Nelson}, D., {Genel}, S., {Vogelsberger}, M., {et~al.} 2015, \mnras, 448, 59

\bibitem[{{Neto} {et~al.}(2007){Neto}, {Gao}, {Bett}, {Cole}, {Navarro},
  {Frenk}, {White}, {Springel}, \& {Jenkins}}]{Neto2007}
{Neto}, A.~F., {Gao}, L., {Bett}, P., {et~al.} 2007, \mnras, 381, 1450

\bibitem[{{Nulsen}(1982)}]{Nulsen1982}
{Nulsen}, P.~E.~J. 1982, \mnras, 198, 1007

\bibitem[{{O'Shea} {et~al.}(2004){O'Shea}, {Bryan}, {Bordner}, {Norman},
  {Abel}, {Harkness}, \& {Kritsuk}}]{O'Shea2004}
{O'Shea}, B.~W., {Bryan}, G., {Bordner}, J., {et~al.} 2004, ArXiv Astrophysics
  e-prints [\eprint{astro-ph/0403044}]

\bibitem[{{Ott} \& {Schnetter}(2003)}]{Ott2003}
{Ott}, F. \& {Schnetter}, E. 2003, ArXiv Physics e-prints
  [\eprint{physics/0303112}]

\bibitem[{{Owen} {et~al.}(2006){Owen}, {Keel}, {Wang}, {Ledlow}, \&
  {Morrison}}]{Owen2006}
{Owen}, F.~N., {Keel}, W.~C., {Wang}, Q.~D., {Ledlow}, M.~J., \& {Morrison},
  G.~E. 2006, \aj, 131, 1974

\bibitem[{{Owers} {et~al.}(2012){Owers}, {Couch}, {Nulsen}, \&
  {Randall}}]{Owers2012}
{Owers}, M.~S., {Couch}, W.~J., {Nulsen}, P.~E.~J., \& {Randall}, S.~W. 2012,
  \apjl, 750, L23

\bibitem[{{Pakmor} {et~al.}(2013){Pakmor}, {Kromer}, {Taubenberger}, \&
  {Springel}}]{Pakmor2013}
{Pakmor}, R., {Kromer}, M., {Taubenberger}, S., \& {Springel}, V. 2013, \apjl,
  770, L8

\bibitem[{{Peng} {et~al.}(2010){Peng}, {Lilly}, {Kova{\v c}}, {Bolzonella},
  {Pozzetti}, {Renzini}, {Zamorani}, {Ilbert}, {Knobel}, {Iovino}, {Maier},
  {Cucciati}, {Tasca}, {Carollo}, {Silverman}, {Kampczyk}, {de Ravel},
  {Sanders}, {Scoville}, {Contini}, {Mainieri}, {Scodeggio}, {Kneib}, {Le
  F{\`e}vre}, {Bardelli}, {Bongiorno}, {Caputi}, {Coppa}, {de la Torre},
  {Franzetti}, {Garilli}, {Lamareille}, {Le Borgne}, {Le Brun}, {Mignoli},
  {Perez Montero}, {Pello}, {Ricciardelli}, {Tanaka}, {Tresse}, {Vergani},
  {Welikala}, {Zucca}, {Oesch}, {Abbas}, {Barnes}, {Bordoloi}, {Bottini},
  {Cappi}, {Cassata}, {Cimatti}, {Fumana}, {Hasinger}, {Koekemoer},
  {Leauthaud}, {Maccagni}, {Marinoni}, {McCracken}, {Memeo}, {Meneux}, {Nair},
  {Porciani}, {Presotto}, \& {Scaramella}}]{Peng2010}
{Peng}, Y.-j., {Lilly}, S.~J., {Kova{\v c}}, K., {et~al.} 2010, \apj, 721, 193

\bibitem[{{Pranger} {et~al.}(2014){Pranger}, {B{\"o}hm}, {Ferrari},
  {Maurogordato}, {Benoist}, {H{\"o}ller}, \& {Schindler}}]{Pranger2014}
{Pranger}, F., {B{\"o}hm}, A., {Ferrari}, C., {et~al.} 2014, \aap, 570, A40

\bibitem[{{Querejeta} {et~al.}(2015){Querejeta}, {Eliche-Moral}, {Tapia},
  {Borlaff}, {Rodr{\'{\i}}guez-P{\'e}rez}, {Zamorano}, \&
  {Gallego}}]{Querejeta2015}
{Querejeta}, M., {Eliche-Moral}, M.~C., {Tapia}, T., {et~al.} 2015, \aap, 573,
  A78

\bibitem[{{Read} {et~al.}(2010){Read}, {Hayfield}, \& {Agertz}}]{Read2010}
{Read}, J.~I., {Hayfield}, T., \& {Agertz}, O. 2010, \mnras, 405, 1513

\bibitem[{{Roediger} \& {Br{\"u}ggen}(2006)}]{Roediger2006a}
{Roediger}, E. \& {Br{\"u}ggen}, M. 2006, \mnras, 369, 567

\bibitem[{{Roediger} \& {Br{\"u}ggen}(2007)}]{Roediger2007}
{Roediger}, E. \& {Br{\"u}ggen}, M. 2007, \mnras, 380, 1399

\bibitem[{{Roediger} \& {Br{\"u}ggen}(2008)}]{Roediger2008}
{Roediger}, E. \& {Br{\"u}ggen}, M. 2008, \mnras, 388, 465

\bibitem[{{Roediger} {et~al.}(2006){Roediger}, {Br{\"u}ggen}, \&
  {Hoeft}}]{Roediger2006}
{Roediger}, E., {Br{\"u}ggen}, M., \& {Hoeft}, M. 2006, \mnras, 371, 609

\bibitem[{{Roediger} \& {Hensler}(2005)}]{Roediger2005}
{Roediger}, E. \& {Hensler}, G. 2005, \aap, 433, 875

\bibitem[{{Ruszkowski} {et~al.}(2014){Ruszkowski}, {Br{\"u}ggen}, {Lee}, \&
  {Shin}}]{Ruszkowski2014}
{Ruszkowski}, M., {Br{\"u}ggen}, M., {Lee}, D., \& {Shin}, M.-S. 2014, \apj,
  784, 75

\bibitem[{{Salpeter}(1955)}]{Salpeter1955}
{Salpeter}, E.~E. 1955, \apj, 121, 161

\bibitem[{{Schawinski} {et~al.}(2014){Schawinski}, {Urry}, {Simmons},
  {Fortson}, {Kaviraj}, {Keel}, {Lintott}, {Masters}, {Nichol}, {Sarzi},
  {Skibba}, {Treister}, {Willett}, {Wong}, \& {Yi}}]{Schawinski2014}
{Schawinski}, K., {Urry}, C.~M., {Simmons}, B.~D., {et~al.} 2014, \mnras, 440,
  889

\bibitem[{{Schindler}(2007)}]{Schindler2007}
{Schindler}, S. 2007, in EAS Publications Series, Vol.~24, EAS Publications
  Series, ed. E.~{Emsellem}, H.~{Wozniak}, G.~{Massacrier}, J.-F. {Gonzalez},
  J.~{Devriendt}, \& N.~{Champavert}, 121--132

\bibitem[{{Schindler} {et~al.}(2005){Schindler}, {Kapferer}, {Domainko},
  {Mair}, {van Kampen}, {Kronberger}, {Kimeswenger}, {Ruffert}, {Mangete}, \&
  {Breitschwerdt}}]{Schindler2005}
{Schindler}, S., {Kapferer}, W., {Domainko}, W., {et~al.} 2005, \aap, 435, L25

\bibitem[{{Scott} {et~al.}(2010){Scott}, {Bravo-Alfaro}, {Brinks}, {Caretta},
  {Cortese}, {Boselli}, {Hardcastle}, {Croston}, \& {Plauchu}}]{Scott2010}
{Scott}, T.~C., {Bravo-Alfaro}, H., {Brinks}, E., {et~al.} 2010, \mnras, 403,
  1175

\bibitem[{{Shu}(1992)}]{Shu1992}
{Shu}, F.~H. 1992, {Physics of Astrophysics, Vol. II} (University Science
  Books)

\bibitem[{{Springel}(2005)}]{Springel2005}
{Springel}, V. 2005, \mnras, 364, 1105

\bibitem[{{Springel}(2010)}]{Springel2010}
{Springel}, V. 2010, \mnras, 401, 791

\bibitem[{{Springel} {et~al.}(2005){Springel}, {Di Matteo}, \&
  {Hernquist}}]{Springel2005a}
{Springel}, V., {Di Matteo}, T., \& {Hernquist}, L. 2005, \mnras, 361, 776

\bibitem[{{Springel} \& {Hernquist}(2003)}]{Springel2003}
{Springel}, V. \& {Hernquist}, L. 2003, \mnras, 339, 289

\bibitem[{{Springel} \& {White}(1999)}]{Springel1999}
{Springel}, V. \& {White}, S.~D.~M. 1999, \mnras, 307, 162

\bibitem[{{Steinhauser} {et~al.}(2012){Steinhauser}, {Haider}, {Kapferer}, \&
  {Schindler}}]{Steinhauser2012}
{Steinhauser}, D., {Haider}, M., {Kapferer}, W., \& {Schindler}, S. 2012, \aap,
  544, A54

\bibitem[{{Sun}(2012)}]{Sun2012}
{Sun}, M. 2012, New Journal of Physics, 14, 045004

\bibitem[{{Tecce} {et~al.}(2010){Tecce}, {Cora}, {Tissera}, {Abadi}, \&
  {Lagos}}]{Tecce2010}
{Tecce}, T.~E., {Cora}, S.~A., {Tissera}, P.~B., {Abadi}, M.~G., \& {Lagos},
  C.~D.~P. 2010, \mnras, 408, 2008

\bibitem[{{Tonnesen} \& {Bryan}(2008)}]{Tonnesen2008}
{Tonnesen}, S. \& {Bryan}, G.~L. 2008, \apjl, 684, L9

\bibitem[{{Tonnesen} \& {Bryan}(2010)}]{Tonnesen2010}
{Tonnesen}, S. \& {Bryan}, G.~L. 2010, \apj, 709, 1203

\bibitem[{{Tonnesen} \& {Bryan}(2012)}]{Tonnesen2012}
{Tonnesen}, S. \& {Bryan}, G.~L. 2012, \mnras, 422, 1609

\bibitem[{{van Leer}(1984)}]{VanLeer1984}
{van Leer}, B. 1984, SIAM J.Sci.Stat. Comput., 5, 1

\bibitem[{{van Leer}(2006)}]{VanLeer2006}
{van Leer}, B. 2006, Communications in Computational Physics, 1, 192

\bibitem[{{Vazza}(2011)}]{Vazza2011}
{Vazza}, F. 2011, \mnras, 410, 461

\bibitem[{{Vogelsberger} {et~al.}(2012){Vogelsberger}, {Sijacki}, {Kere{\v s}},
  {Springel}, \& {Hernquist}}]{Vogelsberger2012}
{Vogelsberger}, M., {Sijacki}, D., {Kere{\v s}}, D., {Springel}, V., \&
  {Hernquist}, L. 2012, \mnras, 425, 3024

\bibitem[{{Vollmer} {et~al.}(2001{\natexlab{a}}){Vollmer}, {Braine},
  {Balkowski}, {Cayatte}, \& {Duschl}}]{Vollmer2001}
{Vollmer}, B., {Braine}, J., {Balkowski}, C., {Cayatte}, V., \& {Duschl}, W.~J.
  2001{\natexlab{a}}, \aap, 374, 824

\bibitem[{{Vollmer} {et~al.}(2001{\natexlab{b}}){Vollmer}, {Cayatte},
  {Balkowski}, \& {Duschl}}]{Vollmer2001a}
{Vollmer}, B., {Cayatte}, V., {Balkowski}, C., \& {Duschl}, W.~J.
  2001{\natexlab{b}}, \apj, 561, 708

\bibitem[{{Wang} {et~al.}(2014){Wang}, {Sales}, {Henriques}, \&
  {White}}]{Wang2014}
{Wang}, W., {Sales}, L.~V., {Henriques}, B.~M.~B., \& {White}, S.~D.~M. 2014,
  \mnras, 442, 1363

\bibitem[{{Weinmann} {et~al.}(2010){Weinmann}, {Kauffmann}, {von der Linden},
  \& {De Lucia}}]{Weinmann2010}
{Weinmann}, S.~M., {Kauffmann}, G., {von der Linden}, A., \& {De Lucia}, G.
  2010, \mnras, 406, 2249

\bibitem[{{Weinmann} {et~al.}(2006){Weinmann}, {van den Bosch}, {Yang}, \&
  {Mo}}]{Weinmann2006}
{Weinmann}, S.~M., {van den Bosch}, F.~C., {Yang}, X., \& {Mo}, H.~J. 2006,
  \mnras, 366, 2

\bibitem[{{Werner} {et~al.}(2008){Werner}, {Durret}, {Ohashi}, {Schindler}, \&
  {Wiersma}}]{Werner2008}
{Werner}, N., {Durret}, F., {Ohashi}, T., {Schindler}, S., \& {Wiersma},
  R.~P.~C. 2008, \ssr, 134, 337

\bibitem[{{Yagi} {et~al.}(2010){Yagi}, {Yoshida}, {Komiyama}, {Kashikawa},
  {Furusawa}, {Okamura}, {Graham}, {Miller}, {Carter}, {Mobasher}, \&
  {Jogee}}]{Yagi2010}
{Yagi}, M., {Yoshida}, M., {Komiyama}, Y., {et~al.} 2010, \aj, 140, 1814

\bibitem[{{Yoshida} {et~al.}(2008){Yoshida}, {Yagi}, {Komiyama}, {Furusawa},
  {Kashikawa}, {Koyama}, {Yamanoi}, {Hattori}, \& {Okamura}}]{Yoshida2008}
{Yoshida}, M., {Yagi}, M., {Komiyama}, Y., {et~al.} 2008, \apj, 688, 918

\bibitem[{{Yurin} \& {Springel}(2014)}]{Yurin2014}
{Yurin}, D. \& {Springel}, V. 2014, \mnras, 444, 62

\bibitem[{{Zaritsky} {et~al.}(1994){Zaritsky}, {Kennicutt}, \&
  {Huchra}}]{Zaritsky1994}
{Zaritsky}, D., {Kennicutt}, Jr., R.~C., \& {Huchra}, J.~P. 1994, \apj, 420, 87

\bibitem[{{Zhang} {et~al.}(2011){Zhang}, {Lagan{\'a}}, {Pierini}, {Puchwein},
  {Schneider}, \& {Reiprich}}]{Zhang2011}
{Zhang}, Y.-Y., {Lagan{\'a}}, T.~F., {Pierini}, D., {et~al.} 2011, \aap, 535,
  A78

\end{thebibliography}

\end{document}